\documentclass[12pt]{article}
\pdfoutput=1
\usepackage[a4paper]{geometry}
 
\usepackage[T1]{fontenc} 
\usepackage{comment}

\usepackage{jheppub,hyperref,float,array,adjustbox,mathtools,physics,xcolor}
\usepackage{lipsum}
\usepackage{booktabs}
\usepackage{pdflscape}
\usepackage{geometry}
\usepackage{fancyhdr}
\usepackage{changepage}
\usepackage[english]{babel}

\usepackage{xcolor,tikz,pgfplots,amsmath,amsfonts}
\usepackage{graphicx}
\usepackage[mathscr]{euscript}
\usetikzlibrary{matrix,calc,positioning,decorations.markings,decorations.pathmorphing,decorations.pathreplacing}
\usetikzlibrary{arrows,cd}
\usetikzlibrary{shapes.misc}
\usetikzlibrary {shapes.geometric}
\usetikzlibrary {shapes.multipart}
\usetikzlibrary{decorations.markings}
\usetikzlibrary{backgrounds}
\usetikzlibrary{shapes.gates.logic.US}
\usetikzlibrary{circuits.ee.IEC}
\usepackage{bbding}

\definecolor{terradisiena}{RGB}{233,116,81}
\definecolor{strisciadipietro}{RGB}{229,204,255}

\usepackage{tikz}
\usetikzlibrary{automata,positioning,calc}
\usetikzlibrary{decorations.markings}
\tikzset{->-/.style={decoration={markings, mark=at position #1 with {\arrow{>}}},postaction={decorate}}}
\tikzset{-<-/.style={decoration={markings, mark=at position #1 with {\arrow{<}}},postaction={decorate}}}
\tikzset{auto shift/.style={auto=right,->, to path={ let \p1=(\tikztostart),\p2=(\tikztotarget), \n1={atan2(\y2-\y1,\x2-\x1)},\n2={\n1+180} in ($(\tikztostart.{\n1})!1mm!270:(\tikztotarget.{\n2})$) -- ($(\tikztotarget.{\n2})!1mm!90:(\tikztostart.{\n1})$) \tikztonodes}}}

\usetikzlibrary {shapes.geometric} 

\newcommand{\Pf}[1]{\,\mathrm{Pf} #1}
\DeclareMathOperator{\USp}{USp}
\DeclareMathOperator{\SO}{SO}
\DeclareMathOperator{\SU}{SU}
\DeclareMathOperator{\UU}{U}

\newcommand{\mi}{\mathrm{i}}

\definecolor{SUcol}{RGB}{235,165,155}
\definecolor{USPcol}{RGB}{185,210,230}
\definecolor{SOcol}{RGB}{165,225,165}

\title{
\begin{center}
Six Easy Pieces: \\
\Large{interplays among dualities in 4d, 3d and 2d}
\end{center}
}

\author[a]{Antonio Amariti,}	
\author[a,b]{Pietro Glorioso,}	
\author[a,b]{Chiara Mascherpa,}
\author[a,b]{and Andrea Zanetti}

\affiliation[a]{INFN, Sezione di Milano, Via Celoria 16, I-20133 Milano, Italy}
\affiliation[b]{Dipartimento di Fisica, Università degli studi di Milano, Via Celoria 16, I-20133}

\emailAdd{antonio.amariti@mi.infn.it}
\emailAdd{pietro.glorioso@mi.infn.it}
\emailAdd{chiara.mascherpa@mi.infn.it}
\emailAdd{andrea.zanetti@mi.infn.it}

\abstract{
In this paper we consider 4d $\mathcal{N}=1$ $\mathrm{SU}(N)$ gauge theories with $N+1$ fundamentals, five antifundamentals and a conjugate two index antisymmetric tensor.
The model has been shown to be in a mixed phase in the IR, splitting in an interacting non-Abelian Coulomb phase and a free magnetic phase.
Through tensor deconfinement, we show that  baryonic deformations  lead to a non-Abelian free magnetic phase. Along the analysis we obtain a duality with symplectic SQCD  that can be further reduced to 3d and 2d. In the 3d case the analysis of the three sphere partition function allows one to obtain dualities between $\mathrm{SU}(N)$ with a two index symmetric tensor and $\mathrm{SO}(N)$ theories. On the other hand, in 2d we recover dualities already known in the literature and propose new ones between special unitary and symplectic gauge theories.  
}

\begin{document}
\maketitle
\flushbottom
\allowdisplaybreaks 

\section{Introduction}

In the last few years many new relations  among supersymmetric dualities with a low amount of supercharges and tensorial matter fields have been found in various dimensions \cite{Pasquetti:2019uop,Benvenuti:2020wpc,Etxebarria:2021lmq,Benvenuti:2021nwt,Bottini:2022vpy,Bajeot:2022lah,Bajeot:2022wmu,Amariti:2022wae,Amariti:2023wts,Amariti:2024sde,Jiang:2024ifv,Amariti:2024gco,Benvenuti:2024glr,Hwang:2024hhy,Amariti:2025jvi,Amariti:2025lem,Jia:2025koz} by applying the technique of tensor deconfinement, pioneered in \cite{Berkooz:1995km,Luty:1996cg,Pouliot:1995me}.
This approach has shown that many dualities (and their confining limits) actually descend from simpler building blocks, corresponding to  dualities between electric and magnetic phases in the absence of tensors. In four dimensions such building blocks correspond to the dualities originally worked out by Seiberg in \cite{Seiberg:1994pq} and Intriligator and Pouliot (IP) in \cite{Intriligator:1995ne}.

The tensor deconfinement approach treats a two index tensor of a gauge group $G_1$ as a composite of another gauge group $G_2$. 
This group $G_2$ is thought as strongly interacting and it is believed to 
s-confine at higher scale relative to the strong coupling scale of $G_1$.
Then at such scale the group $G_1$ is treated as a spectator of the dynamics, \emph{i.e.} as a flavor symmetry for $G_2$. Observe that, when $G_2$ s-confines, it carries a non-vanishing superpotential. If one needs to reproduce the original superpotential, which can either be vanishing or not, for the model with the single group $G_1$, then some care is necessary, and additional superpotential terms should be introduced in the $G_1 \times G_2$ theory\footnote{The process can in principle introduce also spurious symmetries. Here we will avoid this possibility and follow the general prescription worked out in \cite{Bajeot:2022kwt}.}.

At this point, we can focus on the weakly coupled $G_1$ gauge group in the $G_1 \times G_2$ quiver, which after deconfining the tensor corresponds to ordinary SQCD. This node undergoes an ordinary duality.
Various options are possible.
For example  $G_1$ is s-confining and one is left with a dual $G_2$ theory.
Such $G_2$ theory can either be a SQCD model, and can be further dualized, or have tensor matter, and one can try to deconfine the tensor in order to find another dual description.
Another possibility consists of obtaining a $\widetilde G_1 \times G_2$ dual quiver.  In this second case, one can iterate the procedure by dualizing the group $G_2$ finding a $\widetilde G_1 \times \tilde G_2$ quiver.
Remarkably, using this strategy all the s-confining dualities for $\SU(n)$ and $\USp(2n)$ known in the literature have been found to descend from the
s-confining limits of
$\SU(n)$ and $\USp(2n)$ SQCD.

This strategy has also allowed to obtain mixed phases for gauge theories with  two-index tensors. This has been shown to occur in $\SU(n)$ with a conjugate antisymmetric, five fundamentals and $2n+1$ antifundamentals.
The dual phase was originally found by 
\cite{Pouliot:1995me,Terning:1997jj} through tensor deconfinement and sequential dualities and it was then shown in \cite{Csaki:2004uj} that this model\footnote{To be more precise, the analysis of \cite{Csaki:2004uj} regards also cases with a larger amount of fundamentals, but here we focus on the minimal case. See also \cite{Barnes:2005zn} for another example of mixed phase.} is in a mixed phase. It means that the electric theory in the IR splits into an interacting non-Abelian Coulomb phase and a  free magnetic phase.
The claim of \cite{Csaki:2004uj}  was supported by a judicious application of $a$-maximization in presence of accidental symmetries \cite{Kutasov:2003iy} and deconfinement.

In \cite{Craig:2011tx,Craig:2011wj} a classically marginal deformation was considered in the UV picture and its effects on the low energy dynamics have been studied. Depending on the amount of fundamentals such deformation can lead to a runaway, give rise to a confining theory, keep the mixed phases or give origin to a non-Abelian Coulomb phase at the origin of moduli space.

In this paper we consider the models studied in \cite{Pouliot:1995me,Terning:1997jj} and deform them by adding baryonic superpotential  terms,  which are irrelevant in the UV but 
can trigger an RG flow to a consistent IR, in a way compatible with the constraints of the a-theorem. Such dangerously irrelevant \cite{Kutasov:1995ss} baryonic deformations 
lead to a non-Abelian free magnetic phase for generic values of the gauge rank.

Instead of considering the action of these deformation in the dual phase found by \cite{Pouliot:1995me,Terning:1997jj}, here we propose  \emph{ad hoc} deconfinements for the electric theory in presence of the deformation.
We show two alternative way to treat the theory in presence of the deformation.
In the first case, we deconfine the antisymmetric \emph{canonically}, \emph{i.e.} we exchange it with a symplectic gauge node. In the second case, we propose a less canonical deconfinement, which involves a fundamental field. Here, there is still an antisymmetric in the deconfined quiver, but it has been \emph{moved} on the new gauge group.

While the canonical deconfinement requires a separate treatment for the different baryonic deformations, the non-canonical one, when allowed, can be applied for any baryonic deformation under investigation.
In the canonical cases we obtain a dual symplectic IR free SQCD 
after dualizing the original gauge node. On the other hand, the same situation is reproduced by applying the non-canonical deconfinements, by studying the Higgs flow triggered by the baryonic deformation in the deconfined phase.
The dual pictures found in this way correspond to $\USp(2M)$ SQCD models, where the explicit value of $M$ depends on the baryonic deformation at hand. These  $\USp(2M)$ SQCD can be further dualized giving rise to IR free $\USp(2)$ gauge theories. 
These last theories coincide with the ones found by following the fate of the baryonic deformations in the duality found in \cite{Pouliot:1995me,Terning:1997jj}.

Even if it can be potentially interesting \emph{per-se}, the derivation spelled out here becomes crucial when we consider the $S^2$ reduction of these dualities, along the prescription of \cite{Gadde:2015wta}.
Indeed, in this case, it is not possible to find a reduction with non-negative integer $R$ charges in the $\SU(N)$ and in the $\USp(2)$ phase, while it is possible to reduce the duality between $\SU(N)$ and $\USp(2M)$.

We also study the reduction of the duality to 3d by circle compactification, following the ARSW \cite{Aharony:2013dha} prescription, by reducing the identities between the superconformal indices into identities between the squashed three sphere partition functions.  
We then further manipulate the latter by (using the terminology of \cite{Kim:2023qwh,Hayashi:2023boy,Kim:2024vci}) freezing opportunely some of the mass parameters, in order to convert the antisymmetric into symmetric tensors. We then apply the duplication formula, converting the $\SU(N)$ antisymmetric into symmetric tensors and the $\USp(2M)$ into $\SO(K)$ gauge groups. Observe that the latter theories can be further dualized and they give rise to 3d $\SO(3)$ models with vectors, consistently with the reduction of the 4d $\USp(2)$ dual phases and application of the freezing and of the duplication formula to such dual phases.

The paper is organized as follows.
In Section \ref{sec:4d} we study the 4d dualities by adding the baryonic deformations in the $\SU(N)$ case,  distinguishing the even and odd rank cases. Once the deformations are considered, we deconfine the conjugate antisymmetric either in terms of a symplectic  or of a special unitary gauge group.
In each case we arrive to a dual symplectic SQCD model and we check the results  through the analysis of the superconformal index.
We then compare the results with the ones found  in \cite{Pouliot:1995me,Terning:1997jj} and we find that they agree providing that we further dualize the symplectic gauge theories to $\USp(2)$. This is interpreted as the fact that the flow triggered by the baryonic deformations transform the original mixed phases into magnetic free phases.

If we consider the symplectic SQCD duals in the strongly coupled phases we can further reduce to 2d, as explained in Section \ref{sec:2d_dualities} by a twisted compactification on $S^2$, where the twists are along a choice of  integer non-negative $R$ charges.
We consider extensively the case of $\SU(2n)$ with a specific baryonic deformation. In this case 
we perform a detailed analysis, finding all the possible consistent reductions for generic gauge rank (\emph{i.e.} we do not look at possible low rank sporadic cases).
We observe that while some of the 2d dualities obtained through this procedure  have  already been  derived  
from other 4d setups, other dualities are new and they can be derived only in presence of the baryonic deformations in the 4d parents.

In Section \ref{sec:2d_dualities} we study the reduction of the 4d dualities to 3d, focusing on the three sphere partition function. In such case the 4d identities translate into 3d identities where the mass parameters are constrained by the presence of monopole superpotentials. These identities can be further manipulated by freezing some masses to fixed values, compatibly with the constraints imposed by the monopoles. By manipulating the identities with the duplication formula for the one-loop determinant, we convert the antisymmetric into a symmetric tensor. 
The new identities suggest the existence of new dualities with the new field content, between special unitary SQCD with a symmetric tensor and 3d orthogonal SQCD. 
We then provide a proof of these new 3d dualities  via tensor deconfinement.
In Section \ref{Sec:Conclusions} we conclude by summarizing our findings and proposing further directions of research.

For the sake of clarity we collect the main dualities discussed in this work in the tables below. 
We refer the reader to the specific sections for the corresponding explicit discussions.
The 2d case requires more care because the details of the dualities rely on the matter content that survives after the dimensional reduction from the 4d case.
\begin{equation*}
\label{summaryintro}
\begin{aligned}
&
\begin{array}{|c|c|c|}
    \multicolumn{3}{c}{\text{dim = 4\,,\quad}\eta=0,1,2} \\
    \hline
    \text{sec} & G_{\text{ele}} & G_{\text{mag}} \\
    \hline
    \hline
    \ref{su2nsub} & \SU(2n) & \USp(2n-4+2\eta) \\
    \hline
    \ref{sec:4dodd} & \SU(2n+1) & \USp(2n-2+2\eta) \\
    \hline
\end{array}
\\[1.2em]
&
\begin{array}{|c|c|c|}
    \multicolumn{3}{c}{\text{dim}=2} \\
    \hline
    \text{sec} & G_{\text{ele}} & G_{\text{mag}} \\
    \hline
    \hline
    \ref{subsec:2dXX}-\ref{subsec:2dVV} & \SU(2n) & \USp(2n-2) \text{ or LG} \\
    \hline
\end{array}
\end{aligned}
\qquad
\begin{array}{|c|c|c|}
    \multicolumn{3}{c}{\text{dim}=3} \\
    \hline
    \text{sec} & G_{\text{ele}} & G_{\text{mag}} \\
    \hline
    \hline
    \ref{detSeven} & \SU(2n) & \SO(2n-1) \\
    \hline
    \ref{detSodd} & \SU(2n+1) & \SO(2n) \\
    \hline
    \ref{S2nm1even} & \SU(2n) & \SO(2n) \\
    \hline
    \ref{S2nodd} & \SU(2n+1) & \SO(2n+1) \\
    \hline
    \ref{S2nm2even} & \SU(2n) & \SO(2n+1) \\
    \hline
    \ref{S2nm1odd} & \SU(2n+1) & \SO(2n+2) \\
    \hline
\end{array}
\end{equation*}

\section{4d SU/USp dualities}
\label{sec:4d}

In this section we study 4d dualities between $\SU(N)$ and $\USp(2M)$ gauge theories.
The electric gauge theories have a conjugate antisymmetric tensor, $N+1$ fundamentals and five antifundamentals. In the absence of superpotentials these models have an \mbox{$\SU(2) \times \SU(2)$} 
dual quiver description, obtained using tensor deconfinement in \cite{Pouliot:1995me,Terning:1997jj}.

However, as originally observed in \cite{Terning:1997jj} and then largely studied in \cite{Csaki:2004uj}, the duality is neither a standard duality in the conformal window nor a 
duality between an UV free and an IR free model. The electric theory is 
in a mixed phase, splitting into an interacting non-Abelian Coulomb phase and a  free magnetic phase in the IR.
 
Starting from such a duality, more standard dualities have been obtained (with a larger amount of flavors) by studying the effect of a classically marginal UV deformation, which in the flow through the IR can trigger an RG flow.

Here we study the $\SU(N)$ model by adding a different type of deformation, generically irrelevant in the UV. We provide arguments suggesting that this deformation is actually dangerously irrelevant and leads to a standard duality between the original $\SU(N)$ model and an IR free $\SU(2)$ gauge theory.

Additionally, our derivation of the duality isolates yet another  interacting non-Abelian Coulomb phase, corresponding to $\USp(2M)$ with $2M+6$ fundamentals, in addition fo flippers,  where the details on the dual rank $M$ depend on the baryonic deformation added to the original $\SU(N)$ model.
Such dual theory is crucial in the analysis of the reduction to 2d as we will discuss below.

The analysis below distinguishes the case of even and odd $N$. Furthermore for each deformation we will provide two alternative ways (when possible) to study the duality using an auxiliary gauge group. In one case we will deconfine the conjugate antisymmetric through a confining symplectic gauge group, while in the second case we will follow a different procedure, where we will not deconfine any tensor. The auxiliary gauge group in this case corresponds to a flipped version of special unitary confining SQCD, and the original $\SU(N)$ model with the conjugate antisymmetric becomes confining in this quiver description. After confining the original gauge group we observe that the fate of each  baryonic deformation forces an Higgs flow, which gives rise to the expected $\USp(2M)$ with $2M+6$ fundamentals dual description. 
 A further IP duality gives the final $\SU(2)$ gauge theory that one would have obtained directly from the duality of \cite{Pouliot:1995me,Terning:1997jj} in presence of the baryonic deformation.

\subsection{The case of $\SU(2n)$}
\label{su2nsub}
In this case, we distinguish three types of baryonic deformations.
\begin{enumerate}
\item The first deformation corresponds to the Pfaffian operator for the conjugate antisymmetric $\tilde A$ and, in order to use a uniform notation, we denote the superpotential in this case as
  \begin{equation}
    \label{WEevenDef1}
    W_{\text{ele}} =  \tilde{A}^{n} \,,
\end{equation}
where the antisymmetric $\SU(N)$ contractions are understood. In the following the contractions will always be understood unless ambiguities are present.
\item The second baryonic deformation considered in this section is
 \begin{equation}
    \label{WEevenDef2}
    W_{\text{ele}} =  \tilde{A}^{n-1} \tilde{Q}_1 \tilde{Q}_2 \,,
\end{equation}
where we choose arbitarily the two antifundamentals involved in the deformation.
\item The last possible baryonic deformation considered here is
\begin{equation}
    \label{WEevenDef3}
    W_{\text{ele}} =\tilde{A}^{n-2} \tilde{Q}_1 \tilde{Q}_2 \tilde{Q}_3 \tilde{Q}_4\,.
\end{equation}
\end{enumerate}

\subsubsection{Deconfining the conjugate antisymmetric with a symplectic node}

\begin{figure}
    \centering
    \begin{minipage}[b]{0.30\linewidth}
        \centering
        \makebox[\textwidth][c]{
        \begin{tikzpicture}[
            every node/.style={font=\footnotesize},
            box/.style={rectangle, draw, thick},
        ]
        \pgfmathsetmacro{\x}{1}
        \pgfmathsetmacro{\y}{1.5}
        \pgfmathsetmacro{\z}{1.2}
        \pgfmathsetmacro{\delta}{0.28}
        \node[fill=SUcol,circle,draw,thick] (SUgauge) at (0, 0) {$2n$};
        \node[box] (fond) at (\x,-\y) {$2n\!+\!1$};
        \node[box] (afond) at (-\x,-\y) {$5$};
        \draw[->, thick, >=stealth] (SUgauge) -- node[right] {$Q$} (fond.north);
        \draw[<-, thick, >=stealth] (SUgauge) -- node[left] {$\tilde{Q}$} (afond.north);
        \node[box, minimum size=0.2cm] (square1) at (0, \z) {};
        \node[box, minimum size=0.2cm] (square2) at (0, \z+\delta) {};
        \draw[->, thick, >=stealth] (square1) -- node[right] {$\tilde{A}$} (SUgauge);
        \end{tikzpicture}
        }
    \end{minipage}
    \begin{minipage}[b]{0.30\linewidth}
        \centering
        \makebox[\textwidth][c]{
        \begin{tikzpicture}[
            every node/.style={font=\footnotesize},
            box/.style={rectangle, draw, thick},
        ]
        \pgfmathsetmacro{\x}{1}
        \pgfmathsetmacro{\y}{1.5}
        \pgfmathsetmacro{\z}{1.8}
        \node[fill=USPcol,circle,draw,thick] (USPgauge) at (0, \z) {$2n\!-\!4$};
        \node[fill=SUcol,circle,draw,thick] (SUgauge) at (0, 0) {$2n$};
        \node[box] (fond) at (\x,-\y) {$2n\!+\!1$};
        \node[box] (afond) at (-\x,-\y) {$5$};
        \draw[->, thick, >=stealth] (SUgauge) -- node[right] {$Q$} (fond.north);
        \draw[<-, thick, >=stealth] (SUgauge) -- node[left] {$\tilde{Q}$} (afond.north);
        \draw[->, thick, >=stealth] (USPgauge) -- node[right] {$\tilde{P}$} (SUgauge);
        \end{tikzpicture}
        }
    \end{minipage}
    \begin{minipage}[b]{0.30\linewidth}
        \centering
        \makebox[\textwidth][c]{
        \begin{tikzpicture}[
            every node/.style={font=\footnotesize},
            box/.style={rectangle, draw, thick},
        ]
        \pgfmathsetmacro{\x}{1}
        \pgfmathsetmacro{\y}{1.5}
        \pgfmathsetmacro{\w}{1}
        \pgfmathsetmacro{\z}{1.8}
        \node[fill=USPcol,circle,draw,thick] (USPgauge) at (\x, \w) {$2n\!-\!4$};
        \node[box] (fond) at (\x,-\y) {$2n\!+\!1$};
        \node[box] (afond) at (-\x,-\y) {$5$};
        \node[box] (sing) at (-\x,\w) {$1$};
        \draw[->, thick, >=stealth] (USPgauge) -- node[right] {$M_2$} (fond.north);
        \draw[<-, thick, >=stealth] (USPgauge) -- node[above] {$\tilde{B}_2$} (sing);
        \draw[->, thick, >=stealth] (sing) -- node[left] {$\tilde{B}_1$} (afond);
        \draw[->, thick, >=stealth] (fond) -- node[above, xshift=4pt] {$B$} (sing.south east);
        \draw[<-, thick, >=stealth] (fond) -- node[above] {$M_1$} (afond);
        \end{tikzpicture}
        }
    \end{minipage}
    \caption{Quiver representation of the deconfinement of the conjugate antisymmetric in the presence of  the deformation $W_{\text{ele}}=\tilde{A}^n$, using an $\USp(2n-4)$ gauge group, followed by an ordinary Seiberg duality for the $\SU(2n)$ gauge node.}
    \label{UspDecAn}
\end{figure}
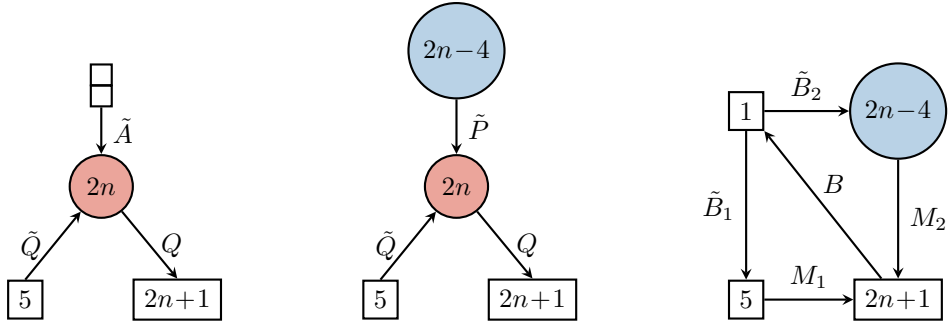

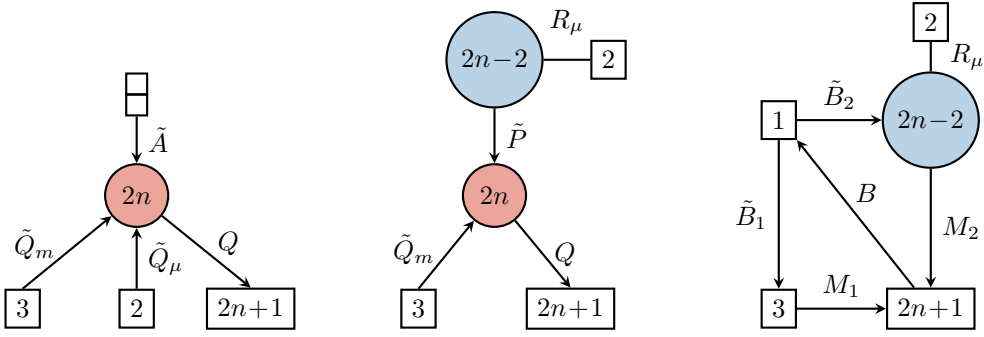
\begin{figure}
    \centering
    \begin{minipage}[b]{0.30\linewidth}
        \centering
        \makebox[\textwidth][c]{
        \begin{tikzpicture}[
            every node/.style={font=\footnotesize},
            box/.style={rectangle, draw, thick},
        ]
        \pgfmathsetmacro{\x}{1.5}
        \pgfmathsetmacro{\y}{1.5}
        \pgfmathsetmacro{\z}{1.2}
        \pgfmathsetmacro{\delta}{0.28}
        \node[fill=SUcol,circle,draw,thick] (SUgauge) at (0, 0) {$2n$};
        \node[box] (fond) at (\x,-\y) {$2n\!+\!1$};
        \node[box] (afond) at (-\x,-\y) {$3$};
        \node[box] (afond2) at (0,-\y) {$2$};
        \draw[->, thick, >=stealth] (SUgauge) -- node[right, yshift=3pt] {$Q$} (fond.north);
        \draw[<-, thick, >=stealth] (SUgauge) -- node[left, yshift=3pt] {$\tilde{Q}_{m}$} (afond.north);
        \draw[<-, thick, >=stealth] (SUgauge) -- node[right] {$\tilde{Q}_{\mu}$} (afond2.north);
        \node[box, minimum size=0.2cm] (square1) at (0, \z) {};
        \node[box, minimum size=0.2cm] (square2) at (0, \z+\delta) {};
        \draw[->, thick, >=stealth] (square1) -- node[right] {$\tilde{A}$} (SUgauge);
        \end{tikzpicture}
        }
    \end{minipage}
    \begin{minipage}[b]{0.30\linewidth}
        \centering
        \makebox[\textwidth][c]{
        \begin{tikzpicture}[
            every node/.style={font=\footnotesize},
            box/.style={rectangle, draw, thick},
        ]
        \pgfmathsetmacro{\x}{1}
        \pgfmathsetmacro{\y}{1.5}
        \pgfmathsetmacro{\z}{1.8}
        \node[fill=USPcol,circle,draw,thick] (USPgauge) at (0, \z) {$2n\!-\!2$};
        \node[fill=SUcol,circle,draw,thick] (SUgauge) at (0, 0) {$2n$};
        \node[box] (fond) at (\x,-\y) {$2n\!+\!1$};
        \node[box] (afond) at (-\x,-\y) {$3$};
        \node[box] (afond2) at (\y,\z) {$2$};
        \draw[->, thick, >=stealth] (SUgauge) -- node[right] {$Q$} (fond.north);
        \draw[<-, thick, >=stealth] (SUgauge) -- node[left,yshift=2pt] {$\tilde{Q}_m$} (afond.north);
        \draw[->, thick, >=stealth] (USPgauge) -- node[right] {$\tilde{P}$} (SUgauge);
        \draw[-, thick, >=stealth] (USPgauge) -- node[above,yshift=5pt] {$R_{\mu}$} (afond2);
        \end{tikzpicture}
        }
    \end{minipage}
    \begin{minipage}[b]{0.30\linewidth}
        \centering
        \makebox[\textwidth][c]{
        \begin{tikzpicture}[
            every node/.style={font=\footnotesize},
            box/.style={rectangle, draw, thick},
        ]
        \pgfmathsetmacro{\x}{1}
        \pgfmathsetmacro{\y}{1.5}
        \pgfmathsetmacro{\w}{1}
        \pgfmathsetmacro{\z}{1.8}
        \node[fill=USPcol,circle,draw,thick] (USPgauge) at (\x, \w) {$2n\!-\!2$};
        \node[box] (fond) at (\x,-\y) {$2n\!+\!1$};
        \node[box] (afond) at (-\x,-\y) {$3$};
        \node[box] (sing) at (-\x,\w) {$1$};
        \node[box] (afond2) at (\x,\z+\x/2) {$2$};
        \draw[->, thick, >=stealth] (USPgauge) -- node[right] {$M_2$} (fond.north);
        \draw[<-, thick, >=stealth] (USPgauge) -- node[above] {$\tilde{B}_2$} (sing);
        \draw[->, thick, >=stealth] (sing) -- node[left] {$\tilde{B}_1$} (afond);
        \draw[->, thick, >=stealth] (fond) -- node[above, xshift=4pt] {$B$} (sing.south east);
        \draw[<-, thick, >=stealth] (fond) -- node[above] {$M_1$} (afond);
        \draw[-, thick, >=stealth] (USPgauge) -- node[right,xshift=3pt] {$R_\mu$} (afond2);
        \end{tikzpicture}
        }
    \end{minipage}
         \caption{Quiver representation of the deconfinement of the conjugate antisymmetric in the presence of  the deformation $W_{\text{ele}}=\tilde{A}^{n-1}\tilde{Q}_1\tilde{Q}_2$, using an $\USp(2n-2)$ gauge group, followed by an ordinary Seiberg duality for the $\SU(2n)$ gauge node.}
    \label{UspDecAnQ2}
\end{figure}

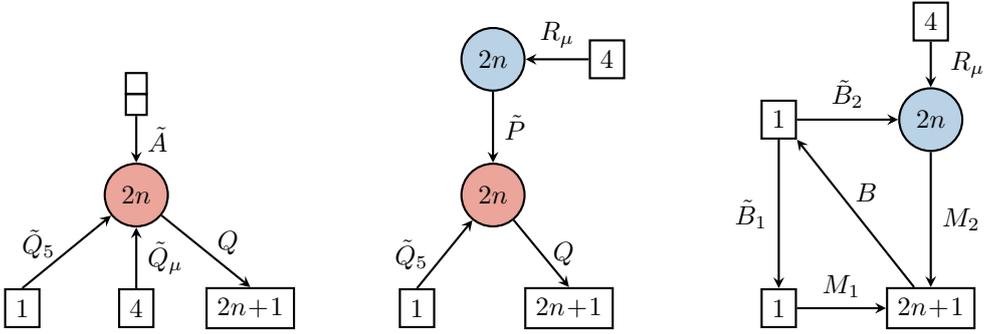
\begin{figure}
    \centering
    \begin{minipage}[b]{0.30\linewidth}
        \centering
        \makebox[\textwidth][c]{
        \begin{tikzpicture}[
            every node/.style={font=\footnotesize},
            box/.style={rectangle, draw, thick},
        ]
        \pgfmathsetmacro{\x}{1.5}
        \pgfmathsetmacro{\y}{1.5}
        \pgfmathsetmacro{\z}{1.2}
        \pgfmathsetmacro{\delta}{0.28}
        \node[fill=SUcol,circle,draw,thick] (SUgauge) at (0, 0) {$2n$};
        \node[box] (fond) at (\x,-\y) {$2n\!+\!1$};
        \node[box] (afond) at (-\x,-\y) {$1$};
        \node[box] (afond2) at (0,-\y) {$4$};
        \draw[->, thick, >=stealth] (SUgauge) -- node[right, yshift=3pt] {$Q$} (fond.north);
        \draw[<-, thick, >=stealth] (SUgauge) -- node[left, yshift=3pt] {$\tilde{Q}_5$} (afond.north);
        \draw[<-, thick, >=stealth] (SUgauge) -- node[right] {$\tilde{Q}_\mu$} (afond2.north);
        \node[box, minimum size=0.2cm] (square1) at (0, \z) {};
        \node[box, minimum size=0.2cm] (square2) at (0, \z+\delta) {};
        \draw[->, thick, >=stealth] (square1) -- node[right] {$\tilde{A}$} (SUgauge);
        \end{tikzpicture}
        }
    \end{minipage}
    \begin{minipage}[b]{0.30\linewidth}
        \centering
        \makebox[\textwidth][c]{
        \begin{tikzpicture}[
            every node/.style={font=\footnotesize},
            box/.style={rectangle, draw, thick},
        ]
        \pgfmathsetmacro{\x}{1}
        \pgfmathsetmacro{\y}{1.5}
        \pgfmathsetmacro{\z}{1.8}
        \pgfmathsetmacro{\w}{0.5}
        \pgfmathsetmacro{\delta}{0.28}
        \node[fill=USPcol,circle,draw,thick] (USPgauge) at (0, \z) {$2n$};
        \node[fill=SUcol,circle,draw,thick] (SUgauge) at (0, 0) {$2n$};
        \node[box] (fond) at (\x,-\y) {$2n\!+\!1$};
        \node[box] (afond) at (-\x,-\y) {$1$};
        \node[box] (afond2) at (\y,\z) {$4$};
        \draw[->, thick, >=stealth] (SUgauge) -- node[right] {$Q$} (fond.north);
        \draw[<-, thick, >=stealth] (SUgauge) -- node[left,xshift=-2pt] {$\tilde{Q}_5$} (afond.north);
        \draw[->, thick, >=stealth] (USPgauge) -- node[right] {$\tilde{P}$} (SUgauge);
        \draw[<-, thick, >=stealth] (USPgauge) -- node[above] {$R_\mu$} (afond2);
        \end{tikzpicture}
        }
    \end{minipage}
    \begin{minipage}[b]{0.30\linewidth}
        \centering
        \makebox[\textwidth][c]{
        \begin{tikzpicture}[
            every node/.style={font=\footnotesize},
            box/.style={rectangle, draw, thick},
        ]
        \pgfmathsetmacro{\delta}{0.28}
        \pgfmathsetmacro{\x}{1}
        \pgfmathsetmacro{\y}{1.5}
        \pgfmathsetmacro{\w}{0.5}
        \pgfmathsetmacro{\z}{1.8}
        \pgfmathsetmacro{\h}{2.17}
        \node[fill=USPcol,circle,draw,thick] (USPgauge) at (\x, \x) {$2n$};
        \node[box] (fond) at (\x,-\y) {$2n\!+\!1$};
        \node[box] (afond) at (-\x,-\y) {$1$};
        \node[box] (sing) at (-\x,\x) {$1$};
        \node[box] (afond2) at (\x,\z+\x/2) {$4$};
        \draw[->, thick, >=stealth] (USPgauge) -- node[right] {$M_2$} (fond.north);
        \draw[<-, thick, >=stealth] (USPgauge) -- node[above] {$\tilde{B}_2$} (sing);
        \draw[->, thick, >=stealth] (sing) -- node[left] {$\tilde{B}_1$} (afond);
        \draw[->, thick, >=stealth] (fond) -- node[above, xshift=4pt] {$B$} (sing.south east);
        \draw[<-, thick, >=stealth] (fond) -- node[above] {$M_1$} (afond);
        \draw[<-, thick, >=stealth] (USPgauge) -- node[right,xshift=3pt] {$R_\mu$} (afond2);
        \end{tikzpicture}
        }
    \end{minipage}
     \caption{Quiver representation of the deconfinement of the conjugate antisymmetric in the presence of  the deformation $W_{\text{ele}}=\tilde{A}^{n-2}\tilde{Q}_1\tilde{Q}_2\tilde{Q}_3\tilde{Q}_4$, using an $\USp(2n)$ gauge group, followed by an ordinary Seiberg duality for the $\SU(2n)$ gauge node.}
    \label{UspDecAnQ4}
\end{figure}

In the following, we study the $\SU(2n)$ model in presence of the deformations 
(\ref{WEevenDef1}), (\ref{WEevenDef2}) and (\ref{WEevenDef3})
by deconfining the conjugate antisymmetric with an $\USp(2m)$ gauge group, where $m=n-2$,
$m=n-1$ and $m=n$ respectively.
The quivers obtained after deconfining the conjugate antisymmetric correspond to the second ones  in Figure 
\ref{UspDecAn}, \ref{UspDecAnQ2} and \ref{UspDecAnQ4} respectively.
The superpotential is vanishing in the first case, while it is given by $W = \sigma R^2$ in the second and in the third case. The difference between such two cases is that the field $\sigma$ is a singlet when we consider the  deformation (\ref{WEevenDef2}), while it is in the antisymmetric representation of the leftover $\SU(4)$ flavor symmetry group when we choose the 
deformation (\ref{WEevenDef3}).
 In each case, we can confine the $\SU(2n)$ gauge theory, because it represents a flipped version of the confining duality for $\SU(2n)$ with $2n+1$ flavors.

The quivers obtained after confining the $\SU(2n)$ nodes correspond to the third ones  in Figure 
\ref{UspDecAn}, \ref{UspDecAnQ2} and \ref{UspDecAnQ4} respectively.
The  superpotentials for these dual theories, obtained after integrating out possible massive deformations, are
\begin{eqnarray}
\label{Wmag1}
W_{\text{mag}}^{(1)} &=&     \det \left(M_1 \,|\, M_2\right) + 
    B \left(M_1 \tilde{B}_1 + M_2 \tilde{B}_2\right),
\\
\label{Wmag2}
W_{\text{mag}}^{(2)} &=& \sigma R_1 R_2 + 
    \det \left(M_1 \,|\, M_2\right) + 
    B \left(M_1 \tilde{B}_1 + M_2 \tilde{B}_2\right),
\\
\label{Wmag3}
W_{\text{mag}}^{(3)} &=&  \sigma_{\mu \nu} R_\mu R_\nu + 
    \det \left(M_1 \,|\, M_2\right) + 
    B \left(M_1 \tilde{B}_1 + M_2 \tilde{B}_2\right),
\end{eqnarray}
where the superscript refers to the numeration of the electric baryonic superpotential deformation, and for shortness we introduced the notation $\det\left(\cdot\vert\cdot\right)$ to indicate a determinant of two-block matrix.
The duality dictionary for these three dualities is specified below.

In each case we can further dualize the models, using the rules of the IP dualities, obtaining an $\USp(2)$ gauge theory with $2m+6$ fundamentals, where $m$ has been specified above. 
It is also possible to study the integral identities associated to the superconformal indices. The derivation is straightforward and we skip the details.

At the level of the superconformal index, we can study the duality as an integral identity 
between the electric and magnetic  theories.
We denote  as $a^2$ the fugacity of the conjugate antisymmetric $\tilde A$, 
$v_{i=1,\dots,5}$ the fugacities of the antifundamentals $\tilde Q$ and $x_{\ell=1,\dots,2n+1}^{-1}$ the fugacities of the fundamentals $Q$.
The requirement of anomaly cancellation reflects into the balancing condition 
\begin{equation}
\label{BC1}
a^{2(2n-2)} \prod_{\ell=1}^{2n+1} x_{\ell}^{-1} \prod_{a=1}^5 v_a = (pq)^2\,.
\end{equation}
The superpotential constraint, on the other hand, fixes a second constraint, different in each case.
\begin{itemize}
\item If we consider the deformation (\ref{WEevenDef1}), the constraint between the fugacities is 
$a^{2n} = pq$.
In this case, the fundamentals $\tilde B_2$ and $M_2$ of the dual $\USp(2n-4)$ model have fugacities  
$a^{2n-5} \prod_{i=1}^5 v_i$ and $a x_{\ell}^{-1}$, respectively. The fugacities of the other singlets 
$B=Q^{2n}$, $M_1 =  Q \tilde Q$ and $\tilde B_1 = \tilde A^{n-1} \tilde Q^4$ are obtained from the duality map.
We summarize in the table below a charge assignation for the electric and the magnetic theories. Notice that in what follows we assigned a trial non-anomalous $R$ charge.
We will focus on the exact $R$ charges in Sec. \ref{amaxculo}.
\begin{equation}
    \begin{array}{c|c|c|c|c|c|}
    &\SU(2n)&\SU(2n+1)&\SU(5)&\UU(1)_{1}&\UU(1)_{R}\\
    \hline
    Q&\square&\overline{\square}&\cdot&1&\frac{2}{n(n+3)}\\
    \tilde Q&\overline{\square}&\cdot&\square&-\frac{2n+1}{5}&\frac{2}{n(n+3)}\\
    \tilde A&\begin{array}{c}\square \vspace{-2.9 mm} \\\square \end{array} &\cdot&\cdot&0&\frac{2}{n}
    \end{array}
\end{equation}
\begin{equation}
    \begin{array}{c|c|c|c|c|c|}
    &\USp(2n-4)&\SU(2n+1)&\SU(5)&\UU(1)_{1}&\UU(1)_{R}\\
    \hline
    M_1&\cdot&\overline{\square}&\square&-\frac{2(n-2)}{5}&\frac{4}{n(n+3)}\\
    M_2&\square&\overline{\square}&\cdot&1&\frac{n+5}{n(n+3)}\\
    B&\cdot&\square&\cdot&2n&\frac{4}{n(n+3)}\\
    \tilde B_1&\cdot&\cdot&\overline{\square}&-\frac{4(2n+1)}{5}&\frac{2(n^2+n-2)}{n+3}\\
    \tilde B_2&\square&\cdot&\cdot&-2n-1&\frac{2n^2+n-5}{n(n+3)}
    \end{array}
\end{equation}
\item If we consider the deformation (\ref{WEevenDef2}), the constraint between the fugacities is  
$ a^{2n-2} v_1 v_2 = pq$.
The fundamentals $\tilde B_2$, $M_2$ and $R$ of the dual $\USp(2n-2)$ model have fugacities  
$a^{2n-5}  v_3 v_4 v_5$, $a x_{\ell}^{-1}$ and $a^{-1} v_{1,2}$, respectively. 
The fugacities of the other singlets 
$B=Q^{2n}$, $M_1 =  Q \tilde Q_a$ (with $3\leq a \leq 5$) and $\tilde B_1 = \tilde A^{n-1} \tilde Q_a \tilde Q_b $ (with $3\leq a <b \leq 5$) are obtained from the duality map.
On the other hand, the singlet $\sigma$ is dual to the operator $\Pf \, \tilde A$. This can be shown by confining back the conjugate antisymmetric and evaluating the equations of motion for the massive fields. The fugacity of $\sigma$ is then  $a^{2n}$ or, equivalently, $pq a^2 v_1^{-1} v_2^{-1}$.
We summarize in the table below a charge assignation for the electric and the magnetic theories.
\begin{equation}
    \begin{array}{c|c|c|c|c|c|c|c|}
    &\SU(2n)&\SU(2n+1)&\SU(3)&\SU(2)&\UU(1)_{1}&\UU(1)_{2}&\UU(1)_{R}\\
    \hline
    Q&\square&\overline{\square}&\cdot&\cdot&1&0&\frac{2}{n(n+3)}\\
    \tilde Q_m&\overline{\square}&\cdot&\square&\cdot&0&1&\frac{2}{n(n+3)}\\
    \tilde Q_\mu&\overline{\square}&\cdot&\dot&\square&\frac{2n+1}{2}&\frac{3}{2}&\frac{2(n+2)}{n(n+3)}\\
    \tilde A&\begin{array}{c}\square \vspace{-2.9 mm} \\\square \end{array} &\cdot&\cdot&\cdot&-\frac{2n+1}{n-1}&-\frac{3}{n-1}&\frac{2(n^2+n-4)}{n(n^2+2n-3)}
    \end{array}
\end{equation}
\begin{equation}
    \begin{array}{c|c|c|c|c|c|c|c|}
    &\USp(2n-2)&\SU(2n+1)&\SU(3)&\SU(2)&\UU(1)_{1}&\UU(1)_{2}&\UU(1)_{R}\\
    \hline
    M_1&\cdot&\overline{\square}&\square&\cdot&1&1&\frac{4}{n(n+3)}\\
    M_2&\square&\overline{\square}&\cdot&\cdot&-\frac{3}{2(n-1)}&-\frac{3}{2(n-1)}&\frac{n(n+3)-6}{n(n+3)(n-1)}\\
    B&\cdot&\square&\cdot&\cdot&2n&0&\frac{4}{n+3}\\
    \tilde B_1&\cdot&\cdot&\overline{\square}&\cdot&-2n-1&-1&\frac{2(n^2+n-2)}{n(n+3)}\\
    \tilde B_2&\square&\cdot&\cdot&\cdot&\frac{3}{2(n-1)}&\frac{3}{2(n-1)}& \frac{2 n^3-n^2-5 n+6}{n (n+3)(n-1) }\\
    R_\mu&\square&\cdot&\cdot&\square&\frac{n(2n+1)}{2(n-1)}&\frac{3n}{2(n-1)}&\frac{n+1}{n^2+2n-3}
    \end{array}
\end{equation}
\item If we consider the deformation (\ref{WEevenDef3}), the constraint between the fugacities is
$a^{2n-4} v_1 v_2 v_3 v_4 = pq$.
The fundamentals $\tilde B_2$, $M_2$ and $R$ of the dual $\USp(2n)$ model have fugacities
$a^{2n-5}  v_3 v_4 v_5$, $a x_{\ell}^{-1}$ and $a^{-1} v_{1,2,3,4}$, respectively. 
The fugacities of the other singlets 
$B=Q^{2n}$, $M_1 =  Q \tilde Q_5$ and $\tilde B_1 = \tilde A^{n} $  are obtained from the duality map.
On the other hand, the singlet $\sigma$ is dual to the operator $\tilde A^{n-1} Q_a Q_b$ with $1\leq a < b \leq 4$. This can be shown by confining back the conjugate antisymmetric and evaluating the equations of motion for the massive fields. The fugacity of $\sigma$ is then  $a^{2n-2} v_a v_b$ or, equivalently, $pq a^2 v_a^{-1} v_b^{-1}$.
We summarize in the table below a charge assignation for the electric and the magnetic theories.
\begin{equation}
    \begin{array}{c|c|c|c|c|c|c|}
    &\SU(2n)&\SU(2n+1)&\SU(4)&\UU(1)_{1}&\UU(1)_{2}&\UU(1)_{R}\\
    \hline
    Q&\square&\overline{\square}&\cdot&1&0&\frac{2}{n(n+3)}\\
    \tilde Q_5&\overline{\square}&\cdot&\cdot&0&1&\frac{2}{n(n+3)}\\
    \tilde Q_\mu&\overline{\square}&\dot&\square&\frac{(2n+1)}{4}&\frac{1}{4}&\frac{2(n^2+n-1)}{n^2(n+3)}\\
    \tilde A&\begin{array}{c}\square \vspace{-2.9 mm} \\\square \end{array} &\cdot&\cdot&-\frac{2n+1}{n-1}&-\frac{1}{n-1}&\frac{2(n^2+n-2)}{n^2(n+3)}
    \end{array}
\end{equation}
\begin{equation}
    \begin{array}{c|c|c|c|c|c|c|c|}
    &\USp(2n)&\SU(2n+1)&\SU(4)&\UU(1)_{1}&\UU(1)_{2}&\UU(1)_{R}\\
    \hline
    M_1&\cdot&\overline{\square}&\cdot&1&1&\frac{4}{n(n+3)}\\
    M_2&\square&\overline{\square}&\cdot&-\frac{3}{2(n-1)}&-\frac{1}{2(n-1)}&\frac{n(n+3)-2}{n^2(n+3)}\\
    B&\cdot&\square&\cdot&2n&0&\frac{4}{n+3}\\
    \tilde B_1&\cdot&\cdot&\cdot&-2n-1&-1&\frac{2(n^2+n-2)}{n(n+3)}\\
    \tilde B_2&\square&\cdot&\cdot&\frac{3}{2(n-1)}&\frac{1}{2(n-1)}& \frac{2+n(2n^2+n-3)}{n^2 (n+3) }\\
    R_\mu&\square&\cdot&\square&\frac{n(2n+1)}{2(n-1)}&\frac{n}{4(n-1)}&\frac{n+1}{n(n+3)}
    \end{array}
\end{equation}
\end{itemize}

\subsubsection{An alternative approach}
\label{alt1}

There is a more exotic way to derive the duality obtained above, which requires adding a confining auxiliary gauge group, but in which the antisymmetric tensor is not deconfined in the process.
The derivation works as follows:

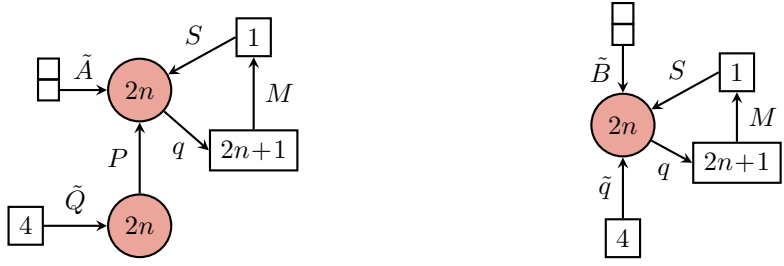
\begin{figure}
    \centering
    \begin{minipage}[b]{0.45\linewidth}
        \centering
        \makebox[\textwidth][c]{
        \begin{tikzpicture}[
            every node/.style={font=\footnotesize},
            box/.style={rectangle, draw, thick},
        ]
        \pgfmathsetmacro{\x}{1}
        \pgfmathsetmacro{\y}{1.5}
        \pgfmathsetmacro{\z}{1.8}
        \pgfmathsetmacro{\a}{1.2}
        \pgfmathsetmacro{\delta}{0.28}
        \node[fill=SUcol,circle,draw,thick] (SUgauge2) at (0, \z) {$2n$};
        \node[fill=SUcol,circle,draw,thick] (SUgauge) at (0, 0) {$2n$};
        \node[box] (fond) at (\y,\x) {$2n\!+\!1$};
        \node[box] (afond) at (\y,0) {$4$};
        \node[box] (sing) at (\y,2.5*\x) {$1$};
        \draw[<-, thick, >=stealth] (SUgauge) -- node[above] {$\tilde{Q}$} (afond);
        \draw[->, thick, >=stealth] (SUgauge) -- node[left] {$P$} (SUgauge2);
        \draw[->, thick, >=stealth] (SUgauge2) -- node[below,xshift=-3pt] {$q$} (fond.west);
        \draw[<-, thick, >=stealth] (SUgauge2) -- node[above,xshift=-3pt] {$S$} (sing.west);
        \draw[->, thick, >=stealth] (fond) -- node[right] {$M$} (sing);
        \node[box, minimum size=0.2cm] (square1) at (-\a, 0) {};
        \node[box, minimum size=0.2cm] (square2) at (-\a, \delta) {};
        \draw[->, thick, >=stealth] (square1) -- node[above] {$\tilde{A}$} (SUgauge);
        \end{tikzpicture}
        }
    \end{minipage}
    \begin{minipage}[b]{0.45\linewidth}
        \centering
        \makebox[\textwidth][c]{
        \begin{tikzpicture}[
            every node/.style={font=\footnotesize},
            box/.style={rectangle, draw, thick},
        ]
        \pgfmathsetmacro{\a}{-1.2}
        \pgfmathsetmacro{\x}{1}
        \pgfmathsetmacro{\y}{1.5}
        \pgfmathsetmacro{\z}{1.8}
        \pgfmathsetmacro{\w}{2.7}
        \pgfmathsetmacro{\v}{-0.15}
        \pgfmathsetmacro{\delta}{0.28}
        \node[fill=SUcol,circle,draw,thick] (SUgauge2) at (0, \y) {$2n$};
        \node[box] (fond) at (\y,\x) {$2n\!+\!1$};
        \node[box] (afond) at (0,0) {$4$};
        \node[box] (sing) at (\y,2.2*\x) {$1$};
        \draw[->, thick, >=stealth] (afond) -- node[left] {$\tilde{q}$} (SUgauge2);
        \draw[->, thick, >=stealth] (SUgauge2) -- node[below,xshift=-3pt] {$q$} (fond.west);
        \draw[<-, thick, >=stealth] (SUgauge2) -- node[above,xshift=-3pt] {$S$} (sing.west);
        \draw[->, thick, >=stealth] (fond) -- node[right] {$M$} (sing);
        \node[box, minimum size=0.2cm] (square1) at (0, \w) {};
        \node[box, minimum size=0.2cm] (square2) at (0, \w+\delta) {};
        \draw[->, thick, >=stealth] (square1) -- node[left] {$\tilde{B}$} (SUgauge2);
        \end{tikzpicture}
        }
    \end{minipage}
    \caption{Alternative deconfinement for the $\SU(2n)$ case in terms of a second $\SU(2n)$ gauge node. In this case we do not deconfine the conjugate antisymmetric $\tilde A$, but rather the fundamentals and the antifundamental $\tilde Q_5$. The next step consists of confining the original $\SU(2n)$ gauge node that here has a conjugate  antisymmetric, $4$ fundamentals and $2n$ antifundamentals \cite{Pouliot:1995me,Csaki:1996zb}. The final quiver, represented on the right of the figure, can be further reduced, once the baryonic deformations are added, because these deformations trigger a partial Higgs flow, breaking $\SU(2n)$ to a symplectic gauge node as explained in the text.}
    \label{UspPincEven}
\end{figure}

\begin{itemize}
\item First, we turn off the baryonic superpotential deformation.
\item Then  we build the $\SU(2n) \times \SU(2n)$ quiver in Figure \ref{UspPincEven}, with superpotential
\begin{equation}
W  = M S q + X P^{2n} + Y q^{2n}\,,
\end{equation}
that coincides with the original one after confining back the $\SU(2n)$ node at the bottom of the figure.
We can verify this by confining the $\SU(2n)$ node in terms of its mesons
$M_1 = q S$ and $M_2 = qP \equiv Q$, its baryon $B = q^{2n}$, and anti-baryons
 $\tilde B_1 = P^{2n}$  and $\tilde B_2 = P^{2n-1} S \equiv \tilde Q_5$.
 The confining superpotential for this node is 
\begin{equation}
W  = M_1^{2n} Q +  M M_1 + X \tilde B_1 + Y B + B\left(M_1 \tilde B_1 + Q \tilde Q_5\right).
\end{equation}
Integrating out the massive fields, the superpotential vanishes, while the flipper $Y$ corresponds to 
the mesonic term $Q \tilde Q_5$ on the electric side. 
\item We observe that the original $\SU(2n)$ node is confining as well \cite{Pouliot:1995me,Csaki:1996zb}, giving rise to an $\SU(2n)$ theory with a conjugate antisymmetric, $2n+1$ fundamentals and five antifundamentals, represented in the second picture in Figure  \ref{UspPincEven}.
While the gauge and field content coincide with the one of the original theory, we have a different superpotential, corresponding to 
\begin{equation}
W  = q M S + Y q^{2n} + \alpha \tilde q^4 \tilde B^{n-2} + \beta \tilde q^2 \tilde B^{n-1} + \gamma \tilde B^{n}\,,
\end{equation}
where the singlets are $\alpha= \tilde A^{n}$, $\beta = \tilde A^{n-1} \tilde Q_{1,\dots,4}^2$ and $\gamma = \tilde A^{n-2} \tilde Q_1 \tilde Q_2 \tilde Q_3 \tilde Q_4$. The four new antifundamentals of $\SU(2n)$ are $\tilde q_{1,\dots,4} = P \tilde Q_{1,\dots,4}$ while the new conjugate antisymmetric is $\tilde B = \tilde A P^2$.
\item So far in the analysis we have not distinguished among the three baryonic deformations.
When the deformations are turned on the electric side their effect consists of an extra superpotential term respectively of the form 
\begin{eqnarray}
\label{Wmag1def}
\Delta W_{\text{mag}}^{(1)} &=&  \alpha  \,,
\\
\label{Wmag2def}
\Delta W_{\text{mag}}^{(2)} &=& \beta_{12} \,,
\\
\label{Wmag3def}
\Delta W_{\text{mag}}^{(3)} &=& \gamma \,.
\end{eqnarray}
\end{itemize}

The superconformal index of the dual phase, before turning on the deformations (\ref{Wmag1def}), (\ref{Wmag2def}) and (\ref{Wmag3def}), can be obtained by following the rules of deconfinement and duality discussed at the field theory level. By applying the associated elementary identities we arrive at
\begin{equation}
    \label{indexevenbeforedef}
    \begin{aligned}
        &
        \Gamma_e\left(a^{2n};a^{2n-4} v_1 v_2 v_3 v_4\right) \prod_{1\leq \mu < \nu \leq 4}\Gamma_e\left(a^{2n-2} v_\mu v_\nu\right)\prod_{\ell=1}^{2n+1} \Gamma_e\left(pq t^{2n} x_\ell v_5^{-1};pq t^{2n} x_{\ell}^{-1} \prod_{j=1}^{2n+1} x_j\right)
        \\
        &
        \times\;I_{\SU(2n)}^{[(2n+1) \square ; 5 \overline \square; 1 \overline{A}]}\left(t^{-1} {\bf x}^{-1};t {\bf v};t^{1-2n} v_5;a^2 t^2\right),
    \end{aligned}
\end{equation}
with the constraints (\ref{BC1}) and 
\begin{equation}
a^{2(2n-2)} t^{2n} \prod_{\mu=1}^4 v_\mu = pq\,.
\end{equation}
For shortness we introduced the integral $I_{\SU(2n)}$ representing the superconformal index of the $\SU(2n)$ gauge theory obtained after the steps discussed above.
The matter content and the corresponding representations are encoded in the square brakets. 
The arguments in $I_{\SU(2n)}$ are separated by a semicolon if they transform under  different representations.
For the sake of clarity we write it explicitly in this case
\begin{equation}
    \label{ISU2n}
    \begin{aligned}
        &I_{\SU(2n)}^{[(2n+1) \square ; 5 \overline \square; 1 \overline{A}]}\left(t^{-1} {\bf x}^{-1};t {\bf v};t^{1-2n} v_5;a^2 t^2\right) = 
        \frac{(p;p)_{\infty}^{2n-1}(q;q)_{\infty}^{2n-1}}{(2n)!}
        \oint \left(\prod_{i=1}^{2n-1} \frac{\dd{z_i}}{2 \pi \mi z_i}\right)
        \\
        &
        \prod_{i=1}^{2n}
        \left(\prod_{\ell=1}^{2n+1}\Gamma_e\left(t^{-1} x_{\ell}^{-1} z_i\right)\right)
        \left(\prod_{\mu=1}^{4}\Gamma_e\left(t v_{\mu} z_i^{-1}\right)\right)
        \Gamma_e\left(t^{1-2n} v_5 z_i^{-1}\right)
        \prod_{j>i}^{2n} \frac{\Gamma_e\left(a^2 t^2 z_i^{-1}z_j^{-1}\right)}{\Gamma_e\left(z_i^{\pm 1} z_j^{\mp 1}\right)}.
    \end{aligned}
\end{equation}
We keep the same notation in the rest of the paper.
Observe that in the first line of \eqref{indexevenbeforedef}, the arguments of the elliptic Gamma functions correspond to the contributions of the singlets $\alpha$, $\gamma$, $\beta$, $Y$ and $M$ respectively.

In each case the deformations trigger an Higgs flow that can be studied along the lines of \cite{Gaiotto:2012xa}.
The details are summarized as follows.
\begin{enumerate}
\item In the first case, the Higgsing is due to the operator $F_{\alpha}$, which gives a vev to the operator $\tilde q_{1} \tilde q_{2} \tilde q_{3} \tilde q_{4} \tilde B^{n-2}$. The gauge group Higgses to $\USp(2n-4)$, yielding the same dual expected from the canonical tensor deconfinement discussed in the subsection above. 
This Higgs flow can be reproduced at the level of the superconformal index through the sequence of poles
\begin{equation}
    \begin{aligned}
        &z_{2I-1} = a^2 t^2 z_{2I}^{-1}\,, \\
        &x_{2n-4+\mu} = t v_{\mu}\,,
    \end{aligned}
    \qquad
    \begin{aligned}
        &I=1,\dots,n-2\,,\\
        &\mu=1,\dots,3\,.
    \end{aligned}
\end{equation}
Enforcing the $\SU(2n)$ constraint and evaluating the index using this sequence of poles, we reduce the $\SU(2n)$ index to the $\USp(2n-4)$ one, as expected from the discussion in the subsection above.
The details of the pinching of the integration contour and of the derivation of the index are identical to the ones extensively discussed in \cite{Amariti:2025zgj}. For this reason, we omit them here and refer the reader to that reference.

\item In the second case, the Higgsing is due to the operator $F_{\beta_{12}}$, that gives a vev to the operator $\tilde q_1 \tilde q_2  \tilde B^{n-1}$. The gauge group Higgses to $\USp(2n-2)$, yielding the same dual expected from the canonical tensor deconfinement discussed in the subsection above.
Again, this Higgs flow can be reproduced at the level of the superconformal index through the sequence of poles
\begin{equation}
    \begin{aligned}
        &z_{2I-1} = a^2 t^2 z_{2I}^{-1}\,, \\
        &x_{2n-1} = t v_{3}\,.
    \end{aligned}
    \qquad
    \begin{aligned}
        &I=1,\dots,n-1\,,\\
        &\phantom{\mu=1,\dots,3\,.}
    \end{aligned}
\end{equation}
Enforcing the $\SU(2n)$ constraint and evaluating the index using
 such sequence of poles, we reduce the $\SU(2n)$ index to the $\USp(2n-2)$
one, as expected from the discussion in the subsection above.
Again, we omit the details of the pinching and of the evaluation of the index.

\item In the third case, the Higgsing is due to the operator $F_{\gamma}$, that gives a vev to the operator $\tilde B^{n}$. The gauge group Higgses to $\USp(2n)$, yielding the same dual expected from the canonical tensor deconfinement discussed in the subsection above.
This Higgs flow can  be reproduced at the level of the superconformal index through the sequence of poles
\begin{equation}
\begin{array}{rl}
z_{2I-1} = a^2 t^2 z_{2I}^{-1}\,, & \quad \quad I=1,\dots,n\,.\\
\end{array}
\end{equation}
By evaluating the index using such sequence of poles, we reduce the $\SU(2n)$ index to the $\USp(2n)$
one, as expected from the discussion in the subsection above.
\end{enumerate}

\subsubsection{Relation with the literature and phases of the dual theories}

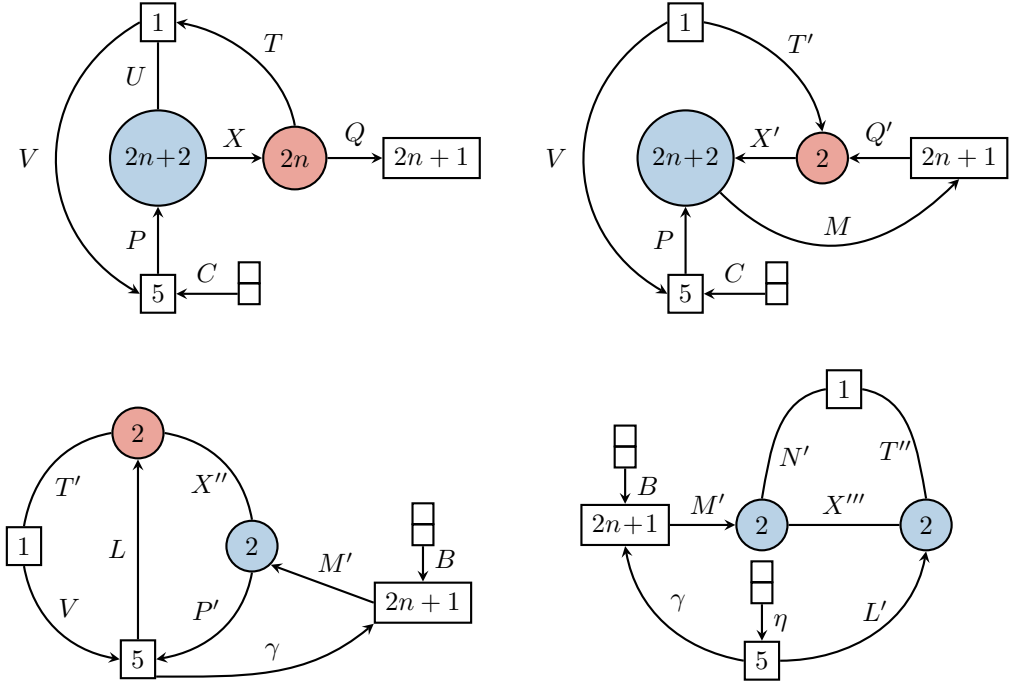
\begin{figure}[h!]
    \centering
        \begin{minipage}[b]{0.45\linewidth}
            \centering
            \makebox[\textwidth][c]{
            \begin{tikzpicture}[
                every node/.style={font=\footnotesize},
                box/.style={rectangle, draw, thick}
            ]
            \pgfmathsetmacro{\x}{1.8}
            \pgfmathsetmacro{\y}{3.5}
            \pgfmathsetmacro{\d}{0.28}
            \node[fill=SUcol,circle,draw,thick] (su) at (0, 0) {$2n$};
            \node[fill=USPcol,circle,draw,thick] (usp) at (-\x, 0) {$2n\!+\!2$};
            \node[box] (r) at (\x, 0) {$2n+1$};
            \node[box] (t) at (-\x, \x) {$1$};
            \node[box] (b) at (-\x, -\x) {$5$};
            \node[box, minimum size=0.2cm] (sq1) at (-\x/3, -\x) {};
            \node[box, minimum size=0.2cm] (sq2) at (-\x/3, -\x+\d) {};
            \draw[<-,thick,>=stealth] (b) -- node[above] {$C$} (sq1);
            \draw[<-,thick,>=stealth] (su) -- node[above] {$X$} (usp);
            \draw[->,thick,>=stealth] (su) -- node[above] {$Q$} (r);
            \draw[->,thick,>=stealth] (b) -- node[left] {$P$} (usp);
            \draw[-,thick,>=stealth] (usp) -- node[left] {$U$} (t);
            \draw[->,thick,>=stealth] (t.west) to[out=-150, in=150, looseness=1.2] (b.west);
            \draw[->,thick,>=stealth] (su.north) to[out=100, in=-10, looseness=0.9] (t.east);
            \node at (-\y, 0) {$V$};
            \node at (-\d, \x-\d) {$T$};
            \end{tikzpicture}
            }
        \end{minipage}
        \begin{minipage}[b]{0.45\linewidth}
            \centering
            \makebox[\textwidth][c]{
            \begin{tikzpicture}[
                every node/.style={font=\footnotesize},
                box/.style={rectangle, draw, thick}
            ]
            \pgfmathsetmacro{\x}{1.8}
            \pgfmathsetmacro{\y}{3.5}
            \pgfmathsetmacro{\d}{0.28}
            \node[fill=SUcol,circle,draw,thick] (su) at (0, 0) {$2$};
            \node[fill=USPcol,circle,draw,thick] (usp) at (-\x, 0) {$2n\!+\!2$};
            \node[box] (r) at (\x, 0) {$2n+1$};
            \node[box] (t) at (-\x, \x) {$1$};
            \node[box] (b) at (-\x, -\x) {$5$};
            \node[box, minimum size=0.2cm] (sq1) at (-\x/3, -\x) {};
            \node[box, minimum size=0.2cm] (sq2) at (-\x/3, -\x+\d) {};
            \draw[<-,thick,>=stealth] (b) -- node[above] {$C$} (sq1);
            \draw[->,thick,>=stealth] (su) -- node[above] {$X'$} (usp);
            \draw[<-,thick,>=stealth] (su) -- node[above] {$Q'$} (r);
            \draw[->,thick,>=stealth] (b) -- node[left] {$P$} (usp);
            \draw[->,thick,>=stealth] (t.west) to[out=-150, in=150, looseness=1.2] (b.west);
            \draw[<-,thick,>=stealth] (su.north) to[out=100, in=-10, looseness=0.9] (t.east);
            \draw[->,thick,>=stealth] (usp.south east) to[out=-45, in=-135, looseness=1.2] (r.south);
            \node at (-\y, 0) {$V$};
            \node at (-\d, \x-\d) {$T'$};
            \node at (0.2, -\x/2) {$M$};
            \end{tikzpicture}
            }
        \end{minipage}
        \\[0.7cm]
        \begin{minipage}[b]{0.45\linewidth}
            \centering
            \makebox[\textwidth][c]{
            \begin{tikzpicture}[
                every node/.style={font=\footnotesize},
                box/.style={rectangle, draw, thick}
            ]
            \pgfmathsetmacro{\x}{1.5}
            \pgfmathsetmacro{\y}{3.5}
            \pgfmathsetmacro{\d}{0.28}
            \node[fill=SUcol,circle,draw,thick] (su) at (0, \x) {$2$};
            \node[fill=USPcol,circle,draw,thick] (usp) at (\x, 0) {$2$};
            \node[box] (r) at (2.5*\x, -\x/2) {$2n+1$};
            \node[box] (l) at (-\x,0) {$1$};
            \node[box] (b) at (0, -\x) {$5$};
            \node[box, minimum size=0.2cm] (sq1) at (2.5*\x, 0.1*\x) {};
            \node[box, minimum size=0.2cm] (sq2) at (2.5*\x, 0.1*\x+\d) {};
            \draw[<-,thick,>=stealth] (r) -- node[right, yshift=2pt] {$B$} (sq1);
            \draw[-,thick,>=stealth] (su.east) to[out=-10, in=100, looseness=1] node[below, xshift=-5pt] {$X''$} (usp.north);
            \draw[->,thick,>=stealth] (usp.south) to[out=-100, in=10, looseness=1] node[above, xshift=-5pt] {$P'$} (b.east);      
            \draw[->,thick,>=stealth] (l.south) to[out=-80, in=170, looseness=1] node[above, xshift=5pt] {$V$} (b.west);
            \draw[-,thick,>=stealth] (l.north) to[out=80, in=-170, looseness=1] node[below, xshift=6pt] {$T'$} (su.west);
            \draw[->,thick,>=stealth] (b) -- node[left] {$L$} (su);
            \draw[<-,thick,>=stealth] (r.south west) to[out=-140, in=0, looseness=1] node[above, xshift=0pt] {$\gamma$} ($ (b.south east) + (0pt,1pt) $);
            \draw[<-,thick,>=stealth] (usp.south east) to node[above, xshift=5pt] {$M'$} (r.west);
            \end{tikzpicture}
            }
        \end{minipage}
        \begin{minipage}[b]{0.45\linewidth}
            \centering
            \makebox[\textwidth][c]{
            \begin{tikzpicture}[
                every node/.style={font=\footnotesize},
                box/.style={rectangle, draw, thick}
            ]
            \pgfmathsetmacro{\x}{1.8}
            \pgfmathsetmacro{\y}{0.9}
            \pgfmathsetmacro{\a}{0}
            \pgfmathsetmacro{\b}{0.9}
            \pgfmathsetmacro{\d}{0.28}
            \node[fill=USPcol,circle,draw,thick] (uspc) at (0, 0) {$2$};
            \node[fill=USPcol,circle,draw,thick] (uspr) at (1.2*\x, 0) {$2$};
            \node[box] (t) at (0.6*\x, \x) {$1$};
            \node[box] (b) at (0, -\x) {$5$};
            \node[box] (l) at (-\x, 0) {$2n\!+\!1$};
            \node[box, minimum size=0.2cm] (sqt1) at (-\x, \y) {};
            \node[box, minimum size=0.2cm] (sqt2) at (-\x, \y+\d) {};
            \draw[<-,thick,>=stealth] (l) -- node[right] {$B$} (sqt1);
            \node[box, minimum size=0.2cm] (sqb1) at (\a, -\b) {};
            \node[box, minimum size=0.2cm] (sqb2) at (\a, -\b+\d) {};
            \draw[<-,thick,>=stealth] (b) -- node[right] {$\eta$} (sqb1);
            \draw[<-,thick,>=stealth] (uspc) -- node[above, xshift=2pt] {$M'$} (l);
            \draw[-,thick,>=stealth] (uspc) -- node[above] {$X'''$} (uspr);
            \draw[<-,thick,>=stealth] (l.south) to[out=-80, in=170, looseness=1] node[above, xshift=5pt] {$\gamma$} (b.west);
            \draw[-,thick,>=stealth] (uspc.north) to[out=80, in=-170, looseness=1] node[below, xshift=6pt, yshift=-2pt] {$N'$} (t.west);
            \draw[-,thick,>=stealth] (t.east) to[out=-10, in=100, looseness=1] node[below, xshift=-5pt] {$T''$} (uspr.north);
            \draw[->,thick,>=stealth] (b.east) to[out=0, in=-100, looseness=1] node[above, xshift=0pt] {$L'$} (uspr.south);
            \end{tikzpicture}
            }
        \end{minipage}
    \caption{Deconfinement and sequential duality steps for $\SU(2n)$ with one conjugate antisymmetric, $2n+1$ fundamentals, $5$ antifundamentals, reproducing the results of  \cite{Terning:1997jj} without introducing fictitious symmetries.}
    \label{fig:TerningDec}
\end{figure}

We conclude this section by discussing the relation between the dualities discussed above and the 
one derived in \cite{Terning:1997jj} for the underformed model.
We reformulate the derivation  of the duality of \cite{Terning:1997jj} by deconfining the antisymmetric tensor without introducing spurious symmetries, using the prescription spelled out in \cite{Bajeot:2022kwt}.
This boils down to deconfine the electric model using the first quiver in Figure \ref{fig:TerningDec}.

The superpotentials for this phase is 
\begin{equation}
W = TUX + PUV + P^2 C\,,
\end{equation}
where $A=X^2$  and $PX = Q$.
We then dualize the $\SU(2n)$ node using the ordinary Seiberg duality, obtaining an $\SU(2)$ gauge group with superpotential
\begin{equation}
W =  P^2 C + U N + PUV + N T' X' + M X' Q'\,,
\end{equation}
where $M=X Q$ and $N = T X$.
Integrating out the massive fields $U$ and $N$, the superpotential becomes
\begin{equation}
W =  P^2 C + PV T' X' + M X' Q'\,,
\end{equation}
associated to the second quiver in Figure \ref{fig:TerningDec}.
We then dualize the $\USp(2n+2)$ gauge group using IP duality. After applying the duality, the gauge group becomes 
$\USp(2)$ and the superpotential for this phase is
\begin{equation}
W =
B {M'}^2 +  
\alpha Q' + 
\alpha M' X''+
\beta C +
\beta {P'}^2 + 
\gamma P' M' + 
L P' X'' + 
L V T' + 
\sigma {X''}^2,
\end{equation}
where
$\alpha=M X'$,
$\beta = P^2$, 
$\gamma=M P$,
$\sigma = {X'}^2$ and $L=X' P$.
After integrating out the massive fields, the superpotential becomes
\begin{equation}
W =
B {M'}^2 +
\gamma P' M' + 
L P' X'' +  
L V T' +  
\sigma {X''}^2,
\end{equation}
and the quiver for this phase corresponds to the third one in Figure  \ref{fig:TerningDec}.
We can further dualize the $\SU(2)$ node on the right of the quiver, treating it as $\USp(2)$ and using the IP duality once again.
We obtain
\begin{equation}
W=
B{M'}^2 + 
\gamma P' M' +
p P' +
p X'''L' +
vV +
v T''L' +
\sigma \rho +
\rho {X'''}^3 +
\eta {L'}^2 +
N' X''' T'',
\end{equation}
where
$p=L X''$, 
$v=LT'$,  
$\rho = {X''}^2$, 
$\eta=L^2$ and
$N'= T' X''$.
After integrating out the massive fields, it becomes
\begin{equation}
W = 
B{M'}^2 +
\gamma  M' X'''L' +
N' X''' T'' +
\eta {L'}^2,
\end{equation}
that coincides with the result of \cite{Terning:1997jj} upon the identifications
$(yy) \leftrightarrow B$,
$y_1 \leftrightarrow M'$,
$(\tilde q q ) \leftrightarrow \gamma$,
$x_4 \leftrightarrow X'''$,
$q_3 \leftrightarrow L'$,
$(z_3 z_1) \leftrightarrow N'$,
$z_2 \leftrightarrow T''$ and
$(q_2 q_2) \leftrightarrow \eta$.
The quiver for this last phase is the fourth one in Figure 
\ref{fig:TerningDec}.

Using this dictionary, we can also read the map between the operators in the chiral ring, see {\bf Formula  2.11} in \cite{Terning:1997jj}.
The three baryonic deformations considered above correspond then to
$A^n \leftrightarrow {X'''}^2$,
$A^{n-1} Q_4 Q_5  \leftrightarrow \eta_{45} $ and
$A^{n-2} Q_2 Q_3 Q_4 Q_5  \leftrightarrow L' T''_1$.
It follows that, in the first case, we can integrate out the bifundamental connecting the two gauge nodes in the last quiver in  Figure \ref{fig:TerningDec}.
The $\USp(2)$ node on the right of the quiver has then six fundamentals and confines. After confining the node, we are left with $\USp(2)$ with $2n+2$ fundamentals in addition to some flippers.
In the second case the $F$-term for the field $\eta_{45}$ Higgses the $\USp(2)$ node on the right, giving a vev to $L_{4}'$ and $L_5'$. After Higgsing this node  we are left again with $\USp(2)$ with $2n+4$ fundamentals in addition to some flippers.
In the last case the fields $L'$ and $T''_1$ are massive, and the  $\USp(2)$ node on the right of the quiver s-confines. We obtain $\USp(2)$ with $2n+6$ fundamentals and flippers.

Observe that each of the three dualities found above has a $\USp(2m)$ with $2m+6$ fundamentals, where $m=n-2$, $m=n-1$, or $m=n$.
A further IP duality lead to the model that was found above starting from the dual theory obtained by  \cite{Terning:1997jj}.
These last correspond to IR theories, implying that the electric model is in a free magnetic phase.

\subsection{The case of $\SU(2n+1)$}
\label{sec:4dodd}
In the odd case we distinguish again three types of baryonic deformations. 
In the following we present each deformation together with the corresponding table of charges:
\begin{enumerate}
\item In the first case the  superpotential is
  \begin{equation}
    \label{WEoddDef1}
    W_{\text{ele}} =   \tilde{A}^{n} \tilde{Q}_1\,.
\end{equation}
The antifundamental involved in the deformation can be chosen arbitrarily.
\begin{equation}
    \begin{array}{c|c|c|c|c|c|c|}
     &\SU(2n+1)&\SU(2n+2)&\SU(4)&\UU(1)_{1}&\UU(1)_{2}&\UU(1)_{R}\\
     \hline
     Q&\square&\overline{\square}&\cdot&1&0&\frac{2}{3}\\
     \tilde Q_m&\overline{\square}&\cdot&\square&0&1&\frac{2}{3}\\
      \tilde Q_1&\overline{\square}&\dot&\cdot&\frac{2(n+1)(n-1)}{n}&\frac{4(n-1)}{n}&\frac{2(2n^2+6n-3)}{3(n-1)}\\
     \tilde A&\begin{array}{c}\square \vspace{-2.9 mm} \\\square \end{array} &\cdot&\cdot&-\frac{2(n+1)}{n}&-\frac{4}{n}&-\frac{2(2n+3)}{3(n-1)}
        \end{array}
\end{equation}
\item The second baryonic deformation considered in this section is
 \begin{equation}
    \label{WEoddDef2}
    W_{\text{ele}} =  
    \tilde{A}^{n-1} 
    \tilde{Q}_1
    \tilde{Q}_2
    \tilde{Q}_3\,,
\end{equation}
where we choose arbitrarily the three antifundamentals involved in the deformation.
\begin{equation}
    \begin{array}{c|c|c|c|c|c|c|c|}
     &\SU(2n+1)&\SU(2n+2)&\SU(2)&\SU(3)&\UU(1)_{1}&\UU(1)_{2}&\UU(1)_{R}\\
     \hline
     Q&\square&\overline{\square}&\cdot&\cdot&1&0&\frac{2}{3}\\
     \tilde Q_m&\overline{\square}&\cdot&\square&\cdot&0&1&\frac{2}{3}\\
      \tilde Q_\mu&\overline{\square}&\cdot&\dot&\square&\frac{2(n-1)(n+1)}{3n}&\frac{2(n-1)}{3n}&\frac{2(2n^2+2n-1)}{9n}\\
     \tilde A&\begin{array}{c}\square \vspace{-2.9 mm} \\\square \end{array} &\cdot&\cdot&\cdot&-\frac{2(n+1)}{n}&-\frac{2}{n}&-\frac{2(2n+1)}{3n}   
      \end{array}
\end{equation}
\item The last possible baryonic deformation considered here is
\begin{equation}
    \label{WEoddDef3}
    W_{\text{ele}} =
    \tilde{A}^{n-2}
    \tilde{Q}_1
    \tilde{Q}_2
    \tilde{Q}_3
    \tilde{Q}_4
    \tilde{Q}_5\,.
\end{equation}
\begin{equation}
    \begin{array}{c|c|c|c|c|c|}
     &\SU(2n+1)&\SU(2n+2)&\SU(5)&\UU(1)_{1}&\UU(1)_{R}\\
     \hline
     Q&\square&\overline{\square}&\cdot&1&\frac{2}{3}\\
     \tilde Q&\overline{\square}&\cdot&\square&\frac{2(n-1)(n+1)}{5n}&\frac{2(2n^2-2n+5)}{15(n+1)}\\
      \tilde A&\begin{array}{c}\square \vspace{-2.9 mm} \\\square \end{array} &\cdot&\cdot&-\frac{2(n+1)}{n}&-\frac{2(2n-1)}{3(n+1)}   
      \end{array}
\end{equation}
\end{enumerate}

\subsubsection{Deconfining the conjugate antisymmetric with a symplectic node}

\begin{figure}
    \centering
    \begin{minipage}[b]{0.30\linewidth}
        \centering
        \makebox[\textwidth][c]{
        \begin{tikzpicture}[
            every node/.style={font=\footnotesize},
            box/.style={rectangle, draw, thick},
        ]
        \pgfmathsetmacro{\x}{1.5}
        \pgfmathsetmacro{\y}{1.5}
        \pgfmathsetmacro{\z}{1.2}
        \pgfmathsetmacro{\delta}{0.28}
        \node[fill=SUcol,circle,draw,thick] (SUgauge) at (0, 0) {$2n\!+\!1$};
        \node[box] (fond) at (\x,-\y) {$2n\!+\!2$};
        \node[box] (afond) at (-\x,-\y) {$4$};
        \node[box] (afond2) at (0,-\y) {$1$};
        \draw[->, thick, >=stealth] (SUgauge) -- node[right, yshift=3pt] {$Q$} (fond.north);
        \draw[<-, thick, >=stealth] (SUgauge) -- node[left, yshift=3pt] {$\tilde{Q}_m$} (afond.north);
        \draw[<-, thick, >=stealth] (SUgauge) -- node[right] {$\tilde{Q}_{1}$} (afond2.north);
        \node[box, minimum size=0.2cm] (square1) at (0, \z) {};
        \node[box, minimum size=0.2cm] (square2) at (0, \z+\delta) {};
        \draw[->, thick, >=stealth] (square1) -- node[right] {$\tilde{A}$} (SUgauge);
        \end{tikzpicture}
        }
    \end{minipage}
    \begin{minipage}[b]{0.30\linewidth}
        \centering
        \makebox[\textwidth][c]{
        \begin{tikzpicture}[
            every node/.style={font=\footnotesize},
            box/.style={rectangle, draw, thick},
        ]
        \pgfmathsetmacro{\x}{1}
        \pgfmathsetmacro{\y}{1.5}
        \pgfmathsetmacro{\z}{1.8}
        \node[fill=USPcol,circle,draw,thick] (USPgauge) at (0, \z) {$2n\!-\!2$};
        \node[fill=SUcol,circle,draw,thick] (SUgauge) at (0, 0) {$2n\!+\!1$};
        \node[box] (fond) at (\x,-\y) {$2n\!+\!2$};
        \node[box] (afond) at (-\x,-\y) {$4$};
        \node[box] (afond2) at (\y,\z) {$1$};
        \draw[->, thick, >=stealth] (SUgauge) -- node[right] {$Q$} (fond.north);
        \draw[<-, thick, >=stealth] (SUgauge) -- node[left,yshift=2pt] {$\tilde{Q}_m$} (afond.north);
        \draw[->, thick, >=stealth] (USPgauge) -- node[right] {$\tilde{P}$} (SUgauge);
        \draw[-, thick, >=stealth] (USPgauge) -- node[above,yshift=5pt] {$R_1$} (afond2);
        \end{tikzpicture}
        }
    \end{minipage}
    \begin{minipage}[b]{0.30\linewidth}
        \centering
        \makebox[\textwidth][c]{
        \begin{tikzpicture}[
            every node/.style={font=\footnotesize},
            box/.style={rectangle, draw, thick},
        ]
        \pgfmathsetmacro{\x}{1}
        \pgfmathsetmacro{\y}{1.5}
        \pgfmathsetmacro{\w}{1}
        \pgfmathsetmacro{\z}{1.8}
        \node[fill=USPcol,circle,draw,thick] (USPgauge) at (\x, \w) {$2n\!-\!2$};
        \node[box] (fond) at (\x,-\y) {$2n\!+\!2$};
        \node[box] (afond) at (-\x,-\y) {$4$};
        \node[box] (sing) at (-\x,\w) {$1$};
        \node[box] (afond2) at (\x,\z+\x/2) {$1$};
        \draw[->, thick, >=stealth] (USPgauge) -- node[right] {$M_2$} (fond.north);
        \draw[<-, thick, >=stealth] (USPgauge) -- node[above] {$\tilde{B}_2$} (sing);
        \draw[->, thick, >=stealth] (sing) -- node[left] {$\tilde{B}_1$} (afond);
        \draw[->, thick, >=stealth] (fond) -- node[above, xshift=4pt] {$B$} (sing.south east);
        \draw[<-, thick, >=stealth] (fond) -- node[above] {$M_1$} (afond);
        \draw[-, thick, >=stealth] (USPgauge) -- node[right,xshift=3pt] {$R_1$} (afond2);
        \end{tikzpicture}
        }
    \end{minipage} 
      \caption{Quiver representation of the deconfinement of the conjugate antisymmetric, in presence of  the deformation $W_{\text{ele}}=\tilde{A}^{n}\tilde{Q}_1$ in terms of an $\USp(2n-2)$ gauge group, followed by ordinary Seiberg duality for the $\SU(2n+1)$ gauge node.}
    \label{UspDecAnQ}
\end{figure}
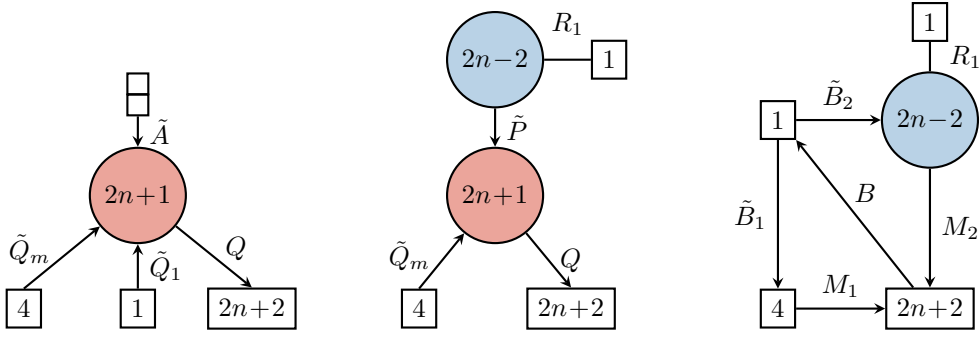

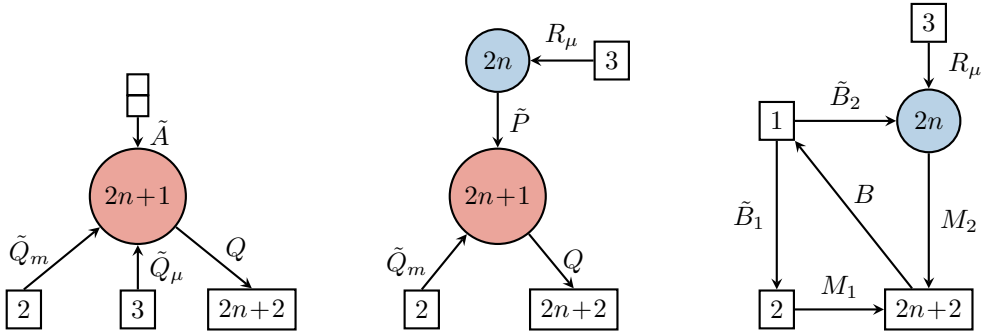
\begin{figure}
    \centering
    \begin{minipage}[b]{0.30\linewidth}
        \centering
        \makebox[\textwidth][c]{
        \begin{tikzpicture}[
            every node/.style={font=\footnotesize},
            box/.style={rectangle, draw, thick},
        ]
        \pgfmathsetmacro{\x}{1.5}
        \pgfmathsetmacro{\y}{1.5}
        \pgfmathsetmacro{\z}{1.2}
        \pgfmathsetmacro{\delta}{0.28}
        \node[fill=SUcol,circle,draw,thick] (SUgauge) at (0, 0) {$2n\!+\!1$};
        \node[box] (fond) at (\x,-\y) {$2n\!+\!2$};
        \node[box] (afond) at (-\x,-\y) {$2$};
        \node[box] (afond2) at (0,-\y) {$3$};
        \draw[->, thick, >=stealth] (SUgauge) -- node[right, yshift=3pt] {$Q$} (fond.north);
        \draw[<-, thick, >=stealth] (SUgauge) -- node[left, yshift=3pt] {$\tilde{Q}_m$} (afond.north);
        \draw[<-, thick, >=stealth] (SUgauge) -- node[right] {$\tilde{Q}_\mu$} (afond2.north);
        \node[box, minimum size=0.2cm] (square1) at (0, \z) {};
        \node[box, minimum size=0.2cm] (square2) at (0, \z+\delta) {};
        \draw[->, thick, >=stealth] (square1) -- node[right] {$\tilde{A}$} (SUgauge);
        \end{tikzpicture}
        }
    \end{minipage}
    \begin{minipage}[b]{0.30\linewidth}
        \centering
        \makebox[\textwidth][c]{
        \begin{tikzpicture}[
            every node/.style={font=\footnotesize},
            box/.style={rectangle, draw, thick},
        ]
        \pgfmathsetmacro{\x}{1}
        \pgfmathsetmacro{\y}{1.5}
        \pgfmathsetmacro{\z}{1.8}
        \pgfmathsetmacro{\w}{0.5}
        \pgfmathsetmacro{\delta}{0.28}
        \node[fill=USPcol,circle,draw,thick] (USPgauge) at (0, \z) {$2n$};
        \node[fill=SUcol,circle,draw,thick] (SUgauge) at (0, 0) {$2n\!+\!1$};
        \node[box] (fond) at (\x,-\y) {$2n\!+\!2$};
        \node[box] (afond) at (-\x,-\y) {$2$};
        \node[box] (afond2) at (\y,\z) {$3$};
        \draw[->, thick, >=stealth] (SUgauge) -- node[right] {$Q$} (fond.north);
        \draw[<-, thick, >=stealth] (SUgauge) -- node[left,xshift=-2pt] {$\tilde{Q}_m$} (afond.north);
        \draw[->, thick, >=stealth] (USPgauge) -- node[right] {$\tilde{P}$} (SUgauge);
        \draw[<-, thick, >=stealth] (USPgauge) -- node[above] {$R_\mu$} (afond2);
        \end{tikzpicture}
        }
    \end{minipage}
    \begin{minipage}[b]{0.30\linewidth}
        \centering
        \makebox[\textwidth][c]{
        \begin{tikzpicture}[
            every node/.style={font=\footnotesize},
            box/.style={rectangle, draw, thick},
        ]
        \pgfmathsetmacro{\delta}{0.28}
        \pgfmathsetmacro{\x}{1}
        \pgfmathsetmacro{\y}{1.5}
        \pgfmathsetmacro{\w}{0.5}
        \pgfmathsetmacro{\z}{1.8}
        \pgfmathsetmacro{\h}{2.17}
        \node[fill=USPcol,circle,draw,thick] (USPgauge) at (\x, \x) {$2n$};
        \node[box] (fond) at (\x,-\y) {$2n\!+\!2$};
        \node[box] (afond) at (-\x,-\y) {$2$};
        \node[box] (sing) at (-\x,\x) {$1$};
        \node[box] (afond2) at (\x,\z+\x/2) {$3$};
        \draw[->, thick, >=stealth] (USPgauge) -- node[right] {$M_2$} (fond.north);
        \draw[<-, thick, >=stealth] (USPgauge) -- node[above] {$\tilde{B}_2$} (sing);
        \draw[->, thick, >=stealth] (sing) -- node[left] {$\tilde{B}_1$} (afond);
        \draw[->, thick, >=stealth] (fond) -- node[above, xshift=4pt] {$B$} (sing.south east);
        \draw[<-, thick, >=stealth] (fond) -- node[above] {$M_1$} (afond);
        \draw[<-, thick, >=stealth] (USPgauge) -- node[right,xshift=3pt] {$R_\mu$} (afond2);
        \end{tikzpicture}
        }
    \end{minipage}
          \caption{Quiver representation of the deconfinement of the conjugate antisymmetric, in presence of  the deformation $W_{\text{ele}}=\tilde{A}^{n-1}\tilde{Q}_1\tilde{Q}_2\tilde{Q}_3$ in terms of an $\USp(2n)$ gauge group, followed by ordinary Seiberg duality for the $\SU(2n+1)$ gauge node.}
    \label{UspDecAnQ3}
\end{figure}

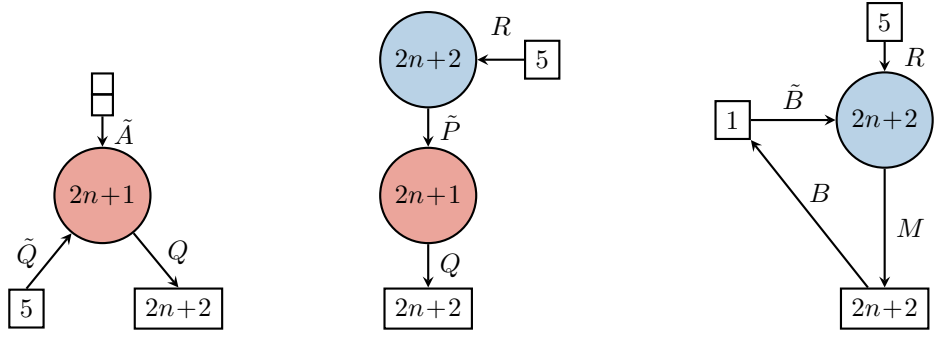
\begin{figure}
    \centering
    \begin{minipage}[b]{0.30\linewidth}
        \centering
        \makebox[\textwidth][c]{
        \begin{tikzpicture}[
            every node/.style={font=\footnotesize},
            box/.style={rectangle, draw, thick},
        ]
        \pgfmathsetmacro{\x}{1}
        \pgfmathsetmacro{\y}{1.5}
        \pgfmathsetmacro{\z}{1.2}
        \pgfmathsetmacro{\delta}{0.28}
        \node[fill=SUcol,circle,draw,thick] (SUgauge) at (0, 0) {$2n\!+\!1$};
        \node[box] (fond) at (\x,-\y) {$2n\!+\!2$};
        \node[box] (afond) at (-\x,-\y) {$5$};
        \draw[->, thick, >=stealth] (SUgauge) -- node[right, yshift=3pt] {$Q$} (fond.north);
        \draw[<-, thick, >=stealth] (SUgauge) -- node[left, yshift=3pt] {$\tilde{Q}$} (afond.north);
        \node[box, minimum size=0.2cm] (square1) at (0, \z) {};
        \node[box, minimum size=0.2cm] (square2) at (0, \z+\delta) {};
        \draw[->, thick, >=stealth] (square1) -- node[right] {$\tilde{A}$} (SUgauge);
        \end{tikzpicture}
        }
    \end{minipage}
    \begin{minipage}[b]{0.30\linewidth}
        \centering
        \makebox[\textwidth][c]{
        \begin{tikzpicture}[
            every node/.style={font=\footnotesize},
            box/.style={rectangle, draw, thick},
        ]
        \pgfmathsetmacro{\x}{1}
        \pgfmathsetmacro{\y}{1.5}
        \pgfmathsetmacro{\z}{1.8}
        \pgfmathsetmacro{\w}{0.5}
        \pgfmathsetmacro{\delta}{0.28}
        \node[fill=USPcol,circle,draw,thick] (USPgauge) at (0, \z) {$2n\!+\!2$};
        \node[fill=SUcol,circle,draw,thick] (SUgauge) at (0, 0) {$2n\!+\!1$};
        \node[box] (fond) at (0,-\y) {$2n\!+\!2$};
        \node[box] (afond2) at (\y,\z) {$5$};
        \draw[->, thick, >=stealth] (SUgauge) -- node[right] {$Q$} (fond.north);
        \draw[->, thick, >=stealth] (USPgauge) -- node[right] {$\tilde{P}$} (SUgauge);
        \draw[<-, thick, >=stealth] (USPgauge) -- node[above,yshift=5pt] {$R$} (afond2);
        \end{tikzpicture}
        }
    \end{minipage}
    \begin{minipage}[b]{0.30\linewidth}
        \centering
        \makebox[\textwidth][c]{
        \begin{tikzpicture}[
            every node/.style={font=\footnotesize},
            box/.style={rectangle, draw, thick},
        ]
        \pgfmathsetmacro{\delta}{0.28}
        \pgfmathsetmacro{\x}{1}
        \pgfmathsetmacro{\y}{1.5}
        \pgfmathsetmacro{\w}{0.5}
        \pgfmathsetmacro{\z}{1.8}
        \pgfmathsetmacro{\h}{2.17}
        \node[fill=USPcol,circle,draw,thick] (USPgauge) at (\x, \x) {$2n\!+\!2$};
        \node[box] (fond) at (\x,-\y) {$2n\!+\!2$};
        \node[box] (sing) at (-\x,\x) {$1$};
        \node[box] (afond2) at (\x,\z+\x/2) {$5$};
        \draw[->, thick, >=stealth] (USPgauge) -- node[right] {$M$} (fond.north);
        \draw[<-, thick, >=stealth] (USPgauge) -- node[above] {$\tilde{B}$} (sing);
        \draw[->, thick, >=stealth] (fond) -- node[above, xshift=4pt] {$B$} (sing.south east);
        \draw[<-, thick, >=stealth] (USPgauge) -- node[right,xshift=3pt] {$R$} (afond2);
        \end{tikzpicture}
        }
    \end{minipage}
            \caption{Quiver representation of the deconfinement of the conjugate antisymmetric, in presence of  the deformation $W_{\text{ele}}=\tilde{A}^{n-2}\tilde{Q}_1\tilde{Q}_2\tilde{Q}_3\tilde{Q}_4\tilde{Q}_5$ in terms of an $\USp(2n+2)$ gauge group, followed by ordinary Seiberg duality for the $\SU(2n+1)$ gauge node.}
    \label{UspDecAnQ5}
\end{figure}

In the following, we study the $\SU(2n+1)$ model in presence of the deformations 
(\ref{WEoddDef1}), (\ref{WEoddDef2}) and (\ref{WEoddDef3})
by deconfining the conjugate antisymmetric with an $\USp(2m)$ gauge group, where $m=n-1$,
$m=n$, and $m=n+1$ respectively.
The quivers obtained after deconfining the conjugate antisymmetric correspond to the second ones  in Figure 
\ref{UspDecAnQ}, \ref{UspDecAnQ3} and \ref{UspDecAnQ5} respectively.
The superpotential vanishes in the first case, while in the second and third case it is given by $W = \sigma R^2$. The difference between such two cases is that, when we consider the deformation (\ref{WEoddDef2}), the field $\sigma$ is an antifundamental of the broken $\SU(3)$ global symmetry,
while it is in the antisymmetric representation of the unbroken $\SU(5)$ flavor symmetry group when we choose the 
deformation (\ref{WEoddDef3}).
In each case we can confine the $\SU(2n+1)$ gauge theory, because it represents a flipped version of the confining duality for $\SU(2n+1)$ with  $2n+2$ flavors.

The quivers obtained after confining the $\SU(2n+1)$ nodes correspond to the third ones  in Figure 
\ref{UspDecAnQ}, \ref{UspDecAnQ3} and \ref{UspDecAnQ5} respectively.
In the following we present for each case the final superpotential, obtained after integrating out the possible massive deformations, together with the corresponding table of charges:
\begin{enumerate}
    \item For the deformation \eqref{WEoddDef1}, we found
    \begin{equation}
        \label{Wmag4}
        W_{\text{mag}}^{(1)} =\det \left(M_1 \,|\, M_2\right) + B \left(M_1 \tilde{B}_1 + M_2 \tilde{B}_2 \right).
    \end{equation}
    \begin{equation}
    \begin{array}{c|c|c|c|c|c|c|c|}
    &\USp(2n-2)&\SU(2n+2)&\SU(4)&\UU(1)_{1}&\UU(1)_{2}&\UU(1)_{R}\\
    \hline
    M_1&\cdot&\overline{\square}&\square&1&1&\frac{4}{3}\\
    M_2&\square&\overline{\square}&\cdot&-\frac{1}{n}&-\frac{2}{n}&-\frac{5}{3(n-1)}\\
    B&\cdot&\square&\cdot&2n+1&0&\frac{2}{3}(2n+1)\\
    \tilde B_1&\cdot&\cdot&\overline{\square}&-2(n+1)&-1&-\frac{4n}{3}\\
    \tilde B_2&\square&\cdot&\cdot&-\frac{2n^2+n-1}{n}&\frac{2}{n}& \frac{1-4n(n-2)}{3(n-1) }\\
    R_1&\square&\cdot&\cdot&n+1&2&\frac{4n^2+14n-3}{3(n-1)}
    \end{array}
    \end{equation}
    \item For the deformation \eqref{WEoddDef2}, we found
    \begin{equation}
        \label{Wmag5}
        W_{\text{mag}}^{(2)} = \left|\epsilon_{\mu \nu \rho}\right| \sigma_{\mu} R_\nu R_\rho + \det \left(M_1 \,|\, M_2\right) + B \left(M_1 \tilde{B}_1 + M_2 \tilde{B}_2 \right).
    \end{equation}
    \begin{equation}
    \begin{array}{c|c|c|c|c|c|c|c|}
    &\USp(2n)&\SU(2n+2)&\SU(2)&\SU(3)&\UU(1)_{1}&\UU(1)_{2}&\UU(1)_{R}\\
    \hline
    M_1&\cdot&\overline{\square}&\square&\cdot&1&1&\frac{4}{3}\\
    M_2&\square&\overline{\square}&\cdot&\cdot&-\frac{1}{n}&-\frac{1}{n}&-\frac{1}{3n}\\
    B&\cdot&\square&\cdot&\cdot&2n+1&0&\frac{2(2n+1)}{3}\\
    \tilde B_1&\cdot&\cdot&\overline{\square}&\cdot&-2(n+1)&-1&-\frac{4n}{3}\\
    \tilde B_2&\square&\cdot&\cdot&\cdot&-\frac{2n^2+n-1}{n}&\frac{1}{n}&- \frac{4n^2-4n-1}{3n }\\
    R_\mu&\square&\cdot&\cdot&\square&n+1&1&\frac{4n^2+10n+1}{9n}
    \end{array}
\end{equation}
    \item For the deformation \eqref{WEoddDef3}, we found
    \begin{equation}
        \label{Wmag6}
        W_{\text{mag}}^{(3)} = \sigma_{\mu \nu} R_\mu R_\nu + \det M + B M \tilde B\,.
    \end{equation}
    \begin{equation}
    \begin{array}{c|c|c|c|c|c|}
    &\USp(2n+2)&\SU(2n+2)&\SU(5)&\UU(1)_{1}&\UU(1)_{R}\\
    \hline
    M_1&\cdot&\overline{\square}&\cdot&\frac{2n^2+5n-2}{5n}&\frac{2(2n^2+3n+10)}{15(n+1)}\\
    M_2&\square&\overline{\square}&\cdot&-\frac{1}{n}&\frac{1}{n+1}\\
    B&\cdot&\square&\cdot&2n+1&\frac{2(2n+1)}{3}\\
    \tilde B_1&\cdot&\cdot&\cdot&-\frac{2(6n^2+5n-1)}{5n}&-\frac{2n(4n+1)}{5(n+1)}\\
    \tilde B_2&\square&\cdot&\cdot&-\frac{2n^2+n-1}{n}&- \frac{4n^2-1}{3(n+1) }\\
    R_\mu&\square&\cdot&\square&\frac{2n+1}{n}&\frac{4n^2+6n+5}{15(n+1)}
    \end{array}
\end{equation}
\end{enumerate}
The superscript in the superpotential refers to the numeration of the electric baryonic superpotential deformation.

We can use the rules of the IP dualities in each case to further dualize the models, obtaining an  $\USp(2)$ gauge theory with $2m+6$ fundamentals, where $m$ has been specified above. 
It is also possible to study the integral identities associated to the superconformal indices. The derivation is straightforward and we omit the details.

\subsubsection{An alternative approach}
\label{alt2}

Following the discussion of section \ref{alt1}, we can also derive the dualities in a less canonical way by adding a confining auxiliary gauge group without deconfining the antisymmetric tensor.
This analysis, for the deformations \eqref{WEevenDef1} and \eqref{WEevenDef2}, allows one to find the operator that Higgses the gauge group, while it does not in the case of \eqref{WEevenDef3}.
Again, we follow the stepwise procedure spelled out in \ref{alt1}.

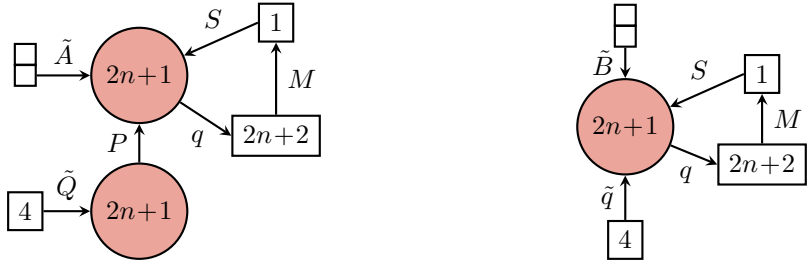
\begin{figure}
    \centering
    \begin{minipage}[b]{0.45\linewidth}
        \centering
        \makebox[\textwidth][c]{
        \begin{tikzpicture}[
            every node/.style={font=\footnotesize},
            box/.style={rectangle, draw, thick},
        ]
        \pgfmathsetmacro{\x}{1}
        \pgfmathsetmacro{\y}{1.5}
        \pgfmathsetmacro{\z}{1.8}
        \pgfmathsetmacro{\a}{1.5}
        \pgfmathsetmacro{\delta}{0.28}
        \node[fill=SUcol,circle,draw,thick] (SUgauge2) at (0, \z) {$2n\!+\!1$};
        \node[fill=SUcol,circle,draw,thick] (SUgauge) at (0, 0) {$2n\!+\!1$};
        \node[box] (fond) at (\z,\x) {$2n\!+\!2$};
        \node[box] (afond) at (\y,0) {$4$};
        \node[box] (sing) at (\z,2.5*\x) {$1$};
        \draw[<-, thick, >=stealth] (SUgauge) -- node[above] {$\tilde{Q}$} (afond);
        \draw[->, thick, >=stealth] (SUgauge) -- node[left] {$P$} (SUgauge2);
        \draw[->, thick, >=stealth] (SUgauge2) -- node[below,xshift=-3pt] {$q$} (fond.west);
        \draw[<-, thick, >=stealth] (SUgauge2) -- node[above,xshift=-3pt] {$S$} (sing.west);
        \draw[->, thick, >=stealth] (fond) -- node[right] {$M$} (sing);
        \node[box, minimum size=0.2cm] (square1) at (-\a, 0) {};
        \node[box, minimum size=0.2cm] (square2) at (-\a, \delta) {};
        \draw[->, thick, >=stealth] (square1) -- node[above] {$\tilde{A}$} (SUgauge);
        \end{tikzpicture}
        }
    \end{minipage}
    \begin{minipage}[b]{0.45\linewidth}
        \centering
        \makebox[\textwidth][c]{
        \begin{tikzpicture}[
            every node/.style={font=\footnotesize},
            box/.style={rectangle, draw, thick},
        ]
        \pgfmathsetmacro{\a}{-1.2}
        \pgfmathsetmacro{\x}{1}
        \pgfmathsetmacro{\y}{1.5}
        \pgfmathsetmacro{\z}{1.8}
        \pgfmathsetmacro{\w}{2.7}
        \pgfmathsetmacro{\v}{-0.15}
        \pgfmathsetmacro{\delta}{0.28}
        \node[fill=SUcol,circle,draw,thick] (SUgauge2) at (0, \y) {$2n\!+\!1$};
        \node[box] (fond) at (\z,\x) {$2n\!+\!2$};
        \node[box] (afond) at (0,0) {$4$};
        \node[box] (sing) at (\z,2.2*\x) {$1$};
        \draw[->, thick, >=stealth] (afond) -- node[left] {$\tilde{q}$} (SUgauge2);
        \draw[->, thick, >=stealth] (SUgauge2) -- node[below,xshift=-3pt] {$q$} (fond.west);
        \draw[<-, thick, >=stealth] (SUgauge2) -- node[above,xshift=-3pt] {$S$} (sing.west);
        \draw[->, thick, >=stealth] (fond) -- node[right] {$M$} (sing);
        \node[box, minimum size=0.2cm] (square1) at (0, \w) {};
        \node[box, minimum size=0.2cm] (square2) at (0, \w+\delta) {};
        \draw[->, thick, >=stealth] (square1) -- node[left] {$\tilde{B}$} (SUgauge2);
        \end{tikzpicture}
        }
    \end{minipage}
    \caption{Alternative deconfinement for the $\SU(2n+1)$ case in terms of a second $\SU(2n+1)$ gauge node. In this case we do not deconfine the conjugate antisymmetric $\tilde A$, but rather the fundamentals and the antifundamental $\tilde Q_5$. The next step consists of confining the original $\SU(2n+1)$ gauge node that in this quiver has a conjugate  antisymmetric, $4$ fundamentals and $2n+1$ antifundamentals \cite{Pouliot:1995me,Csaki:1996zb}. The final quiver, represented on the right of the Figure, can be further reduced once the first two baryonic deformations  \eqref{WEoddDef1} and  \eqref{WEoddDef2} are added. Indeed these deformations trigger a partial Higgs flow, breaking $\SU(2n+1)$ to a symplectic gauge node as explained in the text. On the other hand, the deformation \eqref{WEoddDef3}  gives a self duality and the operator triggering the Higgs flow does not follow from the $F$-terms in this case.}
    \label{UspPincOdd}
\end{figure}

\begin{itemize}
\item First, we turn off the baryonic superpotential deformation.
\item Then we build the $\SU(2n+1) \times \SU(2n+1)$ quiver in Figure \ref{UspPincOdd}, with superpotential
\begin{equation}
W  = M S q + X P^{2n+1} + Y q^{2n+1}\,,
\end{equation}
that coincides with the original one, after confining back the $\SU(2n+1)$ node at the bottom of the figure.
We can verify this by confining the $\SU(2n+1)$ nodes in terms of its mesons
$M_1 = q S$ and $M_2 = P q \equiv Q$, its baryon $B = q^{2n+1}$, and anti-baryons
 $\tilde B_1 = P^{2n+1}$  and $\tilde B_2 = P^{2n} S \equiv \tilde Q_5$.
 The confining superpotential for this node is 
\begin{equation}
W  = M_1^{2n+1} Q +  M M_1 + X \tilde B_1 + Y B + B(M_1 \tilde B_1 + Q \tilde Q_5) \,.
\end{equation}
When we integrate out the massive fields, the superpotential vanishes and we find that the flipper $Y$ corresponds to 
the mesonic term $Q \tilde Q_5$ on the electric side.
\item We then observe that the original $\SU(2n+1)$ node is also confining \cite{Pouliot:1995me,Csaki:1996zb}, giving rise to an $\SU(2n+1)$ theory with a conjugate antisymmetric, $2n+1$ fundamentals and five antifundamentals, represented in the second picture in Figure  \ref{UspPincOdd}.
The gauge and field content coincide with the one of the original theory, but we have a different superpotential, corresponding to 
\begin{equation}
W  = q M S + Y q^{2n+1} + \alpha \tilde q^3 \tilde B^{n-1} + \beta \tilde q \tilde B^{n}\,,
\end{equation}
where the singlets are $\alpha= \tilde A^{n} \tilde Q $ and $\beta = \tilde A^{n-1} \tilde Q^3$. The four new antifundamentals of $\SU(2n+1)$ are $\tilde q_{1,\dots,4} = P \tilde Q_{1,\dots,4} $, while the new conjugate antisymmetric is $\tilde B = \tilde A P^2$.
\item So far in the analysis we have not yet distinguished among the three baryonic deformations.
When the deformations are turned on the electric side their effect consists of an extra superpotential term respectively of the form 
\begin{eqnarray}
\label{Wmag1defodd}
\Delta W_{\text{mag}}^{(1)} &=&  \alpha  \,,
\\
\label{Wmag2defodd}
\Delta W_{\text{mag}}^{(2)} &=& \beta_{123} \,,
\\
\label{Wmag3defodd}
\Delta W_{\text{mag}}^{(3)} &=& \tilde B^{n-2} \tilde q^4 S \,.
\end{eqnarray}
\end{itemize}
In the first two cases the deformations trigger an Higgs flow. The details of these flows are summarized as follows:
\begin{enumerate}
\item In the first case, the Higgsing is due to the operator $F_{\alpha}$, that gives a vev to the operator $\tilde q^{3} \tilde B^{n-2}$. The gauge group Higgses to $\USp(2n-2)$, yielding the same dual expected from the canonical tensor deconfinement discussed in the subsection above.
\item In the second case, the Higgsing is due to the operator $F_{\beta_{123}}$, that gives a vev to the operator $ \tilde q_4  \tilde B^{n}$. The gauge group Higgses to $\USp(2n)$, yielding the same dual expected from the canonical tensor deconfinement discussed in the subsection above.
\item In the third case, the deformation has a self dual behavior, because it is mapped to an analog deformation in the dual phase. The operator triggering the possible Higgsing cannot be extracted in this case from the $F$-term analysis. One should look at the pole structure directly at the level of the index, or find other equivalent phases where the deformation $\tilde A^{n-2} \tilde Q^5$ triggers an Higgs flow. We will skip this case in the following.
\end{enumerate}

\subsubsection{Relation with the literature and phases of the dual theories}
\label{litpouliot}
Similarly to the case of even $N$, we can compare our results for the odd case with the ones in the literature. The odd case in the absence of baryonic deformations has been studied through tensors deconfinement by \cite{Pouliot:1995me}.
In this case, the analysis did not introduce spurious symmetries and we will thus not reproduce it again. The final superpotential for the $\SU(2) \times \USp(2)$ quiver is given in {\bf Formula 3.1} of \cite{Terning:1997jj}.
The deformations $\tilde A^{n} \tilde Q_5$ and $\tilde A^{n-2} \prod_{i=1}^5 \tilde Q_i$
give rise to linear terms in the superpotential, Higgsing an $\USp(2)$ gauge node, and leaving us with the same models found using the other prescription of subsection \ref{sec:4dodd}.

 The deformation $A^{n-1} \tilde Q_3\tilde Q_4\tilde Q_5 $
gives rise to a mass term. One $\USp(2)$  s-confines, 
 and we are again left with the same model found using the other prescription of subsection \ref{sec:4dodd}

Similarly to the even case,  each of the three dualities found at the beginning of this subsection have an $\USp(2m)$ with $2m+6$ fundamentals with either $m=n-1$, $m=n$, or $m=n+1$.
By further applying IP duality, we obtain the model found above starting from the dual IR free theories obtained by  \cite{Pouliot:1995me}. For this reason, in the odd case the electric theory is in a free magnetic phase.

\subsection{Comments on $a$-maximization and $a$-theorem}
\label{amaxculo}

The undeformed $\SU(N)$ theory with one conjugate antisymmetric, $N+1$ fundamentals and 5 antifundamentals has a mixed dynamics in the IR: a non-Abelian Coulomb phase and a free magnetic phase coexist.
This fact is an obstruction for the usual analysis to prove that the deformations \eqref{WEevenDef1}, \eqref{WEevenDef2}, \eqref{WEevenDef3}, \eqref{WEoddDef1}, \eqref{WEoddDef2}, \eqref{WEoddDef3} are dangerously irrelevant.
Indeed, one should start from the undeformed theory, perform the $a$-maximization procedure to obtain the exact superconformal $R$ charges and then compute the corresponding dimensions of the operators under consideration.

However, this procedure assumes that along this RG flow trajectory we reach an IR conformal fixed point, where the scaling dimension of these operator is unambiguously defined. 
Even if we try to do the $a$-maximization we do not obtain real solutions for the extremization problem. 
This suggests that the undeformed theory does not admit a standard interacting SCFT description along this trajectory.

On the other hand, we directly showed in the previous sections that the addition of the baryonic deformations resolves the ambiguity associated with the mixed phase and in each case the deformed theory flows to a free magnetic phase in the IR.
This suggests that such deformations, which are irrelevant in the UV, become relevant in the IR because they effectively change the IR dynamics.

However, one may wonder if these RG flow trajectories are indeed allowed.
Therefore we check if these flows are compatible with the a-theorem.
We start our analysis with the first case: $N=2n$ and $W = \tilde{A}^n$.
The anomaly cancellation of the $R$-symmetry imposes
\begin{equation}
    \label{rano}
    \frac{2n-2}{2}\left(R_{\tilde A}-1\right) + 
    \frac{2n+1}{2}\left(R_{Q}-1\right) + 
    \frac{5}{2}\left(R_{\tilde Q}-1\right) + 
    2n = 0\,,
\end{equation}
while, from the superpotential \eqref{WEevenDef1}, we have that $R_{\tilde A}=\frac{2}{n}$.

In the $a$-maximization procedure, it can happen that some gauge invariant operators violate the unitarity bound having $R$ charge smaller than $\frac{2}{3}$.
The interpretation is that such operators, when they hit the unitarity bound, decouple from the interacting theory becoming free.
This generates new accidental $\UU(1)$ symmetries that mix with the $R$-symmetry and change the exact $R$ charge \cite{Barnes:2004jj}.

Concretely, we look for local maxima of 
\begin{equation}
    \label{atot}
    a(R) = \frac{3}{32} 
    \left(3 \Tr R^3 - \Tr R\right).
\end{equation}
For each operator $\mathcal{O}$ violating the bound, we have to add a free field contribution to the function $a(R)$, and then subtract the contribution of the operator itself \cite{Kutasov:2003iy}:
\begin{equation}
    \label{atotmod}
    a(R) \;\longmapsto\; a(R) + \dim(\mathcal{O})\, g(R)\,,
    \qquad
    g(R) = \frac{3}{32}\left(\frac{2}{9}-\left(3\left(R_{\mathcal{O}}-1\right)^3-\left(R_{\mathcal{O}}-1\right)\right)\right),
\end{equation}
where $\dim(\mathcal{O})$ is the number of indipendent components of $\mathcal{O}$.

One first considers the theory with all its gauge invariant operators, maximizes $a(R)$ and looks for gauge invariant operators whose $R$ charge is less than $\tfrac{2}{3}$.
If there is more than one operator violating the unitarity bound, then one decouples the one with smallest $R$ charge and then performs $a$-maximization again with the modified function $a(R)$ \eqref{atotmod}.
This procedure is iterated until no operators violate the unitarity bound.

\begin{figure}[htbp]
\centering

\begin{tikzpicture}
\begin{axis}[
    width=0.85\textwidth,
    height=0.55\textwidth,
    xlabel={$n$},
    ylabel={$R$},
    xmin=16,
    xmax=110,
    grid=none,
    xtick distance=25,
    ytick distance=0.1,
    tick align=outside,
    legend style={
        at={(0.98,0.98)},
        anchor=north east,
        draw=none,
        fill=white,
        fill opacity=0.85,
        text opacity=1
    },
    legend cell align={left},
]
\definecolor{mustard}{RGB}{218, 165, 32}
\addplot[
    only marks,
    mark=*,
    mark size=2pt,
    color=mustard,
    samples at={17,...,109}
]
{2/x};
\addlegendentry{$R_{\tilde{A}}$}
\definecolor{mutedburgundy}{RGB}{150,65,95}
\addplot[
    only marks,
    mark=*,
    mark size=2pt,
    color=mutedburgundy,
    samples at={17,...,109}
]
{
-(
    105*x^4
    +135*x^3
    +306*x^2
    -5*x^2*sqrt(41*x^4 - 458*x^3 - 147*x^2 + 692*x + 484)
    +48*x
 )
 /
 (
    6*x^2*(25*x^3 + 7*x^2 + 32*x + 8)
 )
};
\addlegendentry{$R_Q$}
\definecolor{verdepetrolio}{RGB}{60,99,113}
\addplot[
    only marks,
    mark=*,
    mark size=2pt,
    color=verdepetrolio,
    samples at={17,...,109}
]
{
 (
    42*x^5
    +195*x^4
    +183*x^3
    +234*x^2
    -2*(x^2*sqrt(41*x^4 - 458*x^3 - 147*x^2 + 692*x + 484) - 24)*x
    -x^2*sqrt(41*x^4 - 458*x^3 - 147*x^2 + 692*x + 484)
 )
 /
 (
    6*x^2*(25*x^3 + 7*x^2 + 32*x + 8)
 )
};
\addlegendentry{$R_{\tilde Q}$}
\end{axis}
\end{tikzpicture}
\caption{Exact $R$ charges for fundamental fields $\tilde{A}$, $Q$, $\tilde{Q}$ in the conformal fixed point reached with the UV deformation $W=\tilde{A}^n$. The $a$-maximization is well defined only for $n \ge 17$. For smaller values of $n$ there is no real extrema for $a(R)$.}
\label{ExactRCharges}
\end{figure}

In the case under exam, we found that the operators that decouple are $Q \tilde{Q}$, $\tilde{A} Q^2$ and $Q^{2n}$.
After decoupling them according to \eqref{atotmod}, we reach an IR conformal fixed point.
We do not write the analytic expressions of the final exact $R$ charges as functions of $n$ because they are not insightful. 
Instead, we plot them in Figure\,\ref{ExactRCharges}.
Notice that in this example the $a$-maximization is well defined only for $n \ge 17$.
Below this value the function $a(R)$ has no real extrema.

\begin{figure}[htbp]
\centering

\begin{tikzpicture}
\begin{axis}[
    width=0.85\textwidth,
    height=0.55\textwidth,
    xlabel={$n$},
    ylabel={$a_{\text{UV}}-a_{\text{IR}}$},
    xmin=16,
    xmax=101,
    xtick distance=25,
    ytick distance=10000,
    grid=none,
    minor tick num=1,
    tick align=outside,
]
\definecolor{verdepetrolio}{RGB}{60,99,113}
\addplot[
    only marks,
    mark=*,
    mark size=2pt,
    color=verdepetrolio
]
coordinates{
    (17, 708.9314237538954)
    (18, 785.678159074405)
    (19, 866.4210185349391)
    (20, 951.1606981989272)
    (21, 1039.897734536689)
    (22, 1132.632548684027)
    (23, 1229.365476341041)
    (24, 1330.0967886343244)
    (25, 1434.8267070538247)
    (26, 1543.5554143769518)
    (27, 1656.2830628029742)
    (28, 1773.0097801049653)
    (29, 1893.7356743463927)
    (30, 2018.4608375413625)
    (31, 2147.1853485262604)
    (32, 2279.9092752350516)
    (33, 2416.632676518396)
    (34, 2557.3556036101327)
    (35, 2702.0781013185433)
    (36, 2850.8002090009136)
    (37, 3003.5219613660583)
    (38, 3160.2433891392666)
    (39, 3320.96451961641)
    (40, 3485.6853771281876)
    (41, 3654.405983431058)
    (42, 3827.126358038009)
    (43, 4003.8465184996808)
    (44, 4184.566480644349)
    (45, 4369.286258783572)
    (46, 4558.005865889109)
    (47, 4750.725313745735)
    (48, 4947.444613083572)
    (49, 5148.163773693201)
    (50, 5352.882804526024)
    (51, 5561.601713212399)
    (52, 5774.320506714042)
    (53, 5991.039191705344)
    (54, 6211.757774598372)
    (55, 6436.476261565869)
    (56, 6665.194658562248)
    (57, 6897.912971340616)
    (58, 7134.631205257755)
    (59, 7375.349365303433)
    (60, 7620.067456128177)
    (61, 7868.785482068953)
    (62, 8121.50344717208)
    (63, 8378.22135521383)
    (64, 8638.939209719877)
    (65, 8903.657014982864)
    (66, 9172.374775078208)
    (67, 9445.092493878666)
    (68, 9721.81017506869)
    (69, 10002.527822157011)
    (70, 10287.245438488986)
    (71, 10575.963027257695)
    (72, 10868.680591514837)
    (73, 11165.398134180037)
    (74, 11466.115658049305)
    (75, 11770.833165802857)
    (76, 12079.550660012392)
    (77, 12392.268143147898)
    (78, 12708.985617583084)
    (79, 13029.703085600428)
    (80, 13354.420549395869)
    (81, 13683.13801108319)
    (82, 14015.855472698117)
    (83, 14352.572936202152)
    (84, 14693.29040348514)
    (85, 15038.007876368604)
    (86, 15386.72535660885)
    (87, 15739.442845899864)
    (88, 16096.160345875025)
    (89, 16456.87785810961)
    (90, 16821.59538412214)
    (91, 17190.312925376524)
    (92, 17563.030483283096)
    (93, 17939.748059200514)
    (94, 18320.465654437568)
    (95, 18705.18327025486)
    (96, 19093.890816597446)
    (97, 19486.607670723195)
    (98, 19883.32451000798)
    (99, 20284.041334920235)
    (100, 20688.758145908923)
};
\end{axis}
\end{tikzpicture}
\caption{The difference $a_{\text{UV}}-a_{\text{IR}}$ as function of $n$ of the electric gauge group $\SU(2n)$. This difference is always positive, confirming that the flow under consideration is compatible with the $a$-theorem.}
\label{atheorem}
\end{figure}

Finally, we compare the IR value of $a(R)$ with the UV value, where all fields have $R$ charge set to $\tfrac{2}{3}$.
Again, the analytic expression of such difference is not very insightful, therefore we just plot it in Figure \ref{atheorem}.
We obtained that $a_{\text{UV}} > a_{\text{IR}}$, confirming, as mentioned before, that this flow is compatible with the $a$-theorem.

We have been very explicit in the discussion of the first deformation \eqref{WEevenDef1} in the even rank case.
However, we applied the same procedure to the other cases, obtaining the same results for the deformations \eqref{WEevenDef2}, \eqref{WEevenDef3}, \eqref{WEoddDef1}, \eqref{WEoddDef2}, \eqref{WEoddDef3}.
In each of these cases, after decoupling the operators $Q \tilde{Q}$, $\tilde{A} Q^2$ and $Q^{2n}$, we reach an IR conformal fixed point and we find that the flow is compatible with the $a$-theorem.

Finally, we stress that, once we include the deformations in the superpotential, this fixes the $R$ charge of the corresponding operator.
We showed that the RG trajectories selected by these deformations are compatible with the $a$-theorem.
Together with the fact that the deformations resolve the mixed phase, this provides evidence that they are dangerously irrelevant.


\section{2d dualities}
\label{sec:2d_dualities}

In this section, we study the reduction to 2d of the 4d dualities studied above.
The reduction follows the prescription of \cite{Gadde:2015wta}, that consists of considering the twisted compactification 
on $S^2$ by turning on an $R$-symmetry flux. There is a choice of $R$ charges that preserves the duality and gives rise to a well defined 2d duality starting from a 4d one, namely by assigning integer, non-negative $R$ charges.
 
 This is motivated by the study of the $S^2 \times T^2$ twisted index \cite{Benini:2015noa}. Supposing the 4d index matches across a duality, 
 the choice of non-negative $R$ charges restricts the matching of the 2d elliptic genera to the zero flux sector.
The matching of the 2d elliptic genera then follows from the matching of the 4d twisted indices  in the zero flux sector.
On the other hand, proving the matching of the elliptic genera by independent arguments provides additional support to our assumption of a matching between the 4d indices.
 In the following we focus on the assignations $R=0,1,2$, since higher charges introduce spurious non-Abelian global symmetries.
 Upon KK reduction to 2d $\mathcal{N}=(0,2)$ \cite{Closset:2013sxa}, a field with $R=0$ survives as a chiral field while a field with $R=2$ survives as a Fermi multiplet.
On the other hand, there are no zero modes for fields with $R=1$, \emph{i.e.} all the modes in the KK tower are massive.

A crucial aspect of the reduction is related to the anomalies, both gauge and global. A 2d  $\mathcal{N}=(0,2)$ field content can be non-anomalous even if the corresponding 4d field content is anomalous. This allows us to eliminate in the reduction (\emph{i.e.} to fix their $R$ charge to 1) different amounts of fundamentals and fundamentals with respect to the 4d case. Furthermore, the 4d constraint that forbids the existence of an axial symmetry can be relaxed in 2d. There are indeed non-anomalous assignments of $R$ charges that allow for the existence of axial symmetries in 2d. Such assignments require the presence of fields with $R$ charge equal to one. On the other hand, in the absence of fields with $R=1$, only $R=0,2$ are allowed and the obstruction to the generation of an axial symmetry is not lifted. This translates into a constraint on the fugacities in the elliptic genus.

Let us explain this point more concretely.
In practice,  we consider  a 4d balancing condition of the type
\begin{equation}
\prod_{a \in \Phi} {\bf u}_{a}^{2T(\rho_a)} = 
(pq)^{
    -\left(
        T(G) + 
        \sum_{a \in \Phi} T(\rho_a)(R_a-1) \right)
    } = 1
\end{equation}
On the l.h.s.  $\Phi$  represents the set of  fields charged under the gauge group of the model and the flavor Cartan fugacities ${\bf u}$ are case dependent.
On the r.h.s. on the other hand we have isolated in the exponent  the anomaly free condition for the $R$-symmetry, and indeed with such choice of charges the balancing condition is 
just $\prod_a {\bf u}_{a} =1$.
An anomaly free assignation of $R$ charges where some fields have $R_{\bar{a}}=1$ gives rise to a balancing condition where the related  ${\bf u}_{\bar{a}}$ 
fugacities are associated to 2d massive fields.
This implies that the balancing condition itself is not a constraint anymore for the fugacities that survive in the elliptic genus, which are the ones related to the 2d massless fields.

This construction is similar to the constraints imposed by the KK monopoles in the 4d/3d reduction.
Such constraints are enforced at non-perturbative level and we do not have a pure field theoretical explanation on their generation (we refer the reader to \cite{
Gadde:2015wta,Sacchi:2020pet,Amariti:2024usp,Amariti:2025jvi} 
for discussions in this direction).

Here, there is a further constraint on the fugacities due to the presence of a 4d superpotential. This forces us to assign the $R$ charges consistently with this constraint as well. If there are fields with $R=1$ involved in the superpotential, this choice lifts such interaction in 2d, otherwise the interactions survives as a 2d $J$-term. This is because one field involved in the superpotential has $R$ charge $R=2$, while all the other fields have $R=0$.

We will in general use a notation with square brackets $[\,\cdot\;,\,\cdot\,]$ to indicate the lift of the 4d superpotential (first entry) and of the 
 4d anomaly cancellation (second entry).
 We use an X to indicate that the relative constraint is lifted and a V to indicate that it still holds in 2d.
In the following we always set $R_{\tilde A}=0$, \emph{i.e.} the models have a 2d conjugate antisymmetric chiral 
$\tilde a$, 
while we indicate the other fields as $\square_\chi$ for the chiral fields 
$q$
in the fundamental representation of the gauge group, $\overline \square_\chi$ for the chiral fields $\tilde q$
in the antifundamental representation of the gauge group, and 
$\square_\psi$ for the fundamental Fermi.

When we apply the prescription of \cite{Gadde:2015wta} four possibilities are allowed, even if they are not realized for each of the six deformations that we studied in 4d. Such four possibilities are
\begin{itemize}
\item 2d models \emph{without} any $J$-terms for the charged matter fields (except for flippers that can in principle be added by hand) and \emph{without} extra non-perturbative constraints on the global symmetry structure. Such possibility is labeled by $[\mathrm{X},\mathrm{X}]$ and it cannot be realized when considering either $\SU(2n)$ with the deformation (\ref{WEevenDef1}) or $\SU(2n+1)$ with the deformation (\ref{WEoddDef1}).
\item 2d models \emph{without} any $J$-terms for the charged matter fields but \emph{with} extra non-perturbative constraints on the global symmetry structure that prevent the generation of axial symmetries. Such possibility is labeled by $[\mathrm{X},\mathrm{V}]$ and, again, it cannot be realized when considering either $\SU(2n)$ with the deformation (\ref{WEevenDef1}) or $\SU(2n+1)$ with the deformation (\ref{WEoddDef1}).
\item 2d models \emph{with} a $J$-term for a conjugate antisymmetric chiral, a fundamental Fermi and possibly other fundamental chirals, and \emph{without} extra non-perturbative constraints on the global symmetry structure.  Such possibility is labeled by $[\mathrm{V},\mathrm{X}]$ and it cannot be realized if we consider $\SU(2n)$ with the deformation (\ref{WEevenDef1}).
\item 2d models \emph{with} a $J$-term for a conjugate antisymmetric chiral, a fundamental Fermi and possibly other fundamental chirals, and \emph{with} extra non-perturbative constraints on the global symmetry structure, preventing the generation of axial symmetries. Such possibility is labeled by $[\mathrm{V},\mathrm{V}]$ and it cannot be realized if we consider $\SU(2n)$ with the deformation (\ref{WEevenDef1}).
\end{itemize}

In the following we will  focus on the realization of these four scenarios in the case of $\SU(2n)$ with the deformation (\ref{WEevenDef2}).
Before providing  a detailed analysis for the various possibilities, we summarize the results in \eqref{summary}.
In the first column we distinguish the lift of the 4d superpotential and of the anomaly. Such lifts depend on the $R$ charge assignation, and for this reason we provide in the following three columns the leftover matter content (excluding the conjugate antisymmetric chiral that is always in the spectrum). The last column refers to the type of 2d duality obtained after tensor deconfinement and sequential dualities. Observe that, from the reduction procedure, we always expect a 2d $\USp(2n-2)$ dual gauge theory.
However, in some cases a further 2d duality is possible, so that the final model becomes a Landau-Ginzburg    (LG) theory.
\begin{equation}
\label{summary}
\begin{array}{c|c|c|c|c}
[ W_{4\mathrm{d}}, R ] & n_{\square_\chi} & n_{\overline \square_\chi}& n_{\psi} &\text{Duality} \\
\hline
\hline
[\mathrm{X},\mathrm{X}]&2n+1&1&0&\USp(2n-2)\\
\hline
[\mathrm{X},\mathrm{X}]&2n&2&0&\text{LG}\\
\hline
[\mathrm{X},\mathrm{X}]&2n-1&3&0&\text{LG}\\
\hline
\hline
[\mathrm{X},\mathrm{V}]&2n+1&2&1&\USp(2n-2)\\
\hline
[\mathrm{X},\mathrm{V}]& 2n&3&1&\USp(2n-2)\\
\hline
\hline
[\mathrm{V},\mathrm{X}]&2n-1&3+1&1&\USp(2n-2)\\
\hline
[\mathrm{V},\mathrm{X}]&2n+1&1+1&1&\USp(2n-2)\\
\hline
[\mathrm{V},\mathrm{X}]&2n&2+1&1&\USp(2n-2)\\
\hline
\hline
[\mathrm{V},\mathrm{V}]&2n&3+1&1+1&\USp(2n-2)\\
\hline
[\mathrm{V},\mathrm{V}]&2n+1&2+1&1+1&\USp(2n-2)\\
\end{array}
\end{equation}
The models of the type $[\mathrm{X},\mathrm{X}]$ have already  been discussed in the literature. The first one is a duality that appeared in \cite{Jiang:2024ifv}, while the other
two cases have been studied in \cite{Amariti:2024usp}.
The models of the type $[\mathrm{X},\mathrm{V}]$ give rise to new  dualities that have not appeared previously in the literature. Observe that the second one in this list has the same field content of  a confining duality that one may in principle think to be derivable from the 4d $\SU(2n)$ confining duality with a conjugate  antisymmetric, $2n$ fundamentals and 
four. antifundamentals, by choosing an $R$-symmetry assignment with one $R$ charge $R=2$ for an antifundamental and $R=0$ for all the other fields. The difference between the duality found here and the confining one that one would obtain with such an assignment  lays in the allowed fugacities for the global symmetries. The choice made here does not allow to further dualize the final $\USp(2n-2)$ gauge theory to a LG model.
The models of type $[\mathrm{V},\mathrm{X}]$ and $[\mathrm{V},\mathrm{V}]$ give rise to 2d  $\SU(n)/\USp(2n-2)$ dualities as well.

In the following, we perform a case by case discussion of the 2d dualities summarized in (\ref{summary}). The generalization to the other $\SU(2n)$ and $\SU(2n+1)$ models is straightforward and we leave the analysis to the interested reader.

\subsection{[X, X] cases: models without charged Fermi fields}
\label{subsec:2dXX}

Here we study models without any charged fundamental Fermi field, where the 4d superpotential and the anomaly constraints are both lifted.
These models are obtained by assigning $R=1$ to the fields $\tilde Q_4$ and $\tilde Q_5$. Such an assignment is compatible with the 
constraints from the anomalies if the $R$ charge  of other two additional fundamental or antifundamental are set to $R=1$.
The possible choices are summarized in the first three  lines of \eqref{summary}.
In the following we will discuss the 2d duality that arises in each case.

\subsubsection*{$2n+1$ fundamental and one antifundamental chirals}

In this case the reduction is performed by assigning $R$ charge equal to one to the fields $\tilde Q_{2,3,4,5}$, while all the other charged fields have $R=0$. 

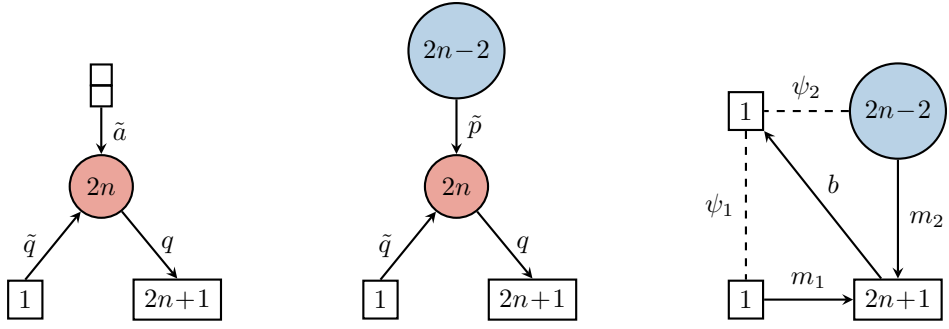
\begin{figure}
    \centering
    \begin{minipage}[b]{0.30\linewidth}
        \centering
        \makebox[\textwidth][c]{
        \begin{tikzpicture}[
            every node/.style={font=\footnotesize},
            box/.style={rectangle, draw, thick},
        ]
        \pgfmathsetmacro{\x}{1}
        \pgfmathsetmacro{\y}{1.5}
        \pgfmathsetmacro{\z}{1.2}
        \pgfmathsetmacro{\delta}{0.28}
        \node[fill=SUcol,circle,draw,thick] (SUgauge) at (0, 0) {$2n$};
        \node[box] (fond) at (\x,-\y) {$2n\!+\!1$};
        \node[box] (afond) at (-\x,-\y) {$1$};
        \draw[->, thick, >=stealth] (SUgauge) -- node[right] {$q$} (fond.north);
        \draw[<-, thick, >=stealth] (SUgauge) -- node[left, xshift=-2pt] {$\tilde{q}$} (afond.north);
        \node[box, minimum size=0.2cm] (square1) at (0, \z) {};
        \node[box, minimum size=0.2cm] (square2) at (0, \z+\delta) {};
        \draw[->, thick, >=stealth] (square1) -- node[right] {$\tilde{a}$} (SUgauge);
        \end{tikzpicture}
        }
    \end{minipage}
    \begin{minipage}[b]{0.30\linewidth}
        \centering
        \makebox[\textwidth][c]{
        \begin{tikzpicture}[
            every node/.style={font=\footnotesize},
            box/.style={rectangle, draw, thick},
        ]
        \pgfmathsetmacro{\x}{1}
        \pgfmathsetmacro{\y}{1.5}
        \pgfmathsetmacro{\z}{1.8}
        \node[fill=USPcol,circle,draw,thick] (USPgauge) at (0, \z) {$2n\!-\!2$};
        \node[fill=SUcol,circle,draw,thick] (SUgauge) at (0, 0) {$2n$};
        \node[box] (fond) at (\x,-\y) {$2n\!+\!1$};
        \node[box] (afond) at (-\x,-\y) {$1$};
        \draw[->, thick, >=stealth] (SUgauge) -- node[right] {$q$} (fond.north);
        \draw[<-, thick, >=stealth] (SUgauge) -- node[left, xshift=-2pt] {$\tilde{q}$} (afond.north);
        \draw[->, thick, >=stealth] (USPgauge) -- node[right] {$\tilde{p}$} (SUgauge);
        \end{tikzpicture}
        }
    \end{minipage}
    \begin{minipage}[b]{0.30\linewidth}
        \centering
        \makebox[\textwidth][c]{
        \begin{tikzpicture}[
            every node/.style={font=\footnotesize},
            box/.style={rectangle, draw, thick},
        ]
        \pgfmathsetmacro{\x}{1}
        \pgfmathsetmacro{\y}{1.5}
        \pgfmathsetmacro{\w}{1}
        \pgfmathsetmacro{\z}{1.8}
        \node[fill=USPcol,circle,draw,thick] (USPgauge) at (\x, \w) {$2n\!-\!2$};
        \node[box] (fond) at (\x,-\y) {$2n\!+\!1$};
        \node[box] (afond) at (-\x,-\y) {$1$};
        \node[box] (sing) at (-\x,\w) {$1$};
        \draw[->, thick, >=stealth] (USPgauge) -- node[right] {$m_2$} (fond.north);
        \draw[-, thick, dashed] (USPgauge) -- node[above] {$\psi_2$} (sing);
        \draw[-, thick, dashed] (sing) -- node[left] {$\psi_1$} (afond);
        \draw[->, thick, >=stealth] (fond) -- node[above, xshift=4pt] {$b$} (sing.south east);
        \draw[<-, thick, >=stealth] (fond) -- node[above] {$m_1$} (afond);
        \end{tikzpicture}
        }
    \end{minipage}
    \caption{Deconfinement of the conjugate two index antisymmetric chiral $\tilde a$ for the $[\mathrm{X},\mathrm{X}]$ case, obtained by fixing $R(\tilde Q_{2,3,4,5})=1$. In the second step the original $\SU(2n)$ node is dualized to a LG, leaving us with an $\USp(2n-2)$ model, represented in the third quiver.}
    \label{2dXX2Qt}
\end{figure}

The field $\sigma$ in the superpotential of the four dimensional dual theory \eqref{Wmag2} is mapped through the duality map to the operator Pf $\tilde a$ on the electric side.
The fields $R_{4,5}$ have $R$ charge 1 and disappear from the dual theory, leaving the $\sigma$ without interaction.
It is then convenient to flip such field on the electric side. This gives rise to a Fermi $\Psi$ that flips the operator Pf $\tilde a$  in the 2d electric theory. The 2d $\mathcal{N}=(0,2)$ electric theory has therefore $J$-term $J_{\Psi} = \text{Pf } \tilde a$.
We then deconfine the conjugate antisymmetric, as in the second quiver of Figure \ref{2dXX2Qt}, and we obtain a model without any $J$ (and $E$) terms and where $\tilde p^2=\tilde  a$.
Then  we dualize the $\SU(2n)$ node, using the duality discussed in detail in {\bf Appendix 2} of \cite{Amariti:2024usp}, obtaining the third quiver in Figure \ref{2dXX2Qt}. The duality map in this case is read from the operator mapping $m_1 = \tilde q q$, $m_2 = \tilde p q$ and $b=q^{2n}$.
The charges of the Fermi $\psi_{1,2}$ on the other hand are read from the $J$-terms $J_{\psi_1} = m_1 b$ and $J_{\psi_2} =m_2 b$. 
Consistently, these $J$-terms are the same we would have obtained by the twisted compactification of the 4d dual superpotential 
(\ref{Wmag2}) by following the 4d duality dictionary from the $R$ charge assignation given above, that corresponds to $\tilde R_{4,5} = 1$
and $R_{\tilde B_2}=2$.
This is the same duality found in \cite{Jiang:2024ifv}, with the only difference of the Fermi flipper $\Gamma$ that is added on the electric side in \cite{Jiang:2024ifv} and that here is flipped and thus, correspondingly, we have its flipper, \emph{i.e.} a chiral singlet $m_1$, on the dual side.
We summarize in the table below the charge assignation for the global symmetries of the first and the last quivers of Figure \ref{Wmag2}.
\begin{equation}
    \begin{array}{c|c|c|c|c|c|}
    &\SU(2n+1)&\UU(1)_q&\UU(1)_{\tilde{q}}&\UU(1)_{\tilde{a}}&\UU(1)_{R}\\
    \hline
    q & \overline{\square} & 1 & 0 & 0 & 0 \\
    \tilde{q} & \cdot & 0 & 1 & 0 & 0 \\
    \tilde{a} & \cdot & 0 & 0 & 2 & 0 \\
    \Psi & \cdot & 0 & 0 & -2n & 1 \\
    \hline
    m_1 & \overline{\square} & 1 & 1 & 0 & 0 \\
    m_2 & \overline{\square} & 1 & 0 & 1 & 0 \\
    b & \square & 2n & 0 & 0 & 0 \\
    \psi_1 & \cdot & -1-2n & -1 & 0 & 1 \\
    \psi_2 & \cdot & -1-2n & 0 & -1 & 1 \\
    \end{array}
\end{equation}
\begin{figure}
    \centering
    \begin{minipage}[b]{0.42\linewidth}
        \centering
        \makebox[\textwidth][c]{
        \begin{tikzpicture}[
            every node/.style={font=\footnotesize},
            box/.style={rectangle, draw, thick},
        ]
        \pgfmathsetmacro{\x}{1}
        \pgfmathsetmacro{\y}{1.5}
        \pgfmathsetmacro{\z}{1.2}
        \pgfmathsetmacro{\delta}{0.28}
        \node[fill=SUcol,circle,draw,thick] (SUgauge) at (0, 0) {$2n$};
        \node[box] (fond) at (\x,-\y) {$2n\!+\!1$};
        \node[box] (afond) at (-\x,-\y) {$1$};
        \draw[->, thick, >=stealth] (SUgauge) -- node[right] {$\sigma x_\mu^{-1}$} (fond.north);
        \draw[<-, thick, >=stealth] (SUgauge) -- node[left, xshift=-2pt] {$\tau$} (afond.north);
        \node[box, minimum size=0.2cm] (square1) at (0, \z) {};
        \node[box, minimum size=0.2cm] (square2) at (0, \z+\delta) {};
        \draw[->, thick, >=stealth] (square1) -- node[right] {$\alpha^2$} (SUgauge);
        \end{tikzpicture}
        }
    \end{minipage}
    \begin{minipage}[b]{0.42\linewidth}
        \centering
        \makebox[\textwidth][c]{
        \begin{tikzpicture}[
            every node/.style={font=\footnotesize},
            box/.style={rectangle, draw, thick},
        ]
        \pgfmathsetmacro{\x}{1}
        \pgfmathsetmacro{\y}{1.5}
        \pgfmathsetmacro{\w}{1}
        \pgfmathsetmacro{\z}{1.8}
        \node[fill=USPcol,circle,draw,thick] (USPgauge) at (\x, \w) {$2n\!-\!2$};
        \node[box] (fond) at (\x,-\y) {$2n\!+\!1$};
        \node[box] (afond) at (-\x,-\y) {$1$};
        \node[box] (sing) at (-\x,\w) {$1$};
        \draw[->, thick, >=stealth] (USPgauge) -- node[right] {$\alpha \sigma x_\mu^{-1}$} (fond.north);
        \draw[-, thick, dashed] (USPgauge) -- node[above] {$\frac{q}{\alpha \sigma^{2n+1}}$} (sing);
        \draw[-, thick, dashed] (sing) -- node[left] {$\frac{q}{\tau \sigma^{2n+1}}$} (afond);
        \draw[->, thick, >=stealth] (fond) -- node[above, xshift=9pt] {$\sigma^{2n} x_{\mu}$} (sing.south east);
        \draw[<-, thick, >=stealth] (fond) -- node[above] {$\sigma \tau x_{\mu}^{-1}$} (afond);
        \end{tikzpicture}
        }
    \end{minipage}
    \caption{Fugacities associated to global symmetries for the initial electric theory and the final magnetic theory. Notice that in the first quiver we also have a Fermi singlet $  \Psi$ whose fugacities are $q \alpha^{-2n}$.}
    \label{2dXX2Qtfug}
\end{figure}
\!\!In Figure \ref{2dXX2Qtfug} we specify the global fugacities for the initial and final theories.
At the level of elliptic genera we have the following integral identity
\begin{equation}
    \label{ellgen}
    \begin{aligned}
        \theta\left(\frac{q}{\alpha^{2n}}\right)
        &\mathcal{I}_{\SU(2n)}^{[(2n+1) \square ; 1 \overline \square; 1 \overline{A}]}\left( \sigma \mathbf{x}^{-1}; \tau; \alpha^2 \right) = \\
        &
        \frac{
            \theta\left(\frac{q}{\sigma^{2n+1}\tau}\right)
        }{
            \prod_{\mu=1}^{2n+1} 
            \theta\left(\sigma \tau x_\mu^{-1}\right)
            \theta\left(\sigma^{2n} x_\mu\right)
        }
        \mathcal{I}_{\USp(2n-2)}^{[(2n+1) \square_{\chi} ; 1 \square_F]}\left( \alpha \sigma \mathbf{x}^{-1}; \frac{q}{\alpha \sigma^{2n+1}} \right).
    \end{aligned}
\end{equation}
The special function $\theta$ is defined as $\theta(z;q) = (z;q)_{\infty} (q z^{-1};q)_{\infty}$. For shortness we keep implicit the second argument that is always $q$.
Here, as in the previous section for the case of the superconformal index, we introduced a short notation for the gauge integrals.
Again, in this notation the matter content and the corresponding representations are encoded in the square brakets. 
The arguments in $\mathcal{I}_{\SU(2n)}$ and $\mathcal{I}_{\USp(2n-2)}$ are separated by a semicolon if they transform under  different representations.
On the l.h.s. we have the elliptic genus of the $\SU(2n)$ gauge theory with $2n+1$ fundamental chirals, one antifundamental chiral and one conjugate antisymmetric
\begin{equation}
    \begin{aligned}
        &
        \mathcal{I}_{\SU(2n)}^{[(2n+1) \square ; 1 \overline \square; 1 \overline{A}]}\left( \sigma \mathbf{x}^{-1}; \tau; \alpha^2 \right) = \\
        &
        \frac{(q;q)_{\infty}^{2(2n-1)}}{(2n)!}
        \oint_{\text{JK}}\!\!
        \left(\prod_{i=1}^{2n-1}\frac{\dd{z_i}}{2 \pi \mi z_i}\right)
        \prod_{i=1}^{2n}
        \frac{
            \prod_{j>i}^{2n} \theta\!\left(z_i^{\pm 1}z_j^{\mp 1}\right)
        }{
            \theta\!\left(\tau z_i^{-1}\right)
            \prod_{\mu=1}^{2n+1}\theta\!\left(\sigma x_\mu^{-1} z_i\right)
            \prod_{j>i}^{2n} \theta\!\left(\alpha^2 z_i^{-1}z_j^{-1}\right)
        }.
    \end{aligned}
\end{equation}
On the r.h.s. we have the elliptic genus of the $\USp(2n-2)$ gauge theory with $2n+1$ fundamental chirals and one fundamental Fermi
\begin{equation}
    \begin{aligned}
        &
        \mathcal{I}_{\USp(2n-2)}^{[(2n+1) \square_{\chi} ; 1 \square_F]}\left( \alpha \sigma \mathbf{x}^{-1}; \frac{q}{\alpha \sigma^{2n+1}} \right) = \\
        &
        \frac{(q;q)_{\infty}^{2n-2}}{2^{n-1}(n-1)!}
        \oint_{\text{JK}}\!\!
        \left(\prod_{i=1}^{n-1}\frac{\dd{z_i}}{2 \pi \mi z_i}\right)
        \prod_{i=1}^{n-1}
        \frac{
            \theta\!\left(\frac{q}{\alpha \sigma^{2n+1}} z_i^{\pm 1}\right)
            \theta\!\left(z_i^{\pm 2}\right)
            \prod_{j>i}^{n-1}
            \theta\!\left(z_i^{\pm 1}z_j^{\pm 1}\right)
        }{
            \prod_{\mu=1}^{2n+1} \theta\!\left(\alpha \sigma  x_\mu^{-1} z_i^{\pm 1}\right)
        }.
    \end{aligned}
\end{equation}

In this first example we have been very explicit in showing the global charges assignation and the identity between the elliptic genera.
In order to avoid too many repetitions, in the following cases we only write the duality dictionary.
From this,  one can directly obtain the global fugacities of the final magnetic theory, the table of global charges and the corresponding identity between the elliptic genera.


\subsubsection*{$2n$ fundamental and two antifundamental chirals}

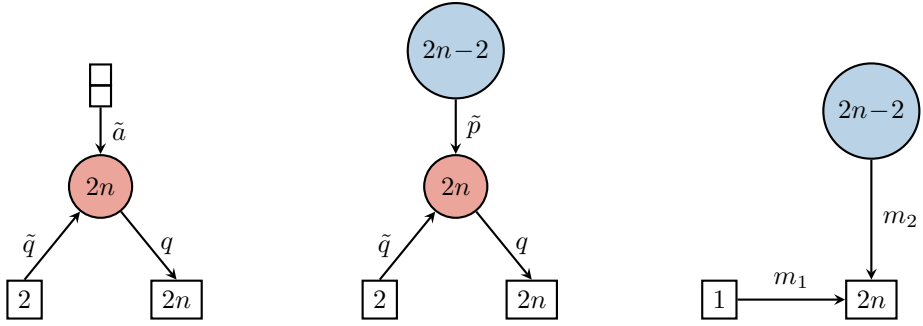
\begin{figure}
    \centering
    \begin{minipage}[b]{0.30\linewidth}
        \centering
        \makebox[\textwidth][c]{
        \begin{tikzpicture}[
            every node/.style={font=\footnotesize},
            box/.style={rectangle, draw, thick},
        ]
        \pgfmathsetmacro{\x}{1}
        \pgfmathsetmacro{\y}{1.5}
        \pgfmathsetmacro{\z}{1.2}
        \pgfmathsetmacro{\delta}{0.28}
        \node[fill=SUcol,circle,draw,thick] (SUgauge) at (0, 0) {$2n$};
        \node[box] (fond) at (\x,-\y) {$2n$};
        \node[box] (afond) at (-\x,-\y) {$2$};
        \draw[->, thick, >=stealth] (SUgauge) -- node[right] {$q$} (fond.north);
        \draw[<-, thick, >=stealth] (SUgauge) -- node[left, xshift=-2pt] {$\tilde{q}$} (afond.north);
        \node[box, minimum size=0.2cm] (square1) at (0, \z) {};
        \node[box, minimum size=0.2cm] (square2) at (0, \z+\delta) {};
        \draw[->, thick, >=stealth] (square1) -- node[right] {$\tilde{a}$} (SUgauge);
        \end{tikzpicture}
        }
    \end{minipage}
    \begin{minipage}[b]{0.30\linewidth}
        \centering
        \makebox[\textwidth][c]{
        \begin{tikzpicture}[
            every node/.style={font=\footnotesize},
            box/.style={rectangle, draw, thick},
        ]
        \pgfmathsetmacro{\x}{1}
        \pgfmathsetmacro{\y}{1.5}
        \pgfmathsetmacro{\z}{1.8}
        \node[fill=USPcol,circle,draw,thick] (USPgauge) at (0, \z) {$2n\!-\!2$};
        \node[fill=SUcol,circle,draw,thick] (SUgauge) at (0, 0) {$2n$};
        \node[box] (fond) at (\x,-\y) {$2n$};
        \node[box] (afond) at (-\x,-\y) {$2$};
        \draw[->, thick, >=stealth] (SUgauge) -- node[right] {$q$} (fond.north);
        \draw[<-, thick, >=stealth] (SUgauge) -- node[left, xshift=-2pt] {$\tilde{q}$} (afond.north);
        \draw[->, thick, >=stealth] (USPgauge) -- node[right] {$\tilde{p}$} (SUgauge);
        \end{tikzpicture}
        }
    \end{minipage}
    \begin{minipage}[b]{0.30\linewidth}
        \centering
        \makebox[\textwidth][c]{
        \begin{tikzpicture}[
            every node/.style={font=\footnotesize},
            box/.style={rectangle, draw, thick},
        ]
        \pgfmathsetmacro{\x}{1}
        \pgfmathsetmacro{\y}{1.5}
        \pgfmathsetmacro{\w}{1}
        \pgfmathsetmacro{\z}{1.8}
        \node[fill=USPcol,circle,draw,thick] (USPgauge) at (\x, \w) {$2n\!-\!2$};
        \node[box] (fond) at (\x,-\y) {$2n$};
        \node[box] (afond) at (-\x,-\y) {$1$};
        \draw[->, thick, >=stealth] (USPgauge) -- node[right] {$m_2$} (fond.north);
        \draw[<-, thick, >=stealth] (fond) -- node[above] {$m_1$} (afond);
        \end{tikzpicture}
        }
    \end{minipage}
        \caption{Deconfinement of the conjugate two index antisymmetric chiral $\tilde a$ for the 
       $[\mathrm{X},\mathrm{X}]$ case, obtained by fixing  $R(Q_{2n+1})=R(\tilde Q_{3,4,5})=1$. In the second step the original $\SU(2n)$ node is dualized to a LG, leaving us with an $\USp(2n-2)$ model, represented in the third quiver. This last case can be further dualized to a LG as explained in the text.}
    \label{2dXXQQt}
\end{figure}

The second case of type $[\mathrm{X},\mathrm{X}]$ is obtained  by assigning $R$ charge equal to one to the fields $\tilde Q_{3,4,5}$ and $Q_{2n+1}$,
while all the other charged fields have $R=0$.
Again, we flip the field $\sigma$ by adding a Fermi singlet on the electric side with the interaction $J_{\Psi} = \text{Pf } \tilde a$.
We then deconfine the conjugate antisymmetric as in the second quiver of Figure \ref{2dXXQQt} and obtain a model without any $J$ (and $E$) terms and where $\tilde p^2=\tilde  a$.
Then we dualize the $\SU(2n)$ node, using the duality discussed in detail in {\bf Appendix 1} of \cite{Amariti:2024usp}, obtaining the third quiver in Figure \ref{2dXXQQt}. The duality map in this case is read from the operator mapping $m_1 = \tilde q q$, $m_2 = \tilde p q$, $b=q^{2n}$ and $\tilde b_1 = \tilde q^2 \tilde p^{2n-2}$. In this case there is also a Fermi field $\lambda$ with $J_\lambda = \det (m_1|m_2) + b \tilde b_1$.
Consistently, this $J$-term is the same we would have obtained by the twisted compactification of the 4d dual superpotential 
(\ref{Wmag2}), by following the 4d duality dictionary from the $R$ charge assignation given above.
We conclude by observing that this last $\USp(2n-2)$ model can be further dualized to a LG model, given in terms of an antisymmetric contraction $\hat{a} =m_2^2$ with an extra interaction $J_{\hat \Psi} = \Pf \hat{a}$. Furthermore, in the final LG model, the $J$-terms for the Fermi field  $\lambda$ becomes $J_\lambda = m_1 \hat{a}^{n-1} + b \tilde b_1$. Consistently, this is the same duality found in {\bf Section 3.1} of \cite{Amariti:2024usp} with the Fermi flipper $\Psi$ on the electric side

\subsubsection*{$2n-1$ fundamental and three antifundamental chirals}

\begin{figure}
    \centering
    \begin{minipage}[b]{0.30\linewidth}
        \centering
        \makebox[\textwidth][c]{
        \begin{tikzpicture}[
            every node/.style={font=\footnotesize},
            box/.style={rectangle, draw, thick},
        ]
        \pgfmathsetmacro{\x}{1}
        \pgfmathsetmacro{\y}{1.5}
        \pgfmathsetmacro{\z}{1.2}
        \pgfmathsetmacro{\delta}{0.28}
        \node[fill=SUcol,circle,draw,thick] (SUgauge) at (0, 0) {$2n$};
        \node[box] (fond) at (\x,-\y) {$2n\!-\!1$};
        \node[box] (afond) at (-\x,-\y) {$3$};
        \draw[->, thick, >=stealth] (SUgauge) -- node[right] {$q$} (fond.north);
        \draw[<-, thick, >=stealth] (SUgauge) -- node[left, xshift=-2pt] {$\tilde{q}$} (afond.north);
        \node[box, minimum size=0.2cm] (square1) at (0, \z) {};
        \node[box, minimum size=0.2cm] (square2) at (0, \z+\delta) {};
        \draw[->, thick, >=stealth] (square1) -- node[right] {$\tilde{a}$} (SUgauge);
        \end{tikzpicture}
        }
    \end{minipage}
    \begin{minipage}[b]{0.30\linewidth}
        \centering
        \makebox[\textwidth][c]{
        \begin{tikzpicture}[
            every node/.style={font=\footnotesize},
            box/.style={rectangle, draw, thick},
        ]
        \pgfmathsetmacro{\x}{1}
        \pgfmathsetmacro{\y}{1.5}
        \pgfmathsetmacro{\z}{1.8}
        \node[fill=USPcol,circle,draw,thick] (USPgauge) at (0, \z) {$2n\!-\!2$};
        \node[fill=SUcol,circle,draw,thick] (SUgauge) at (0, 0) {$2n$};
        \node[box] (fond) at (\x,-\y) {$2n\!-\!1$};
        \node[box] (afond) at (-\x,-\y) {$3$};
        \draw[->, thick, >=stealth] (SUgauge) -- node[right] {$q$} (fond.north);
        \draw[<-, thick, >=stealth] (SUgauge) -- node[left, xshift=-2pt] {$\tilde{q}$} (afond.north);
        \draw[->, thick, >=stealth] (USPgauge) -- node[right] {$\tilde{p}$} (SUgauge);
        \end{tikzpicture}
        }
    \end{minipage}
    \begin{minipage}[b]{0.30\linewidth}
        \centering
        \makebox[\textwidth][c]{
        \begin{tikzpicture}[
            every node/.style={font=\footnotesize},
            box/.style={rectangle, draw, thick},
        ]
        \pgfmathsetmacro{\x}{1}
        \pgfmathsetmacro{\y}{1.5}
        \pgfmathsetmacro{\w}{1}
        \pgfmathsetmacro{\z}{1.8}
        \node[fill=USPcol,circle,draw,thick] (USPgauge) at (\x, \w) {$2n\!-\!2$};
        \node[box] (fond) at (\x,-\y) {$2n\!-\!1$};
        \node[box] (afond) at (-\x,-\y) {$1$};
        \node[box] (sing) at (-\x,\w) {$1$};
        \draw[->, thick, >=stealth] (USPgauge) -- node[right] {$m_2$} (fond.north);
        \draw[<-, thick, >=stealth] (USPgauge) -- node[above] {$\tilde{b}_2$} (sing);
        \draw[->, thick, >=stealth] (sing) -- node[left] {$\tilde{b}_1$} (afond);
        \draw[-, thick, dashed] (fond) -- node[above, xshift=4pt] {$\psi$} (sing.south east);
        \draw[<-, thick, >=stealth] (fond) -- node[above] {$m_1$} (afond);
        \end{tikzpicture}
        }
    \end{minipage}
     \caption{Deconfinement of the conjugate two index antisymmetric chiral $\tilde a$ for the $[\mathrm{X},\mathrm{X}]$ case, obtained by fixing  $R(Q_{2n,2n+1})=R(\tilde Q_{4,5})=1$. In the second step the original $\SU(2n)$ node is dualized to a LG, leaving us with an $\USp(2n-2)$ model, represented in the third quiver. Again, this last case can be further dualized to a LG as explained in the text.}
    \label{2dXX2Q}
\end{figure}
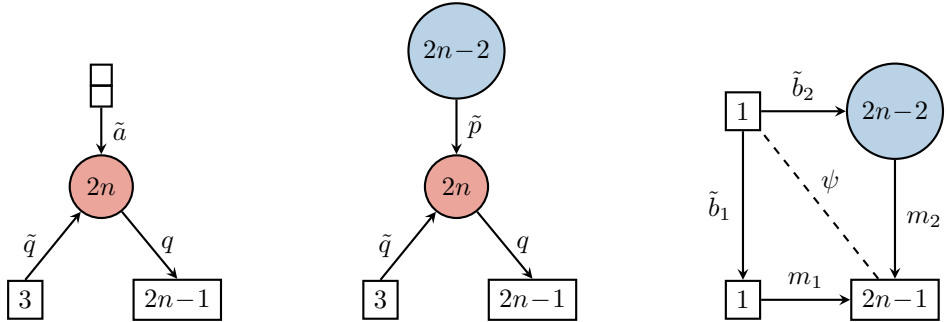

The last case of type $[\mathrm{X},\mathrm{X}]$ is obtained  by assigning $R$ charge equal to one to the fields $\tilde Q_{4,5}$ and $Q_{2n,2n+1}$,
while all the other charged fields have $R=0$.
Again, we flip the field $\sigma$ adding a Fermi singlet on the electric side with the interaction $J_{\Psi} = \text{Pf } \tilde a$.
Then we deconfine the conjugate antisymmetric, as in the second quiver of Figure \ref{2dXX2Q} and we obtain a model without any $J$ (and $E$) terms and where $\tilde p^2=\tilde  a$.
Then we dualize the $\SU(2n)$ node, using the duality discussed in detail in {\bf Appendix 1} of \cite{Amariti:2024usp}, obtaining the third quiver in Figure \ref{2dXX2Q}. The duality map in this case is read from the operator mapping $m_1 = \tilde q q$, $m_2 = \tilde p q$, $\tilde b_1=\tilde q^3 \tilde p^{2n-3}$ and $\tilde b_2=\tilde q^2 \tilde p^{2n-2}$. In this case there is also a Fermi field $\psi$ with $J_\psi =  m_1 \tilde b_1+ m_2 \tilde b_2$.
Consistently, this $J$-term is the same of the we would have obtained by the twisted compactification of the 4d dual superpotential 
(\ref{Wmag2}) by following the 4d duality dictionary from the $R$ charge assignation given above.
We conclude by observing that this last $\USp(2n-2)$ model can be further dualized to a LG model, given in terms of an antisymmetric contraction $\hat{a} =m_2^2$, while the gauge invariant contraction $\varphi=m_2 \tilde b_2 $ is set to $\varphi = m_1 \tilde b_1$ by the $J$-term $J_\psi$.
The final LG model has a further Fermi field $\Lambda$ with $J_{\Lambda}  = \hat{a}^{n-1} m_1 \tilde b_1$.
Consistently, this is the same duality found in {\bf Section 3.3} of \cite{Amariti:2024usp} with the Fermi flipper $\Psi$ on the electric side.

\subsection{[X, V] cases: models with a Fundamental Fermi}
\label{subsec:2dXV}
Here we study models with a fundamental Fermi field, where the 4d superpotential does not give rise to a 2d $J$-term.
The models are obtained by assigning $R=1$ to the fields $\tilde Q_4$ and $\tilde Q_5$. Such an assignment is compatible with the 
constraint from the anomalies provided that the $R$ charge  of an additional fundamental or antifundamental is set to $R=2$.
The two possibilities are summarized in the fourth and fifth line of \eqref{summary}.
In the following we will discuss the 2d duality that arises in each case.

\subsubsection*{$2n$ fundamental and  three antifundamental chirals}

\begin{figure}
    \centering
    \begin{minipage}[b]{0.30\linewidth}
        \centering
        \makebox[\textwidth][c]{
        \begin{tikzpicture}[
            every node/.style={font=\footnotesize},
            box/.style={rectangle, draw, thick},
        ]
        \pgfmathsetmacro{\x}{1}
        \pgfmathsetmacro{\y}{1.5}
        \pgfmathsetmacro{\z}{1.2}
        \pgfmathsetmacro{\delta}{0.28}
        \node[fill=SUcol,circle,draw,thick] (SUgauge) at (0, 0) {$2n$};
        \node[box] (fond) at (\x,-\y) {$2n$};
        \node[box] (afond) at (-\x,-\y) {$3$};
        \node[box] (fermi) at (0,-\y) {$1$};
        \draw[->, thick, >=stealth] (SUgauge) -- node[right] {$q$} (fond.north);
        \draw[<-, thick, >=stealth] (SUgauge) -- node[left, xshift=-2pt] {$\tilde{q}$} (afond.north);
        \draw[-, thick, dashed] (SUgauge) -- node[right] {$\eta$} (fermi);
        \node[box, minimum size=0.2cm] (square1) at (0, \z) {};
        \node[box, minimum size=0.2cm] (square2) at (0, \z+\delta) {};
        \draw[->, thick, >=stealth] (square1) -- node[right] {$\tilde{a}$} (SUgauge);
        \end{tikzpicture}
        }
    \end{minipage}
    \begin{minipage}[b]{0.30\linewidth}
        \centering
        \makebox[\textwidth][c]{
        \begin{tikzpicture}[
            every node/.style={font=\footnotesize},
            box/.style={rectangle, draw, thick},
        ]
        \pgfmathsetmacro{\x}{1}
        \pgfmathsetmacro{\y}{1.5}
        \pgfmathsetmacro{\z}{1.8}
        \node[fill=USPcol,circle,draw,thick] (USPgauge) at (0, \z) {$2n\!-\!2$};
        \node[fill=SUcol,circle,draw,thick] (SUgauge) at (0, 0) {$2n$};
        \node[box] (fond) at (\x,-\y) {$2n$};
        \node[box] (afond) at (-\x,-\y) {$3$};
        \node[box] (fermi) at (0,-\y) {$1$};
        \draw[->, thick, >=stealth] (SUgauge) -- node[right] {$q$} (fond.north);
        \draw[<-, thick, >=stealth] (SUgauge) -- node[left, xshift=-2pt] {$\tilde{q}$} (afond.north);
        \draw[->, thick, >=stealth] (USPgauge) -- node[right] {$\tilde{p}$} (SUgauge);
        \draw[-, thick, dashed] (SUgauge) -- node[right] {$\eta$} (fermi);
        \end{tikzpicture}
        }
    \end{minipage}
    \begin{minipage}[b]{0.30\linewidth}
        \centering
        \makebox[\textwidth][c]{
        \begin{tikzpicture}[
            every node/.style={font=\footnotesize},
            box/.style={rectangle, draw, thick},
        ]
        \pgfmathsetmacro{\x}{1}
        \pgfmathsetmacro{\y}{1.5}
        \pgfmathsetmacro{\w}{1.8}
        \pgfmathsetmacro{\z}{1.8}
        \node[fill=USPcol,circle,draw,thick] (USPgauge) at (\x, \w) {$2n\!-\!2$};
        \node[box] (fond) at (\x,-\y) {$2n$};
        \node[box] (afond) at (-\x,-\y) {$3$};
        \node[box] (sing) at (0,0) {$1$};
        \node[box] (culo) at (-\x,\w) {$1$};
        \draw[->, thick, >=stealth] (USPgauge) -- node[right] {$m_2$} (fond.north);
        \draw[<-, thick, >=stealth] (USPgauge) -- node[right,xshift=0pt,yshift=-3pt] {$\tilde{b}_2$} (sing);
        \draw[->, thick, >=stealth] (sing) -- node[left, xshift=2pt,yshift=5pt] {$\tilde{b}_1$} (afond);
        \draw[<-, thick, >=stealth] (fond) -- node[above] {$m_1$} (afond);
        \draw[->,thick,>=stealth] (culo) -- node[left] {$b$} (sing);
        \draw[-, thick, dashed] (fond) -- node[above, xshift=4pt] {$\lambda$} (sing.south east);
        \draw[-, thick, dashed] (culo) -- node[left] {$\psi_1$} (afond);
        \draw[-, thick, dashed] (culo) -- node[above] {$\psi_2$} (USPgauge);
        \end{tikzpicture}
        }
    \end{minipage}
      \caption{Deconfinement of the conjugate two index antisymmetric chiral $\tilde a$ for the $[\mathrm{X},\mathrm{V}]$ case, obtained by fixing $R(Q_{2n+1})=2$, $R(\tilde Q_{4,5})=1$. In the second step the original $\SU(2n)$ node is dualized to a LG, leaving us with an $\USp(2n-2)$ model, represented in the third quiver.}
    \label{2dXVQ}
\end{figure}
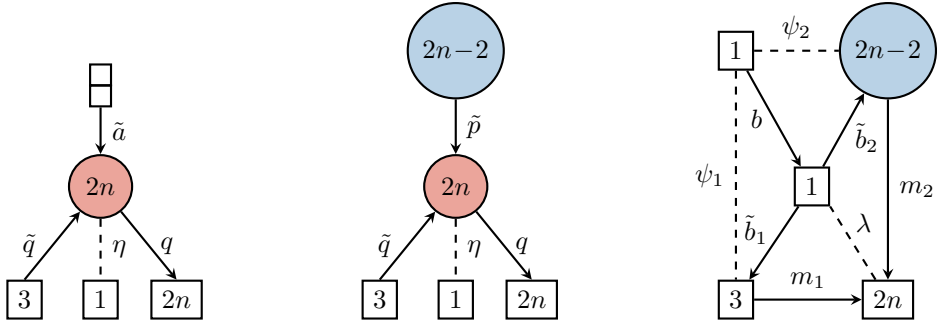

We start by assigning $R=2$ to a fundamental field.
In this case, the 4d constraint on the global symmetry structure imposed by anomaly cancellation is not lifted in 2d, and this implies that we need to focus on the
global symmetry structure by imposing this constraint. The constraint propagates when we deconfine the conjugate antisymmetric and further dualize the $\SU(2n)$ gauge node.
We further continue flipping the pfaffian operator $\mathrm{Pf}\,\tilde a$ on the electric side, \emph{i.e.} we consider a Fermi singlet $\Psi$ in the original $\SU(2n)$ theory with $J_\Psi=\text{Pf}\,\tilde a$.
The charged field content is described by the first quiver in Figure \ref{2dXVQ}.
Then we deconfine the conjugate antisymmetric as in the second quiver of Figure \ref{2dXVQ}, where the duality dictionary imposes 
$\tilde a = \tilde p^2$. The theory at this level has vanishing $J$ and $E$ terms.
Observe that the constraint on the global symmetries of the original model also appears as a constraint on the global symmetry of this quiver. 

This model and its dual phase have not been studied in full details in the literature (see for example the comment in {\bf Footnote 5} of \cite{Amariti:2025jvi}).
The dual phase in this case can be obtained reducing $N_f=N_c+1$ 4d SQCD by assigning $R=2$ to an antifundamental and $R=0$ to all the other fields. 
For the case at hand it follows that the final model is given by the third quiver  in Figure \ref{2dXVQ}. The duality dictionary in this case is given by 
$b=q^{2n}$,
$\tilde b_1 = \tilde p^{2n-2}\tilde q^2$,
$\tilde b_2= \tilde p^{2n-3}\tilde q^3$,
$\psi_1 = \eta \tilde p$,
$\psi_2 = \eta \tilde q$,
$\lambda = \eta q^{2n-1}$,
$m_1 = \tilde p q$
and 
$m_2 = \tilde q q$.
The $J$-terms of this dual phase are
\begin{equation}
\label{final_J}
J_{\psi_1} = m_1^2 m_2^{2n-2} + b \tilde b_1\,, \quad
J_{\psi_2} = m_1^3 m_2^{2n-3} + b \tilde b_2\,, \quad
J_{\lambda} = m_1 \tilde b_1+m_2 \tilde b_2\,.
\end{equation}

The constraint on the global charges reflects through the duality map to constraints on the charges for the  $\USp(2n-2)$ gauge theory. However, in this case the constraints do not allow the $\USp(2n-2)$ gauge theory to be reduced to a LG model, even if its field content is compatible with the one discussed in {\bf Section 2.1} of \cite{Amariti:2025jvi}. This is because the constraint on the charges necessary to dualize the model to a LG theory is different from the one obtained here for the $\USp(2n-2)$ dual gauge theory.

We conclude by observing that also in this case the $J$-terms in \eqref{final_J} correspond to those that one would have obtained by following the fate of the superpotential \eqref{Wmag2} with the $R$ charge assignation discussed above.

\subsubsection*{$2n+1$ fundamental and  two antifundamental chirals}
\begin{figure}
    \centering
    \begin{minipage}[b]{0.30\linewidth}
        \centering
        \makebox[\textwidth][c]{
        \begin{tikzpicture}[
            every node/.style={font=\footnotesize},
            box/.style={rectangle, draw, thick},
        ]
        \pgfmathsetmacro{\x}{1}
        \pgfmathsetmacro{\y}{1.5}
        \pgfmathsetmacro{\z}{1.2}
        \pgfmathsetmacro{\delta}{0.28}
        \node[fill=SUcol,circle,draw,thick] (SUgauge) at (0, 0) {$2n$};
        \node[box] (fond) at (\x,-\y) {$2n\!+\!1$};
        \node[box] (afond) at (-\x,-\y) {$2$};
        \node[box] (fermi) at (0,-\y) {$1$};
        \draw[->, thick, >=stealth] (SUgauge) -- node[right] {$q$} (fond.north);
        \draw[<-, thick, >=stealth] (SUgauge) -- node[left, xshift=-2pt] {$\tilde{q}$} (afond.north);
        \draw[-, thick, dashed] (SUgauge) -- node[right] {$\eta$} (fermi);
        \node[box, minimum size=0.2cm] (square1) at (0, \z) {};
        \node[box, minimum size=0.2cm] (square2) at (0, \z+\delta) {};
        \draw[->, thick, >=stealth] (square1) -- node[right] {$\tilde{a}$} (SUgauge);
        \end{tikzpicture}
        }
    \end{minipage}
    \begin{minipage}[b]{0.30\linewidth}
        \centering
        \makebox[\textwidth][c]{
        \begin{tikzpicture}[
            every node/.style={font=\footnotesize},
            box/.style={rectangle, draw, thick},
        ]
        \pgfmathsetmacro{\x}{1}
        \pgfmathsetmacro{\y}{1.5}
        \pgfmathsetmacro{\z}{1.8}
        \node[fill=USPcol,circle,draw,thick] (USPgauge) at (0, \z) {$2n\!-\!2$};
        \node[fill=SUcol,circle,draw,thick] (SUgauge) at (0, 0) {$2n$};
        \node[box] (fond) at (\x,-\y) {$2n\!+\!1$};
        \node[box] (afond) at (-\x,-\y) {$2$};
        \node[box] (fermi) at (0,-\y) {$1$};
        \draw[->, thick, >=stealth] (SUgauge) -- node[right] {$q$} (fond.north);
        \draw[<-, thick, >=stealth] (SUgauge) -- node[left, xshift=-2pt] {$\tilde{q}$} (afond.north);
        \draw[->, thick, >=stealth] (USPgauge) -- node[right] {$\tilde{p}$} (SUgauge);
        \draw[-, thick, dashed] (SUgauge) -- node[right] {$\eta$} (fermi);
        \end{tikzpicture}
        }
    \end{minipage}
    \begin{minipage}[b]{0.30\linewidth}
        \centering
        \makebox[\textwidth][c]{
        \begin{tikzpicture}[
            every node/.style={font=\footnotesize},
            box/.style={rectangle, draw, thick},
        ]
        \pgfmathsetmacro{\x}{1}
        \pgfmathsetmacro{\y}{1.5}
        \pgfmathsetmacro{\z}{2.5}
        \node[fill=USPcol,circle,draw,thick] (USPgauge) at (\z, 0) {$2n\!-\!2$};
        \node[box] (fond) at (0,-\y) {$2n\!+\!1$};
        \node[box] (afond) at (\x,0) {$2$};
        \node[box] (top) at (0,\y) {$1$};
        \node[box] (left) at (-\x,0) {$1$};
        \draw[->, thick, >=stealth] (afond) -- node[right, yshift=-2pt] {$m_1$} (fond);
        \draw[->, thick, >=stealth] (USPgauge.south west) -- node[below, xshift=2pt] {$m_2$} (fond.east);
        \draw[->, thick, >=stealth] (fond) -- node[left] {$b$} (top);
        \draw[->, thick, >=stealth] (top) -- node[left, yshift=7pt, xshift=2pt] {$\tilde{b}_1$} (left);
        \draw[-, thick, dashed] (left) -- node[left,yshift=-2pt] {$\lambda$} (fond);
        \draw[-, thick, dashed] (top) -- node[right, xshift=-2pt, yshift=5pt] {$\psi_1$} (afond);
        \draw[-, thick, dashed] (top.east) -- node[right, xshift=-2pt, yshift=5pt] {$\psi_2$} (USPgauge.north west);
        \end{tikzpicture}
        }
    \end{minipage}
          \caption{Deconfinement of the conjugate two index antisymmetric chiral $\tilde a$ for the 
       $[\mathrm{X},\mathrm{V}]$ case, obtained by fixing $R(\tilde Q_{3})=2$, $R(\tilde Q_{4,5})=1$. In the second step the original $\SU(2n)$ node is dualized to a LG, leaving us with an $\USp(2n-2)$ model, represented in the third quiver.}
    \label{2dXVt}
\end{figure}
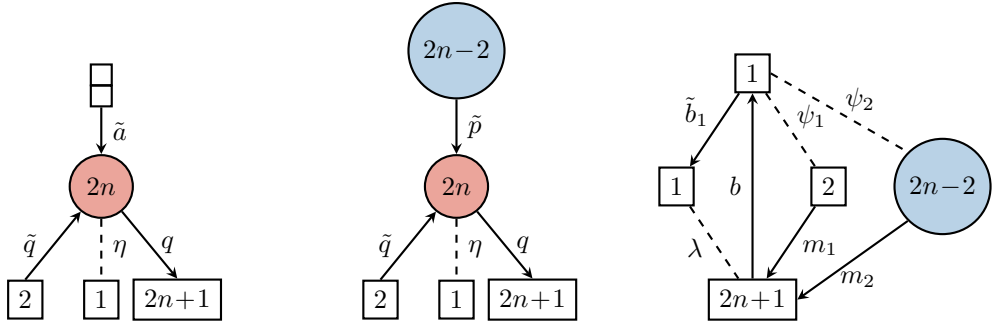

The other $[\mathrm{X},\mathrm{V}]$ case is found by assigning $R=2$ to an antifundamental. It this case, we are left with the first
 quiver in Figure \ref{2dXVt}, in addition to the $J$-term $J_{\Psi} = \text{Pf } \tilde a$.
 The discussion is identical to the one performed above except for a conjugation of the representation in the second quiver in Figure \ref{2dXVt}. After deconfining, we obtain the conjugate antisymmetric $\tilde a = \tilde p^2$.
 
 The final $\USp(2n-2)$ model is found by dualizing the $\SU(2n)$ model, and the duality dictionary in this case is
$m_2 = \tilde p q$, $m_1=\tilde q q$, $b= q^{2n}$, $\tilde b_1 = \tilde p^{2n-2} \tilde q^2$,
$\lambda = \eta q$, $\psi_1 = \eta \tilde p^{2n-2} \tilde q$ and $\psi_2= \eta \tilde p^{2n-3} \tilde q^2$.
The $J$-terms of this dual phase are
\begin{equation}
\label{final_J2}
J_\lambda = m_1^2 m_2^{2n-2} + b \tilde b_1\,, \quad
J_{\psi_1} = b m_1\,, \quad
J_{\psi_2} = b m_2\,.
\end{equation}

Again, in this case the constraints on the global charges of the dual theory  do not allow to reduce the $\USp(2n-2)$ gauge theory to a LG model, even if its field content is compatible with those discussed in {\bf Section 2.1} of \cite{Amariti:2025jvi}. 
Furthermore, one may naively think of obtaining a confining duality from the original $\SU(2n)$ theory, since it can be derived by reducing the 4d confining gauge theory of \cite{Csaki:1996zb} with $2n$ fundamentals and four antifundamentals, by fixing  $R=2$ for a fundamental. Nevertheless, this is not the case, because the constraint on the global charges, that would result in such way, are different from those we have obtained here.

Again, also in this case the $J$-terms in \eqref{final_J2} correspond to those one would have obtained by following the fate of the superpotential \eqref{Wmag2} with the $R$ charge assignation discussed above.

\subsection{[V, X] cases: one fundamental Fermi with a $J$-term }
\label{subsec:2dVX}

The cases studied below have a charged fundamental Fermi field in the $\SU(2n)$ phase, but differently from the $[\mathrm{X},\mathrm{V}]$ cases in this case there is a $J$-term for such Fermi.
These models are obtained by assigning $R=2$ to the field $\tilde Q_5$ and $R=0$ to the field 
$\tilde  Q_4$ (or viceversa).
This charge assignation keeps the 4d $\mathcal{N}=1$ superpotential as a 2d 
$\mathcal{N}=(0,2)$ $J$-term. On the other hand, the constraint from the anomaly cancellation is removed, because we can fix $R=1$ in three ways: for two fundamentals, for two antifundamentals, or for a fundamental and an antifundamental. In the following we will consider the three possibilities separately.

\subsection*{$2n-1$ fundamental and 4 antifundamental chirals}

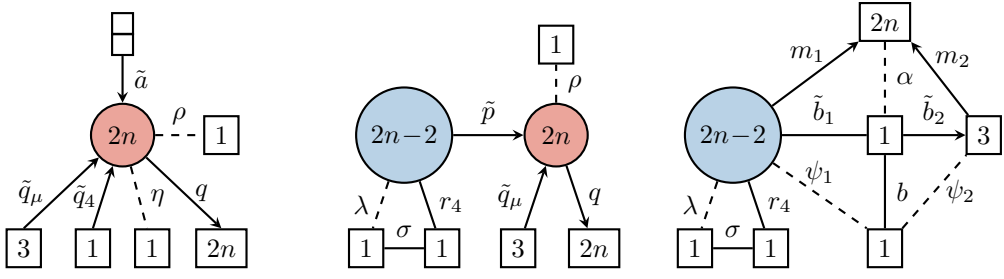
\begin{figure}
    \centering
    \begin{minipage}[b]{0.30\linewidth}
        \centering
        \makebox[\textwidth][c]{
        \begin{tikzpicture}[
            every node/.style={font=\footnotesize},
            box/.style={rectangle, draw, thick},
        ]
        \pgfmathsetmacro{\x}{1.3}
        \pgfmathsetmacro{\y}{1.5}
        \pgfmathsetmacro{\z}{1.2}
        \pgfmathsetmacro{\delta}{0.28}
        \node[fill=SUcol,circle,draw,thick] (SUgauge) at (0, 0) {$2n$};
        \node[box] (fond) at (\x,-\y) {$2n\!-\!1$};
        \node[box] (afond) at (-\x,-\y) {$3$};
        \node[box] (fermi) at (0.3*\x,-\y) {$1$};
        \node[box] (afond2) at (-0.3*\x,-\y) {$1$};
        \draw[->, thick, >=stealth] (SUgauge) -- node[right] {$q$} (fond.north);
        \draw[<-, thick, >=stealth] (SUgauge) -- node[left, xshift=-2pt] {$\tilde{q}_\mu$} (afond.north);
        \draw[<-, thick, >=stealth] (SUgauge) -- node[left, xshift=2pt] {$\tilde{q}_4$} (afond2.north);
        \draw[-, thick, dashed] (SUgauge) -- node[right] {$\eta$} (fermi);
        \node[box, minimum size=0.2cm] (square1) at (0, \z) {};
        \node[box, minimum size=0.2cm] (square2) at (0, \z+\delta) {};
        \draw[->, thick, >=stealth] (square1) -- node[right] {$\tilde{a}$} (SUgauge);
        \end{tikzpicture}
        }
    \end{minipage}
    \begin{minipage}[b]{0.30\linewidth}
        \centering
        \makebox[\textwidth][c]{
        \begin{tikzpicture}[
            every node/.style={font=\footnotesize},
            box/.style={rectangle, draw, thick},
        ]
        \pgfmathsetmacro{\x}{1}
        \pgfmathsetmacro{\y}{1.5}
        \pgfmathsetmacro{\z}{0.5}
        \node[fill=USPcol,circle,draw,thick] (USPgauge) at (-\x, \y) {$2n\!-\!2$};
        \node[fill=SUcol,circle,draw,thick] (SUgauge) at (\x, \y) {$2n$};
        \node[box] (fond) at (\x+\z,0) {$2n\!-\!1$};
        \node[box] (afond) at (\x-\z,0) {$3$};
        \node[box] (singl) at (-\x-\z,0) {$1$};
        \node[box] (singr) at (-\x+\z,0) {$1$};
        \draw[->, thick, >=stealth] (USPgauge) -- node[above] {$\tilde{p}$} (SUgauge);
        \draw[->, thick, >=stealth] (SUgauge) -- node[right] {$q$} (fond);
        \draw[<-, thick, >=stealth] (SUgauge) -- node[left] {$\tilde{q}_\mu$} (afond);
        \draw[-, thick, >=stealth] (USPgauge) -- node[right] {$r_4$} (singr);
        \draw[-, thick, >=stealth] (singl) -- node[above] {$\sigma$} (singr);
        \draw[-, thick, dashed] (USPgauge) -- node[left] {$\lambda$} (singl);
        \end{tikzpicture}
        }
    \end{minipage}
    \begin{minipage}[b]{0.30\linewidth}
        \centering
        \makebox[\textwidth][c]{
        \begin{tikzpicture}[
            every node/.style={font=\footnotesize},
            box/.style={rectangle, draw, thick},
        ]
        \pgfmathsetmacro{\x}{1}
        \pgfmathsetmacro{\y}{1.5}
        \pgfmathsetmacro{\z}{0.5}
        \node[fill=USPcol,circle,draw,thick] (USPgauge) at (-\x, \y) {$2n\!-\!2$};
        \node[box] (fond) at (\x, \y) {$2n\!-\!1$};
        \node[box] (afond) at (\x+\z,0) {$3$};
        \node[box] (singrr) at (\x-\z,0) {$1$};
        \node[box] (singl) at (-\x-\z,0) {$1$};
        \node[box] (singr) at (-\x+\z,0) {$1$};
        \draw[->, thick, >=stealth] (USPgauge) -- node[above] {$m_2$} (fond);
        \draw[->, thick, >=stealth] (afond) -- node[right] {$m_1$} (fond);
        \draw[->, thick, >=stealth] (singrr) -- node[above] {$\tilde{b}_1$} (afond);
        \draw[->, thick, >=stealth] (singrr) -- node[above,xshift=4pt,yshift=-2pt] {$\tilde{b}_2$} (USPgauge);
        \draw[-, thick, >=stealth] (USPgauge) -- node[right] {$r_4$} (singr);
        \draw[-, thick, >=stealth] (singl) -- node[above] {$\sigma$} (singr);
        \draw[-, thick, dashed] (USPgauge) -- node[left] {$\lambda$} (singl);
        \draw[-, thick, dashed] (singrr) -- node[left] {$\psi$} (fond);
        \end{tikzpicture}
        }
    \end{minipage}
        \caption{Deconfinement of the conjugate two index antisymmetric chiral $\tilde a$ for the $[\mathrm{V},\mathrm{V}]$ case, obtained by fixing $ R(Q_{2n,2n+1})=1$, $R(\tilde Q_{5})=2$. In the second step the original $\SU(2n)$ node is dualized to a LG, leaving us with an $\USp(2n-2)$ model, represented in the third quiver.}
    \label{2dVX2Q}
\end{figure}

The first $[\mathrm{V},\mathrm{X}]$ case is found by assigning $R=2$ to $\tilde Q_{5}$ and $R=1$ to $Q_{2n,2n+1}$.
In this case, we are left with the first
quiver in Figure \ref{2dVX2Q}, in addition to the $J$-term $J_{\eta} = \tilde a^{n-1} \tilde q_4$.
We then deconfine the conjugate antisymmetric using the confining duality of {\bf Section 2.1 } of \cite{Amariti:2025jvi}. We therefore obtain an $\USp(2n-2)$ gauge theory as in the second quiver in Figure \ref{2dVX2Q}.
In this case, the duality dictionary is
$\tilde p^2 = \tilde a^2$ and
$\tilde p \rho = \eta$ and  $r_4 \tilde p = \tilde q_4$ and there is a $J$-term $J_{\rho} = \sigma r_4$.
The singlet $\sigma$ corresponds on the electric side to the operator $\mathrm{\Pf}\, \tilde a$, and it can be shown by confining back the $\USp(2n-2)$ gauge node and solving the equations of motions for the linear $J$-term involving the field $\sigma$. This is the $J$-term for a Fermi field $\mu= r_4 \rho$ and we have $J_{\mu} = \sigma+\text{Pf } \tilde a$.

Observe that the deconfinement is valid if we impose a constraint on the allowed charges of chirals and the Fermi of the $\USp(2n-2)$ gauge group, preventing the generation of an axial symmetry.
This constraint however does not propagate after we dualize the $\SU(2n)$ model to a LG, arriving to the third quiver of Figure \ref{2dVX2Q} using the duality of {\bf Appendix A.2} of \cite{Amariti:2024usp}. The duality dictionary in this case is
$m_2 = \tilde p q$, $m_1 = \tilde q_\mu q$, $\tilde b_1 = \tilde p^{2n-2} \tilde q_\mu^2$ and 
$\tilde b_2  = \tilde p^{2n-3} \tilde q_\mu^3$. There is a Fermi field $\psi$ with  $J$-term 
$J_{\psi} = m_1 \tilde b_1 + m_2 \tilde b_2$ in addition to $J_{\rho}$.
Again, this duality can be obtained  from the 4d dual one by following the duality dictionary on the $R$ charge assignation.

\subsection*{$2n$ fundamental and 3 antifundamental chirals}

The second $[\mathrm{V},\mathrm{X}]$ case is found by assigning $R=2$ to $\tilde Q_{5}$ and $R=1$ to $Q_{2n+1}$ and $\tilde Q_3$.
It this case, we are left with the first quiver in Figure \ref{2dVXQQt}, in addition to the $J$-term $J_{\eta} = \tilde a^{n-1} \tilde q_4$.
Again, we deconfine the conjugate antisymmetric using the confining duality of {\bf Section 2.1} of \cite{Amariti:2025jvi}, obtaining   an $\USp(2n-2)$ gauge theory as in the second quiver in Figure \ref{2dVXQQt}.
The duality dictionary is again
$\tilde p^2 = \tilde a^2$,
$\tilde p \rho = \eta$ and  $r_4 \tilde p = \tilde q_4$, with $J_{\rho} = \sigma r_4$.
The singlet $\sigma$ is still identified with the operator $\Pf\, \tilde a$ of the electric theory.

In this case, the $\SU(2n)$ theory is dual to the LG model discussed in  {\bf Appendix A.1} of \cite{Amariti:2024usp} and in the dual phase the constraint forbidding the axial symmetry for the $\USp(2n-2)$ model is lifted.
The final $\USp(2n-2)$ theory is represented  in the third quiver of Figure \ref{2dVXQQt}
and the duality dictionary in this case is
$m_2 = \tilde p q$, $m_1 = \tilde q_\mu q$, $\tilde b = \tilde p^{2n-2} \tilde q_\mu^2$ and 
$b  =q^{2n}$. 
There is a Fermi field $\psi$ with  $J$-term 
$J_{\psi} = \det (m_1|m_2) +  b \tilde b$ in addition to $J_{\rho}$.
Also in this case this duality can be obtained  from the 4d dual one by following the duality dictionary on the $R$ charge assignation.

\begin{figure}
    \centering
    \begin{minipage}[b]{0.30\linewidth}
        \centering
        \makebox[\textwidth][c]{
        \begin{tikzpicture}[
            every node/.style={font=\footnotesize},
            box/.style={rectangle, draw, thick},
        ]
        \pgfmathsetmacro{\x}{1.3}
        \pgfmathsetmacro{\y}{1.5}
        \pgfmathsetmacro{\z}{1.2}
        \pgfmathsetmacro{\delta}{0.28}
        \node[fill=SUcol,circle,draw,thick] (SUgauge) at (0, 0) {$2n$};
        \node[box] (fond) at (\x,-\y) {$2n$};
        \node[box] (afond) at (-\x,-\y) {$2$};
        \node[box] (fermi) at (0.3*\x,-\y) {$1$};
        \node[box] (afond2) at (-0.3*\x,-\y) {$1$};
        \draw[->, thick, >=stealth] (SUgauge) -- node[right] {$q$} (fond.north);
        \draw[<-, thick, >=stealth] (SUgauge) -- node[left, xshift=-2pt] {$\tilde{q}_\mu$} (afond.north);
        \draw[<-, thick, >=stealth] (SUgauge) -- node[left, xshift=2pt] {$\tilde{q}_4$} (afond2.north);
        \draw[-, thick, dashed] (SUgauge) -- node[right] {$\eta$} (fermi);
        \node[box, minimum size=0.2cm] (square1) at (0, \z) {};
        \node[box, minimum size=0.2cm] (square2) at (0, \z+\delta) {};
        \draw[->, thick, >=stealth] (square1) -- node[right] {$\tilde{a}$} (SUgauge);
        \end{tikzpicture}
        }
    \end{minipage}
    \begin{minipage}[b]{0.30\linewidth}
        \centering
        \makebox[\textwidth][c]{
        \begin{tikzpicture}[
            every node/.style={font=\footnotesize},
            box/.style={rectangle, draw, thick},
        ]
        \pgfmathsetmacro{\x}{1}
        \pgfmathsetmacro{\y}{1.5}
        \pgfmathsetmacro{\z}{0.5}
        \node[fill=USPcol,circle,draw,thick] (USPgauge) at (-\x, \y) {$2n\!-\!2$};
        \node[fill=SUcol,circle,draw,thick] (SUgauge) at (\x, \y) {$2n$};
        \node[box] (fond) at (\x+\z,0) {$2n$};
        \node[box] (afond) at (\x-\z,0) {$2$};
        \node[box] (singl) at (-\x-\z,0) {$1$};
        \node[box] (singr) at (-\x+\z,0) {$1$};
        \draw[->, thick, >=stealth] (USPgauge) -- node[above] {$\tilde{p}$} (SUgauge);
        \draw[->, thick, >=stealth] (SUgauge) -- node[right] {$q$} (fond);
        \draw[<-, thick, >=stealth] (SUgauge) -- node[left] {$\tilde{q}_\mu$} (afond);
        \draw[-, thick, >=stealth] (USPgauge) -- node[right] {$r_4$} (singr);
        \draw[-, thick, >=stealth] (singl) -- node[above] {$\sigma$} (singr);
        \draw[-, thick, dashed] (USPgauge) -- node[left] {$\lambda$} (singl);
        \end{tikzpicture}
        }
    \end{minipage}
    \begin{minipage}[b]{0.30\linewidth}
        \centering
        \makebox[\textwidth][c]{
        \begin{tikzpicture}[
            every node/.style={font=\footnotesize},
            box/.style={rectangle, draw, thick},
        ]
        \pgfmathsetmacro{\x}{1}
        \pgfmathsetmacro{\y}{1.5}
        \pgfmathsetmacro{\z}{0.5}
        \node[fill=USPcol,circle,draw,thick] (USPgauge) at (-\x, \y) {$2n\!-\!2$};
        \node[box] (fond) at (\x, \y) {$2n$};
        \node[box] (afond) at (\x,0) {$2$};
        \node[box] (singl) at (-\x-\z,0) {$1$};
        \node[box] (singr) at (-\x+\z,0) {$1$};
        \draw[->, thick, >=stealth] (USPgauge) -- node[above] {$m_2$} (fond);
        \draw[->, thick, >=stealth] (afond) -- node[right] {$m_1$} (fond);
        \draw[-, thick, >=stealth] (USPgauge) -- node[right] {$r_4$} (singr);
        \draw[-, thick, >=stealth] (singl) -- node[above] {$\sigma$} (singr);
        \draw[-, thick, dashed] (USPgauge) -- node[left] {$\lambda$} (singl);
        \end{tikzpicture}
        }
    \end{minipage}  
        \caption{Deconfinement of the conjugate two index antisymmetric chiral $\tilde a$ for the 
       $[\mathrm{V},\mathrm{X}]$ case, obtained by fixing $R(Q_{2n+1})=R(\tilde Q_3)=1$, $R(\tilde Q_{5})=2$. In the second step the original $\SU(2n)$ node is dualized to a LG, leaving us with an $\USp(2n-2)$ model, represented in the third quiver.}
    \label{2dVXQQt}
\end{figure}

\subsection*{$2n+1$ fundamental and 2 antifundamental chirals}

The last $[\mathrm{V},\mathrm{X}]$ case is found by assigning $R=2$ to $\tilde Q_{5}$ and $R=1$ to $\tilde Q_{2,3}$.
 It this case, we are left with the first
 quiver in Figure \ref{2dVX2Qt}, in addition to the $J$-term $J_{\eta} = \tilde a^{n-1} \tilde q_4$.
The conjugate antisymmetric is still deconfined using  the confining duality of {\bf Section 2.1 } of \cite{Amariti:2025jvi}, and the corresponding $\USp(2n-2) \times \SU(2n)$ quiver is represented in  
Figure \ref{2dVX2Qt}, with the same duality dictionary of the other two $[\mathrm{V},\mathrm{X}]$ cases discussed above.

The $\SU(2n)$ theory is dual to the LG model discussed in  {\bf Appendix A.2} of \cite{Amariti:2024usp} and in the dual phase the constraint forbidding the axial symmetry for the $\USp(2n-2)$ model is lifted.
The final $\USp(2n-2)$ theory is represented  in the third quiver of Figure \ref{2dVX2Qt},
and the duality dictionary in this case is
$m_1 = \tilde q_\mu q$, $m_2 = \tilde p q$ and $b = q^{2n}$.
There are two Fermi fields $\psi$ with  $J$-terms
$J_{\psi_1} =b m_1$ and $J_{\psi_2} =b m_2$ in addition to $J_{\rho}$.
Again, this duality can be obtained  from the 4d dual one by following the duality dictionary on the $R$ charge assignation.

\begin{figure}
    \centering
    \begin{minipage}[b]{0.30\linewidth}
        \centering
        \makebox[\textwidth][c]{
        \begin{tikzpicture}[
            every node/.style={font=\footnotesize},
            box/.style={rectangle, draw, thick},
        ]
        \pgfmathsetmacro{\x}{1.3}
        \pgfmathsetmacro{\y}{1.5}
        \pgfmathsetmacro{\z}{1.2}
        \pgfmathsetmacro{\delta}{0.28}
        \node[fill=SUcol,circle,draw,thick] (SUgauge) at (0, 0) {$2n$};
        \node[box] (fond) at (\x,-\y) {$2n\!+\!1$};
        \node[box] (afond) at (-\x,-\y) {$1$};
        \node[box] (fermi) at (0.3*\x,-\y) {$1$};
        \node[box] (afond2) at (-0.3*\x,-\y) {$1$};
        \draw[->, thick, >=stealth] (SUgauge) -- node[right] {$q$} (fond.north);
        \draw[<-, thick, >=stealth] (SUgauge) -- node[left, xshift=-2pt] {$\tilde{q}_1$} (afond.north);
        \draw[<-, thick, >=stealth] (SUgauge) -- node[left, xshift=2pt] {$\tilde{q}_4$} (afond2.north);
        \draw[-, thick, dashed] (SUgauge) -- node[right] {$\eta$} (fermi);
        \node[box, minimum size=0.2cm] (square1) at (0, \z) {};
        \node[box, minimum size=0.2cm] (square2) at (0, \z+\delta) {};
        \draw[->, thick, >=stealth] (square1) -- node[right] {$\tilde{a}$} (SUgauge);
        \end{tikzpicture}
        }
    \end{minipage}
    \begin{minipage}[b]{0.30\linewidth}
        \centering
        \makebox[\textwidth][c]{
        \begin{tikzpicture}[
            every node/.style={font=\footnotesize},
            box/.style={rectangle, draw, thick},
        ]
        \pgfmathsetmacro{\x}{1}
        \pgfmathsetmacro{\y}{1.5}
        \pgfmathsetmacro{\z}{0.5}
        \node[fill=USPcol,circle,draw,thick] (USPgauge) at (-\x, \y) {$2n\!-\!2$};
        \node[fill=SUcol,circle,draw,thick] (SUgauge) at (\x, \y) {$2n$};
        \node[box] (fond) at (\x+\z,0) {$2n\!+\!1$};
        \node[box] (afond) at (\x-\z,0) {$1$};
        \node[box] (singl) at (-\x-\z,0) {$1$};
        \node[box] (singr) at (-\x+\z,0) {$1$};
        \draw[->, thick, >=stealth] (USPgauge) -- node[above] {$\tilde{p}$} (SUgauge);
        \draw[->, thick, >=stealth] (SUgauge) -- node[right] {$q$} (fond);
        \draw[<-, thick, >=stealth] (SUgauge) -- node[left] {$\tilde{q}_1$} (afond);
        \draw[-, thick, >=stealth] (USPgauge) -- node[right] {$r_4$} (singr);
        \draw[-, thick, >=stealth] (singl) -- node[above] {$\sigma$} (singr);
        \draw[-, thick, dashed] (USPgauge) -- node[left] {$\lambda$} (singl);
        \end{tikzpicture}
        }
    \end{minipage}
    \begin{minipage}[b]{0.30\linewidth}
        \centering
        \makebox[\textwidth][c]{
        \begin{tikzpicture}[
            every node/.style={font=\footnotesize},
            box/.style={rectangle, draw, thick},
        ]
        \pgfmathsetmacro{\x}{1}
        \pgfmathsetmacro{\y}{1.5}
        \pgfmathsetmacro{\z}{0.5}
        \node[fill=USPcol,circle,draw,thick] (USPgauge) at (-\x, \y) {$2n\!-\!2$};
        \node[box] (fond) at (\x, \y) {$2n\!+\!1$};
        \node[box] (afond) at (\x+\z,0) {$1$};
        \node[box] (singrr) at (\x-\z,0) {$1$};
        \node[box] (singl) at (-\x-\z,0) {$1$};
        \node[box] (singr) at (-\x+\z,0) {$1$};
        \draw[->, thick, >=stealth] (USPgauge) -- node[above] {$m_2$} (fond);
        \draw[->, thick, >=stealth] (afond) -- node[right] {$m_1$} (fond);
        \draw[-, thick, dashed] (singrr) -- node[above] {$\psi_1$} (afond);
        \draw[-, thick, dashed] (singrr) -- node[above,xshift=4pt,yshift=-2pt] {$\psi_2$} (USPgauge);
        \draw[-, thick, >=stealth] (USPgauge) -- node[right] {$r_4$} (singr);
        \draw[-, thick, >=stealth] (singl) -- node[above] {$\sigma$} (singr);
        \draw[-, thick, dashed] (USPgauge) -- node[left] {$\lambda$} (singl);
        \draw[->, thick, >=stealth] (singrr) -- node[left] {$b$} (fond);
        \end{tikzpicture}
        }
    \end{minipage}
       \caption{Deconfinement of the conjugate two index antisymmetric chiral $\tilde a$ for the 
       $[\mathrm{V},\mathrm{X}]$ case, obtained by fixing  $R(\tilde Q_{2,3})=1$, $R(\tilde Q_{5})=2$. In the second step the original $\SU(2n)$ node is dualized to a LG, leaving us with an $\USp(2n-2)$ model, represented in the third quiver.}
    \label{2dVX2Qt}
\end{figure}

\subsection{[V, V] cases: two fundamentals and one $J$-term}
\label{subsec:2dVV}

The cases studied below have two charged fundamental Fermi fields in the $\SU(2n)$ phase, one of which interacts in the $\SU(2n)$ phase through a $J$-term.
These models are obtained by assigning $R=2$ to the field $\tilde Q_5$ and $R=0$ to the field 
$\tilde  Q_4$ (or viceversa).
This charge assignation keeps the 4d $\mathcal{N}=1$ superpotential as a 2d 
$\mathcal{N}=(0,2)$ $J$-term. The constraint from the anomaly cancellation is maintained in this case, because we fix $R=2$ either for one fundamental or for one antifundamental.
In the following we will discuss the two possibilities separately.

\subsection*{$2n$ fundamentals and $4$ antifundamentals}

The first $[\mathrm{V},\mathrm{V}]$ case is obtained by assigning $R=2$ to $\tilde Q_{5}$ and $ Q_{2n+1}$.
In this case, we are left with the first
quiver in Figure \ref{2dVVQ}, in addition to the $J$-term $J_{\eta} = \tilde a^{n-1} \tilde q_4$.
Furthermore, in this case, the global charges do not allow for the existence of an axial symmetry.
The conjugate antisymmetric is still deconfined using the confining duality of {\bf Section 2.1} of \cite{Amariti:2025jvi}, and the corresponding $\USp(2n-2) \times \SU(2n)$ quiver is represented in  
Figure \ref{2dVVQ}, with the same duality dictionary of the $[\mathrm{V},\mathrm{X}]$ cases  above.

The $\SU(2n)$ model can be further dualized to a LG model and the duality is that of the $[\mathrm{X},\mathrm{V}]$ cases above, which is
obtained reducing $N_f=N_c+1$ 4d SQCD by assigning $R=2$ to an antifundamental and $R=0$ to all the other fields. 

For the case at hand, it follows that the final model is given by the third quiver  in Figure \ref{2dVVQ}, with the duality dictionary 
$m_1 =\tilde p q$,
$m_2 =\tilde q_\mu q  $,
$b = q^{2n}$,
$\tilde b_1 = \tilde p^{2n-3} \tilde q_\mu^3$,
$\tilde b_2= \tilde p^{2n-2} \tilde q_\mu^2$,
$\alpha =\rho \tilde q^{2n-1} $,
$\psi_1 = \tilde p \rho$,
and
$\psi_2 = \tilde q_\mu \rho$
.
The $J$-terms of the final $\USp(2n-2)$ dual are
\begin{equation}
J_{\psi_1} = b \tilde b_1+m_1^{2n-3} m_2^{3}\,,\quad
J_{\psi_2} = b \tilde b_2+m_1^{2n-2} m_2^{2}\,,\quad
J_{\alpha} =  m_1 \tilde b_1+m_2 \tilde b_2\,,
\end{equation}
in addition to $J_{\lambda}= \sigma r_4$.
As in the various cases discussed above, this duality can be obtained  from the twisted reduction on $S^2$  of the 4d dual phase by following the duality dictionary on the $R$ charge assignation.

\begin{figure}
    \centering
    \begin{minipage}[b]{0.30\linewidth}
        \centering
        \makebox[\textwidth][c]{
        \begin{tikzpicture}[
            every node/.style={font=\footnotesize},
            box/.style={rectangle, draw, thick},
        ]
        \pgfmathsetmacro{\x}{1.3}
        \pgfmathsetmacro{\y}{1.5}
        \pgfmathsetmacro{\z}{1.2}
        \pgfmathsetmacro{\delta}{0.28}
        \node[fill=SUcol,circle,draw,thick] (SUgauge) at (0, 0) {$2n$};
        \node[box] (fond) at (\x,-\y) {$2n$};
        \node[box] (afond) at (-\x,-\y) {$3$};
        \node[box] (fermi) at (0.3*\x,-\y) {$1$};
        \node[box] (afond2) at (-0.3*\x,-\y) {$1$};
        \node[box] (fermi2) at (\x,0) {$1$};
        \draw[->, thick, >=stealth] (SUgauge) -- node[right] {$q$} (fond.north);
        \draw[<-, thick, >=stealth] (SUgauge) -- node[left, xshift=-2pt] {$\tilde{q}_\mu$} (afond.north);
        \draw[<-, thick, >=stealth] (SUgauge) -- node[left, xshift=2pt] {$\tilde{q}_4$} (afond2.north);
        \draw[-, thick, dashed] (SUgauge) -- node[right] {$\eta$} (fermi);
        \draw[-, thick, dashed] (SUgauge) -- node[above] {$\rho$} (fermi2);
        \node[box, minimum size=0.2cm] (square1) at (0, \z) {};
        \node[box, minimum size=0.2cm] (square2) at (0, \z+\delta) {};
        \draw[->, thick, >=stealth] (square1) -- node[right] {$\tilde{a}$} (SUgauge);
        \end{tikzpicture}
        }
    \end{minipage}
    \begin{minipage}[b]{0.30\linewidth}
        \centering
        \makebox[\textwidth][c]{
        \begin{tikzpicture}[
            every node/.style={font=\footnotesize},
            box/.style={rectangle, draw, thick},
        ]
        \pgfmathsetmacro{\x}{1}
        \pgfmathsetmacro{\y}{1.5}
        \pgfmathsetmacro{\z}{0.5}
        \node[fill=USPcol,circle,draw,thick] (USPgauge) at (-\x, \y) {$2n\!-\!2$};
        \node[fill=SUcol,circle,draw,thick] (SUgauge) at (\x, \y) {$2n$};
        \node[box] (fond) at (\x+\z,0) {$2n$};
        \node[box] (afond) at (\x-\z,0) {$3$};
        \node[box] (singl) at (-\x-\z,0) {$1$};
        \node[box] (singr) at (-\x+\z,0) {$1$};
        \node[box] (fermi2) at (\x,1.8*\y) {$1$};
        \draw[->, thick, >=stealth] (USPgauge) -- node[above] {$\tilde{p}$} (SUgauge);
        \draw[->, thick, >=stealth] (SUgauge) -- node[right] {$q$} (fond);
        \draw[<-, thick, >=stealth] (SUgauge) -- node[left] {$\tilde{q}_\mu$} (afond);
        \draw[-, thick, >=stealth] (USPgauge) -- node[right] {$r_4$} (singr);
        \draw[-, thick, >=stealth] (singl) -- node[above] {$\sigma$} (singr);
        \draw[-, thick, dashed] (USPgauge) -- node[left] {$\lambda$} (singl);
        \draw[-, thick, dashed] (SUgauge) -- node[right] {$\rho$} (fermi2);
        \end{tikzpicture}
        }
    \end{minipage}
    \begin{minipage}[b]{0.30\linewidth}
        \centering
        \makebox[\textwidth][c]{
        \begin{tikzpicture}[
            every node/.style={font=\footnotesize},
            box/.style={rectangle, draw, thick},
        ]
        \pgfmathsetmacro{\x}{1}
        \pgfmathsetmacro{\y}{1.5}
        \pgfmathsetmacro{\z}{0.5}
        \node[fill=USPcol,circle,draw,thick] (USPgauge) at (-\x, \y) {$2n\!-\!2$};
        \node[box] (singl) at (-\x-\z,0) {$1$};
        \node[box] (singr) at (-\x+\z,0) {$1$};
        \node[box] (fond) at (\x,2*\y) {$2n$};
        \node[box] (singc) at (\x,\y) {$1$};
        \node[box] (singb) at (\x,0) {$1$};
        \node[box] (afond) at (2.3*\x,\y) {$3$};
        \draw[-, thick, >=stealth] (USPgauge) -- node[right] {$r_4$} (singr);
        \draw[-, thick, >=stealth] (singl) -- node[above] {$\sigma$} (singr);
        \draw[-, thick, dashed] (USPgauge) -- node[left] {$\lambda$} (singl);
        \draw[-, thick, >=stealth] (USPgauge) -- node[above] {$\tilde{b}_1$} (singc);
        \draw[->, thick, >=stealth] (USPgauge) -- node[above,xshift=-3pt] {$m_1$} (fond);
        \draw[-, thick, dashed] (USPgauge) -- node[above] {$\psi_1$} (singb.north west);
        \draw[-, thick, dashed] (singb) -- node[right] {$\psi_2$} (afond);
        \draw[-, thick, dashed] (fond) -- node[right] {$\alpha$} (singc);
        \draw[<-, thick, >=stealth] (fond.south east) -- node[above,xshift=5pt] {$m_2$} (afond);
        \draw[->, thick, >=stealth] (singc) -- node[above] {$\tilde{b}_2$} (afond);
        \draw[-, thick, >=stealth] (singc) -- node[right] {$b$} (singb);
        \end{tikzpicture}
        }
    \end{minipage}
     \caption{Deconfinement of the conjugate two index antisymmetric chiral $\tilde a$ for the $[\mathrm{V},\mathrm{V}]$ case, obtained by fixing  $R(Q_{2n+1})=R(\tilde Q_{5})=2$. In the second step the original $\SU(2n)$ node is dualized to a LG, leaving us with an $\USp(2n-2)$ model, represented in the third quiver.}
    \label{2dVVQ}
\end{figure}

\subsection*{$2n+1$ fundamentals and $3$ antifundamentals}

\begin{figure}
    \centering
    \begin{minipage}[b]{0.30\linewidth}
        \centering
        \makebox[\textwidth][c]{
        \begin{tikzpicture}[
            every node/.style={font=\footnotesize},
            box/.style={rectangle, draw, thick},
        ]
        \pgfmathsetmacro{\x}{1.3}
        \pgfmathsetmacro{\y}{1.5}
        \pgfmathsetmacro{\z}{1.2}
        \pgfmathsetmacro{\delta}{0.28}
        \node[fill=SUcol,circle,draw,thick] (SUgauge) at (0, 0) {$2n$};
        \node[box] (fond) at (\x,-\y) {$2n\!+\!1$};
        \node[box] (afond) at (-\x,-\y) {$2$};
        \node[box] (fermi) at (0.3*\x,-\y) {$1$};
        \node[box] (afond2) at (-0.3*\x,-\y) {$1$};
        \node[box] (fermi2) at (\x,0) {$1$};
        \draw[->, thick, >=stealth] (SUgauge) -- node[right] {$q$} (fond.north);
        \draw[<-, thick, >=stealth] (SUgauge) -- node[left, xshift=-2pt] {$\tilde{q}_\mu$} (afond.north);
        \draw[<-, thick, >=stealth] (SUgauge) -- node[left, xshift=2pt] {$\tilde{q}_4$} (afond2.north);
        \draw[-, thick, dashed] (SUgauge) -- node[right] {$\eta$} (fermi);
        \draw[-, thick, dashed] (SUgauge) -- node[above] {$\rho$} (fermi2);
        \node[box, minimum size=0.2cm] (square1) at (0, \z) {};
        \node[box, minimum size=0.2cm] (square2) at (0, \z+\delta) {};
        \draw[->, thick, >=stealth] (square1) -- node[right] {$\tilde{a}$} (SUgauge);
        \end{tikzpicture}
        }
    \end{minipage}
    \begin{minipage}[b]{0.30\linewidth}
        \centering
        \makebox[\textwidth][c]{
        \begin{tikzpicture}[
            every node/.style={font=\footnotesize},
            box/.style={rectangle, draw, thick},
        ]
        \pgfmathsetmacro{\x}{1}
        \pgfmathsetmacro{\y}{1.5}
        \pgfmathsetmacro{\z}{0.5}
        \node[fill=USPcol,circle,draw,thick] (USPgauge) at (-\x, \y) {$2n\!-\!2$};
        \node[fill=SUcol,circle,draw,thick] (SUgauge) at (\x, \y) {$2n$};
        \node[box] (fond) at (\x+\z,0) {$2n\!+\!1$};
        \node[box] (afond) at (\x-\z,0) {$2$};
        \node[box] (singl) at (-\x-\z,0) {$1$};
        \node[box] (singr) at (-\x+\z,0) {$1$};
        \node[box] (fermi2) at (\x,1.8*\y) {$1$};
        \draw[->, thick, >=stealth] (USPgauge) -- node[above] {$\tilde{p}$} (SUgauge);
        \draw[->, thick, >=stealth] (SUgauge) -- node[right] {$q$} (fond);
        \draw[<-, thick, >=stealth] (SUgauge) -- node[left] {$\tilde{q}_\mu$} (afond);
        \draw[-, thick, >=stealth] (USPgauge) -- node[right] {$r_4$} (singr);
        \draw[-, thick, >=stealth] (singl) -- node[above] {$\sigma$} (singr);
        \draw[-, thick, dashed] (USPgauge) -- node[left] {$\lambda$} (singl);
        \draw[-, thick, dashed] (SUgauge) -- node[right] {$\rho$} (fermi2);
        \end{tikzpicture}
        }
    \end{minipage}
    \begin{minipage}[b]{0.30\linewidth}
        \centering
        \makebox[\textwidth][c]{
        \begin{tikzpicture}[
            every node/.style={font=\footnotesize},
            box/.style={rectangle, draw, thick},
        ]
        \pgfmathsetmacro{\x}{1}
        \pgfmathsetmacro{\y}{1.5}
        \pgfmathsetmacro{\z}{0.5}
        \pgfmathsetmacro{\w}{0.7}
        \node[fill=USPcol,circle,draw,thick] (USPgauge) at (-\x, \y) {$2n\!-\!2$};
        \node[box] (singl) at (-\x-\z,0) {$1$};
        \node[box] (singr) at (-\x+\z,0) {$1$};
        \node[box] (fond) at (0.8*\x,2*\y) {$2n\!+\!1$};
        \node[box] (singc) at (\x+\w,\y) {$2$};
        \node[box] (singb) at (0.8*\x,0) {$1$};
        \node[box] (afond) at (2.3*\x+\w,\y) {$1$};
        \draw[-, thick, >=stealth] (USPgauge) -- node[right] {$r_4$} (singr);
        \draw[-, thick, >=stealth] (singl) -- node[above] {$\sigma$} (singr);
        \draw[-, thick, dashed] (USPgauge) -- node[left] {$\lambda$} (singl);
        \draw[->, thick, >=stealth] (USPgauge) -- node[above,xshift=-3pt] {$m_2$} (fond.south west);
        \draw[-, thick, dashed] (USPgauge) -- node[below,xshift=0pt] {$\psi_2$} (singb.north west);
        \draw[->, thick, >=stealth] (fond) -- node[left] {$b$} (singb);
        \draw[->, thick, >=stealth] (singc.north) -- node[right,xshift=-2pt] {$m_1$} (fond.south east);
        \draw[-, thick, dashed] (singc.south) -- node[right,xshift=0pt] {$\psi_1$} (singb.north east);
        \draw[->, thick, >=stealth] (singb.east) -- node[below,xshift=0pt] {$\tilde{b}_1$} (afond.south);
        \draw[-, thick, dashed] (fond.east) -- node[above,xshift=0pt] {$\nu$} (afond.north);
        \end{tikzpicture}
        }
    \end{minipage}
      \caption{Deconfinement of the conjugate two index antisymmetric chiral $\tilde a$ for the $[\mathrm{V},\mathrm{V}]$ case, obtained by fixing  $R(\tilde Q_{3,5})=2$. In the second step the original $\SU(2n)$ node is dualized to a LG, leaving us with an $\USp(2n-2)$ model, represented in the third quiver.}
    \label{2dVVQt}
\end{figure}
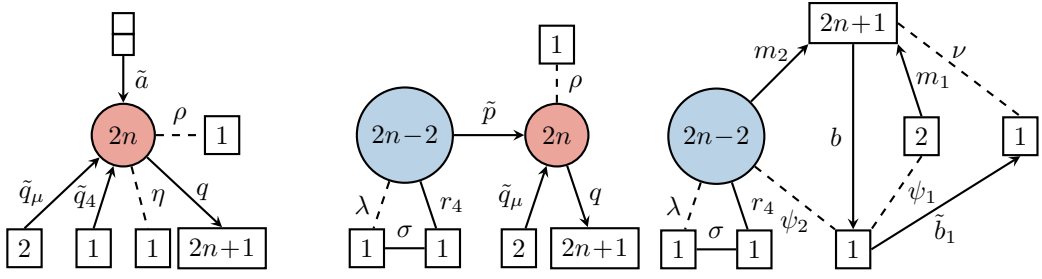

The second $[\mathrm{V},\mathrm{V}]$ case is found by assigning $R=2$ to $\tilde Q_{5}$ and $ \tilde Q_{3}$.
It this case, we are left with the first quiver in Figure \ref{2dVVQt} in addition to the $J$-term $J_{\eta} = \tilde a^{n-1} \tilde q_4$.
Again, in this case the global charges do not allow for the existence of an axial symmetry and
the conjugate antisymmetric is  deconfined  as above, using   the confining duality of {\bf Section 2.1 } of \cite{Amariti:2025jvi}.
The corresponding $\USp(2n-2) \times \SU(2n)$ quiver is represented in  
Figure \ref{2dVVQt}, with the same duality dictionary of the $[\mathrm{V},\mathrm{X}]$ cases and of the first $[\mathrm{V},\mathrm{V}]$  case studied above.
The $\SU(2n)$ model can be further dualized to a LG model  using the duality which is obtained reducing $N_f=N_c+1$ 4d SQCD by assigning $R=2$ to a fundamental, and $R=0$ to all the other fields.

For the case at hand, it follows that the final model is given by the third quiver  in Figure \ref{2dVVQt}, with the duality dictionary 
$m_1 =\tilde q_\mu q $,
$m_2 =\tilde p q$,
$b = q^{2n}$,
$\nu = \rho q$,
$\psi_1 = \rho \tilde q_\mu \tilde p^{2n-2}$,
$\psi_2= \rho  q_\mu^2 \tilde p^{2n-3}$ and
$\tilde b_1 =\tilde p^{2n-2} \tilde q_\mu $.
The $J$-terms of the final $\USp(2n-2)$ dual are
\begin{equation}
J_\nu= m_1^{2} m_2^{2n-2} + b \tilde b_1\,,\quad
J_{\psi_1} = m_1 b\,,\quad
J_{\psi_2} = m_2 b\,,
\end{equation}
in addition to $J_{\lambda}= \sigma r_4$.
The final duality can be obtained  directly from the twisted reduction on $S^2$ of the 4d dual phase by following the duality dictionary on the $R$ charge assignation.

%
%
%
%
%
\section{Effective reduction to 3d and duplication formula}
\label{sec:3deffective}
%
%
%
%
%
%

Here we study the reduction of the dualities obtained in Section
 \ref{sec:4d} at the level of the superconformal index, obtaining the identities among the three sphere partition functions.
 Once such identities are established, we fix the values of some mass parameters by freezing them at some specific value, and then we apply the duplication formula for the hyperbolic Gamma functions, finding new identities between $\SU(N)$ theories with a symmetric tensor and $\SO(N)$ theories.
 
Even if the analysis is performed distinguishing the parity of the gauge rank, we  observe that after the application of the duplication formula, the new dual pairs do not necessitate such a distinction anymore. Therefore, we provide an independent proof of the dualities (and of the integral identities) using tensor deconfinement.

%
%
%
%
%
%
\subsection{$\SU(2n)$ with $W=\det \tilde S$}
\label{detSeven}
%
%
%
%
%
%
Here, we consider the duality between $\SU(2n)$ with $2n+1$ fundamentals,
five antifundamentals, a conjugate antisymmetric $\tilde A$, superpotential 
$W = \tilde A^{n-1} \tilde Q_4 \tilde Q_5$, and $\USp(2n-2)$ with $2n+4$ fundamentals.
We can reduce the duality on $S^1$ by adding the contribution of the KK superpotential, avoiding the generation of an axial symmetry for both phases.
At the level of the three sphere partition function we have the following identity 
\footnote{Here we adopt the same notation for the various representations of the charged matter used in the discussion of the superconformal index. On the RHS of \eqref{3deffZ1} we have separated with the comma the real masses for matter fields in the same representation of the gauge group.}
\begin{equation}
    \label{3deffZ1}
    \begin{aligned}
        &Z_{\SU(2n)}^{[(2n+1)\square ;5 \overline{\square} ;1 \overline{\text A}]} (\vec \mu;\vec \nu;\tau_{\tilde A})
        =
        Z_{\USp(2n-2)}^{[(2n+4) \square]}\left(\frac{\tau_{\tilde A}}{2}+\vec \mu,\left(n-\frac{3}{2}\right) \tau_{\tilde A} + \sum_{a=1}^{3} \nu_a,\nu_{4,5}-\frac{\tau_{\tilde A}}{2}\right)
        \\
        &\times
        \Gamma_h \left(n \tau_{\tilde A} \right)\prod_{a=1}^{2n+1} \prod_{b=1}^3 \Gamma_h(\mu_a + \nu_{b})\prod_{a=1}^{2n+1}\Gamma_h \left(\sum_{c=1}^{2n+1}  \mu_c- \mu_a\!\right)
        \!\!\!\prod_{1\leq a <b \leq 3} 
        \!\!\!\Gamma_h \left( (n-1) \tau_{\tilde A} + \nu_a + \nu_b \right),
    \end{aligned}
\end{equation}
This identity holds provided that 
the two relations
\begin{equation}
\label{3deffBC1}
(2n-2) \tau_{\tilde A} + \sum_{a=1}^{2n+1}\mu_a+\sum_{b=1}^5 \nu_b = 4 \omega
\,,\quad
(n-1) \tau_{\tilde A} + \nu_4 + \nu_5 = 2 \omega\,,
\end{equation}
hold. The first relation is imposed by the KK monopole superpotential, while the second one is imposed by the tree level superpotential involving the 3d chiral fields.

We now fix the mass parameters in terms of the squashing parameters of $S_b^3$, 
$\omega_{1}=i b$ and $\omega_2=ib^{-1}$ (with $\omega = \frac{\omega_1+\omega_2}{2}$) as
\begin{equation}
\label{freezing1}
\nu_3 =  \frac{\tau_{\tilde S}}{2}
\,,\quad
\nu_4 = \frac{\omega_1}{2} + \frac{\tau_{\tilde S}}{2}
\,,\quad
\nu_5 = \frac{\omega_2}{2} + \frac{\tau_{\tilde S}}{2}\,,
\end{equation}
where we are also re-defining $\tau_{\tilde A}$ as  $\tau_{\tilde S}$,
in order to have a more direct interpretation of the results after the application of the duplication formula 
\begin{equation}
\label{duplication}
\Gamma_h(2 x) = 
\Gamma_h(x)
\Gamma_h\left( x+\frac{\omega_1}{2}\right)
\Gamma_h\left( x+\frac{\omega_2}{2}\right)
\Gamma_h\left( x+\omega\right).
\end{equation}
In this way we arrive to the relation
\begin{eqnarray}
\label{3dduplic1}
&&
Z_{\SU(2n)}^{[(2n+2) \square ;2 \overline{\square}; 1 \overline{\text S}]} 
\left(\vec \mu,\omega-\frac{\tau_{\tilde S}}{2};\vec \nu;\tau_{\tilde S}\right)
= 
Z_{\SO(2n-1)}^{[(2n+2) \text{V}]}
\left(\frac{\tau_{\tilde S}}{2}+\vec \mu,\left(n-1\right) \tau_{\tilde S} + \nu_1+\nu_2\right)
\nonumber
\\
&&
\times \,\prod_{b=1}^2 \Gamma_h\left(\omega-\frac{\tau_{\tilde S}}{2} + \nu_{b}\right)
\prod_{a=1}^{2n+1} \Gamma_h\left(\mu_a + \nu_{b}\right)
\prod_{a=1}^{2n+1}\Gamma_h \left(\sum_{c=1}^{2n+1}  \mu_c- \mu_a\right),
\end{eqnarray}
which holds provided that 
the mass parameters satisfy the relations
\begin{equation}
\label{3duplBC1}
\left(n-\frac{1}{2}\right) \tau_{\tilde S} + \sum_{a=1}^{2n+1}\mu_a+\nu_1+\nu_2 = 3 \omega\,,\quad
n \tau_{\tilde S} = \omega\,.
\end{equation}
The first relation is compatible with the constraints enforced by the monopole superpotentials $Y_{\SU(2n-2)}^{(\text{bare})}$, while the second one is compatible with the superpotential deformation $\det \tilde S$. A further constraint on the mass parameters of the $\SU(2n)$ fundamentals concerns the $2n+2$-th mass parameter, which is fixed to $\omega-\frac{\tau_{\tilde S}}{2}$. Denoting the corresponding fields as $Q_{\tilde S}$ the final superpotential of the electric theory is 
\begin{equation}
W = \det \tilde S  +\tilde S Q_{\tilde S}^2+Y_{\SU(2n-2)}^{(\text{bare})}\,.
\end{equation}
On the dual side, we observe, using the arguments of the hyperbolic Gamma functions, the presence of  three singlets 
$L= \tilde Q Q_{\tilde S} $,
$B = Q^{2n}$
and $M = Q \tilde Q$. Furthermore, by distinguishing the dual  $2n+2$ vectors 
as $2n+1$ fields $\varphi$ and one field $u$, the constraint on the mass parameters is compatible with a superpotential
\begin{equation}
\label{Wdualafterdupldet}
W = Y_{\SO(2n-1)}^{+}+ B u \varphi+\varphi^{2n-1} M^2+B L M\,.
\end{equation}
Observe that the linear monopole superpotentials is required to enforce the constraints on the mass parameters of the  $2n+2$ vectors $\varphi$ and $u$
\begin{equation}
m_u + \sum_{i=1}^{2n+1} m_{\varphi_i} = 3\omega\,,
\end{equation}
which holds because of the constraints (\ref{3duplBC1}).
\\
\\
We will prove below that this duality can be derived by deconfining the $\SU(2n)$ conjugate symmetric $\tilde S$ using an $\SO(2n-1)$ gauge theory and then by confining the original $\SU(2n)$ gauge group. Such a procedure will be spelled out after discussing the $\SU(2n+1)$ case with the same superpotential. Indeed, as  we observe below,  the results for the even and for the odd cases can be unified.
%
%
%
%
%
%
\subsection{$\SU(2n+1)$ with $W=\det \tilde S$}
\label{detSodd}
%
%
%
%
%
%
Here, we consider the duality between $\SU(2n+1)$ with $2n+2$ fundamentals,
five antifundamentals, a conjugate antisymmetric $\tilde A$, superpotential 
$W = \tilde A^{n-1} \tilde Q_3 \tilde Q_4 \tilde Q_5$, and $\USp(2n)$ with $2n+5$ fundamentals.
We can reduce the duality on $S^1$ by adding the contribution of the KK superpotential, avoiding the generation of an axial symmetry for both the phases.
At the level of the three sphere partition function we have the following identity 
\begin{equation}
    \label{3deffZ2}
    \begin{aligned}
        &
        Z_{\SU(2n+1)}^{[(2n+2)\square ;5 \overline{\square};1 \overline{\text A}]} (\vec \mu;\vec \nu;\tau_{\tilde A})
        =
        \prod_{b=1}^{5}\Gamma_h \left(n \tau_{\tilde A} +\nu_b\right)
        \prod_{a=1}^{2n+2} \Gamma_h\left(\mu_a + \nu_{1,2},\sum_{c=1}^{2n+1}  \mu_c- \mu_a\right) 
        \\
        &\times
        Z_{\USp(2n)}^{[(2n+6)\square]}\left(\frac{\tau_{\tilde A}}{2}+\vec \mu,\left(n-\frac{1}{2}\right) \tau_{\tilde A} + \nu_1+\nu_2,\nu_{3,4,5}-\frac{\tau_{\tilde A}}{2}\right).
    \end{aligned}
\end{equation}
This identity holds provided that 
the two relations
\begin{equation}
\label{3deffBC2}
(2n-1) \tau_{\tilde A} + \sum_{a=1}^{2n+2}\mu_a+\sum_{b=1}^5 \nu_b = 4 \omega
\,,\quad
(n-1) \tau_{\tilde A} +  \nu_3 +  \nu_4 +  \nu_5 = 2 \omega\,,
\end{equation}
hold. The first relation is imposed by the KK monopole superpotential, while the second one is imposed by the tree level superpotential involving the 3d chiral fields.

We now fix the mass parameters as in formula (\ref{freezing1}),
where again we  re-define $\tau_{\tilde A}$ as  $\tau_{\tilde S}$.
In this way, by applying the duplication formula \eqref{duplication}  we arrive to the relation
\begin{eqnarray}
\label{3dduplic2}
&&
Z_{\SU(2n+1)}^{[(2n+3)\square;2\overline{\square};1 \overline{\text S}]}
 \left(\vec \mu,\omega-\frac{\tau_{\tilde S}}{2};\vec \nu;\tau_{\tilde S}\right)= Z_{\SO(2n)}^{[(2n+3) \square]}\left(\frac{\tau_{\tilde S}}{2}+\vec \mu,\left(n-\frac{1}{2}\right) \tau_{\tilde S} + \nu_1+\nu_2\right)
\nonumber
\\
&&
\times \,\prod_{b=1}^2 \Gamma_h\left(\omega-\frac{\tau_{\tilde S}}{2} + \nu_{b}\right)
\prod_{a=1}^{2n+2} \Gamma_h\left(\mu_a + \nu_{b},\sum_{c=1}^{2n+2}  \mu_c- \mu_a\right),
\end{eqnarray}
which holds provided that 
the mass parameters satisfy the relations
\begin{equation}
\label{3duplBC2}
\left(2n+\frac{1}{2}\right) \tau_{\tilde S} + \sum_{a=1}^{2n+2}\mu_a+\sum_{b=1}^2 \nu_b = 3 
\omega\,,\quad
\left(n+\frac{1}{2}\right) \tau_{\tilde S}  = \omega\,.
\end{equation}

The first relation is compatible with the constraints enforced by the monopole superpotentials $Y_{\SU(2n-1)}^{(\text{bare})}$, while the second one is compatible with the superpotential deformation $\det \tilde S$. A further constraint on the mass parameters of the $\SU(2n+1)$ fundamentals concerns the $2n+3$-th mass parameter, which is fixed to $\omega-\frac{\tau_{\tilde S}}{2}$. Denoting the associated fields as $Q_{\tilde S}$ the final superpotential of the electric theory is 
\begin{equation}
W = \det \tilde S  +\tilde S Q_{\tilde S}^2+Y_{\SU(2n-1)}^{(\text{bare})}\,.
\end{equation}
On the dual side, we observe, using the arguments of the hyperbolic Gamma functions, the presence of  three singlets 
$L= \tilde Q Q_{\tilde S} $,
$B = Q^{N}$
and $M = Q \tilde Q$, formally identical to the ones obtained for the $\SU(2n)$ case above. Furthermore, by distinguishing the dual  $2n+3$ vectors 
as $2n+2$ fields $\varphi$ and one field $u$, the constraint on the mass parameters are compatible with the superpotential (\ref{Wdualafterdupldet}).
\\
\\
For this reason, we propose the following unified duality that holds for generic $N$: 
\begin{center}
\begin{tabular}{|ccc|}
\hline
$\SU(N)$ with $1$  $\overline{\square \!  \square}$ $\tilde S$ & & $\SO(N-1)$ with $N+1$ $\square$ $\varphi$, $1$ $\square$ $u$  \\
$N+1$ $\square$ $Q$, $1$  $\square$ $Q_{\tilde S} $, $2$ $\overline \square $  $\tilde Q$  & $\leftrightarrow$  & Singlets  $L= \tilde Q Q_{\tilde S} $,
$B = Q^{N}$, $M = Q \tilde Q$ \\
$W = \det \tilde S  +\tilde S Q_{\tilde S}^2+Y_{\SU(N-2)}^{(\text{bare})}$
&& $W = Y_{\SO(N-1)}^{+} + B u \varphi+\varphi^{N-1} M^2+B L M$\\
\hline
\end{tabular}
\end{center}
Below we will provide a physical derivation of such a duality in terms of other fundamental dualities by deconfining the conjugate symmetric tensor in terms of an $\SO(N-1)$ gauge group and then by confining the $\SU(N)$ gauge group. In this way, we will be able to reproduce the expected dual superpotential as well.

%
%
%
%
%
%
\subsection{$\SU(2n)$ with $W= \tilde S^{2n-1} \tilde Q_2^2$}
\label{S2nm1even}
%
%
%
%
%
%

Here, we consider the duality between $\SU(2n)$ with $2n+1$ fundamentals,
five antifundamentals, a conjugate antisymmetric $\tilde A$, superpotential 
$W = \tilde A^{n-1} \tilde Q_2 \tilde Q_3 \tilde Q_4 \tilde Q_5$, and $\USp(2n)$ with $2n+6$ fundamentals.
We can reduce the duality on $S^1$ by adding the contribution of the KK superpotential, avoiding the generation of an axial symmetry for both the phases.
At the level of the three sphere partition function we have the following identity 
\begin{eqnarray}
\label{3deffZ3}
&&
Z_{\SU(2n)}^{[(2n+1) \square ;5 \overline {\square} ;1\overline{\text A} ]} (\vec \mu;\vec \nu;\tau_{\tilde A})= 
Z_{\USp(2n)}^{[(2n+6) \square]}\left(\frac{\tau_{\tilde A}}{2}+\vec \mu,\left(n-\frac{1}{2}\right) \tau_{\tilde A} + \nu_1,\nu_{2,\dots,5}-\frac{\tau_{\tilde A}}{2}\right)
\nonumber 
\\
&&
\times \,
\Gamma_h \left(n \tau_{\tilde A} \right)
\prod_{a=1}^{2n+1} \Gamma_h\left(\mu_a + \nu_{1},\sum_{c=1}^{2n+1}  \mu_c- \mu_a\right)
\prod_{2\leq a <b \leq 5} \Gamma_h \left( (n-1) \tau_{\tilde A} + \nu_a + \nu_b \right).
\nonumber 
\\
\end{eqnarray}
This identity holds provided that 
the two relations
\begin{equation}
\label{3deffBC3}
(2n-2) \tau_{\tilde A} + \sum_{a=1}^{2n+1}\mu_a+\sum_{b=1}^5 \nu_b = 4 \omega
\,,\quad
(n-2) \tau_{\tilde A} + \nu_2 + \nu_3 + \nu_4 + \nu_5 = 2 \omega\,,
\end{equation}
hold. The first relation is imposed by the KK monopole superpotential, while the second one is imposed by the tree level superpotential involving the 3d chiral fields.

We now fix the mass parameters as in formula (\ref{freezing1}),
where again we  re-define $\tau_{\tilde A}$ as  $\tau_{\tilde S}$.
In this way, by applying the duplication formula (\ref{duplication})  we arrive to the relation
\begin{eqnarray}
\label{3dduplic3}
&&
\!\!\!\!
Z_{\SU(2n)}^{[(2n+2)\square ;2 \overline \square ; 1 \overline{\text A}]}\! 
\left(\vec \mu;\omega\!-\!\frac{\tau_{\tilde S}}{2};\vec \nu;\tau_{\tilde A} \right) \!=\! Z_{\SO(2n)}^{[(2n+3) \text{V} ]} \! \left(\frac{\tau_{\tilde S}}{2}+\vec \mu,\left(n\!-\!\frac{1}{2}\right) \tau_{\tilde S} + \nu_1,\nu_{2}\!-\!\frac{\tau_{\tilde S}}{2}\! \right)
\nonumber
\\
&&
\times\,
\Gamma_h \left(2n \tau_{\tilde S} \right)
\prod_{a=1}^{2n+1} \Gamma_h\left(\mu_a + \nu_{1},\sum_{c=1}^{2n+1}  \mu_c- \mu_a\right),
\end{eqnarray}
which holds provided that 
the mass parameters satisfy the relations
\begin{equation}
\label{3duplBC3}
\left(2n-\frac{1}{2}\right) \tau_{\tilde S} + \sum_{a=1}^{2n+1}\mu_a+\sum_{b=1}^2 \nu_b = 3 \omega\,,\quad
(2n-1) \tau_{\tilde S} + 2\nu_2  =  2 \omega\,.
\end{equation}

The first relation is compatible with the constraints enforced by the monopole superpotentials $Y_{\SU(2n-2)}^{(\text{bare})}$, while the second one is compatible with the superpotential deformation $\tilde S^{2n-1} \tilde Q_2^2$. A further constraint on the mass parameters of the $\SU(2n)$ fundamentals concerns the $2n+2$-th mass parameter, which is fixed to $\omega-\frac{\tau_{\tilde S}}{2}$. Denoting the associated fields as $Q_{\tilde S}$, the final superpotential of the electric theory is 
\begin{equation}
W =\tilde S^{2n-1} \tilde Q_2^2  +\tilde S Q_{\tilde S}^2+Y_{\SU(2n-2)}^{(\text{bare})}\,.
\end{equation}
On the dual side, we observe, using the arguments of the hyperbolic Gamma functions, the presence of  three singlets 
$\sigma= \det \tilde S $,
$B = Q^{2n}$
and $M = Q \tilde Q_1$. Furthermore, by distinguishing the dual  $2n+3$ vectors 
as $2n+1$ fields $\varphi$, one field $u_1$ and one field $u_2$, the constraint on the mass parameters are compatible with a superpotential
\begin{equation}
\label{WdualafterduplS2nm2culo}
W = Y_{\SO(2n-1)}^{+}+ \sigma u_2^2 + B u_1 \varphi+\varphi^{2n} M\,.
\end{equation}
Observe that the linear monopole superpotentials is required to enforce the constraints on the mass parameters of the  $2n+3$ vectors $\varphi$ and $u_{1,2}$
\begin{equation}
m_{u_1} +m_{u_2} + \sum_{i=1}^{2n+1} m_{\varphi_i} =3 \omega\,,
\end{equation}
which holds because of the constraints (\ref{3duplBC3}).
\\
\\
We will prove below that this duality can be derived by deconfining the $\SU(2n)$ conjugate symmetric $\tilde S$ using an $\SO(2n)$ gauge theory and then by confining the original $\SU(2n)$ gauge group. Such a procedure will be spelled out after discussing the $\SU(2n+1)$ case with the same superpotential. Indeed, as  we observe below,  the results for the even and for the odd cases can be unified.

%
%
%
%
%
%
\subsection{$\SU(2n+1)$ with $W= \tilde S^{2n} \tilde Q_2^2$}
\label{S2nodd}
%
%
%
%
%
%

Here, we consider the duality between $\SU(2n+1)$ with $2n+2$ fundamentals,
five antifundamentals, a conjugate antisymmetric $\tilde A$, superpotential 
$W = \tilde A^{n-1} \tilde Q_3 \tilde Q_4 \tilde Q_5$, and $\USp(2n)$ with $2n+5$ fundamentals.
We can reduce the duality on $S^1$ by adding the contribution of the KK superpotential, avoiding the generation of an axial symmetry for both the phases.
At the level of the three sphere partition function we have the  identity (\ref{3deffZ2})
which holds provided the validity of  (\ref{3deffBC2}).

We now fix the mass parameters as
\begin{equation}
\label{freezing2}
\nu_2 =  \frac{\tau_{\tilde S}}{2}
\,,\quad
\nu_4 = \frac{\omega_1}{2} + \frac{\tau_{\tilde S}}{2}
\,,\quad
\nu_5 = \frac{\omega_2}{2} + \frac{\tau_{\tilde S}}{2}\,,
\end{equation}
where we are also re-defining $\tau_{\tilde A}$ as  $\tau_{\tilde S}$
in order to have a more direct interpretation of the results after the application of the duplication formula (\ref{duplication}).
Furthermore, for the ease of notation we re-define the leftover free parameter $\nu_3$ using the substitution $\nu_3 \rightarrow \nu_2$. 

In this way we arrive to the relation
\begin{eqnarray}
\label{3dduplic4}
&&
 Z_{\SU(2n+1)}^{[(2n+3)\square ;2\overline{\square} ;1\overline{\text{A}}]} \left(\vec \mu;\omega-\frac{\tau_{\tilde S}}{2};\vec \nu;\tau_{\tilde A} \right)= Z_{\SO(2n+1)}^{[(2n+4)\text{V}]}\left(\frac{\tau_{\tilde S}}{2}+\vec \mu,n \tau_{\tilde S} + \nu_1,\nu_{2}-\frac{\tau_{\tilde S}}{2}\right)
\nonumber
\\
&& \times\,
\Gamma_h \left((2n+1) \tau_{\tilde S} \right)
\prod_{a=1}^{2n+2} \Gamma_h\left(\mu_a + \nu_{1},\sum_{c=1}^{2n+2}  \mu_c- \mu_a\right),
\end{eqnarray}
which holds provided that 
the mass parameters satisfy the relations
\begin{equation}
\label{3duplBC4}
\left(2n+\frac{1}{2}\right) \tau_{\tilde S} + \sum_{a=1}^{N+1}\mu_a+\sum_{b=1}^2 \nu_b = 3 \omega
\,,\quad
2n \tau_{\tilde S} +  2 \nu_2 = 2 \omega\,.
\end{equation}

The first relation is compatible with the constraints enforced by the monopole superpotentials $Y_{\SU(2n-1)}^{(\text{bare})}$, while the second one is compatible with the superpotential deformation $ \tilde S^{2n} \tilde Q_2^2$.

A further constraint on the mass parameters of the $\SU(2n+1)$ fundamentals concerns the $2n+3$-th mass parameter, which is fixed to $\omega-\frac{\tau_{\tilde S}}{2}$. Denoting the associated fields as $Q_{\tilde S}$, the final superpotential of the electric theory is 
\begin{equation}
W = \tilde S^{2n} \tilde Q_2^2 +\tilde S Q_{\tilde S}^2+Y_{\SU(2n-1)}^{(\text{bare})}\,.
\end{equation}
On the dual side, we observe, using the arguments of the hyperbolic Gamma functions, the presence of  three singlets 
$\sigma= \det \tilde S $,
$B = Q^{2n+1}$
and $M = Q \tilde Q_1$, formally identical to the ones obtained for the $\SU(2n)$ case above. Furthermore, by distinguishing the dual  $2n+4$ vectors 
as $2n+2$ fields $\varphi$, one field $u_1$ and one field $u_2$, the constraint on the mass parameters are compatible with the superpotential (\ref{Wdualafterdupldet}).
\\
\\
For this reason we propose the following unified duality holding for generic $N$: 
\begin{center}
\begin{tabular}{|ccc|}
\hline
$\SU(N)$ with $1$  $\overline{\square \!  \square}$ $\tilde S$ & & $\SO(N)$ with $N+1$ $\square$ $\varphi$, $2$ $\square$ $u_{1,2}$  \\
$N+1$ $\square$ $Q$, $1$  $\square$ $Q_{\tilde S} $, $2$ $\overline \square $  $\tilde Q$  & $\leftrightarrow$  & Singlets  $\sigma=\det \tilde S $,
$B = Q^{N}$, $M = Q \tilde Q_1$ \\
$W = \tilde S^{N-1} \tilde Q_2^2 +\tilde S Q_{\tilde S}^2+Y_{\SU(N-2)}^{(\text{bare})}$
&& $W = Y_{\SO(N)}^{+} + \sigma u_2^2 + B u_1 \varphi+\varphi^N M$\\
\hline
\end{tabular}
\end{center}
Below we will provide a physical derivation of such a duality in terms of other fundamental dualities by deconfining the conjugate symmetric tensor in terms of an $\SO(N)$ gauge group and then by confining the $\SU(N)$ gauge group. In this way we will be able to reproduce the expected dual superpotential as well.

%
%
%
%
%
%
\subsection{$\SU(2n)$ with $W= \tilde S^{2n-2} \tilde Q_1^2\tilde Q_2^2$}
\label{S2nm2even}
%
%
%
%
%
%

Here, we consider the duality between $\SU(2n)$ with $2n+1$ fundamentals,
five antifundamentals, a conjugate antisymmetric $\tilde A$, superpotential 
$W = \tilde A^{n-2} \tilde Q_1 \tilde Q_2\tilde Q_3 \tilde Q_4$, and $\USp(2n)$ with $2n+6$ fundamentals.
We can reduce the duality on $S^1$ by adding the contribution of the KK superpotential, avoding the generation of an axial symmetry for both the phases.
At the level of the three sphere partition function we have the following identity 
\begin{eqnarray}
\label{3deffZ5}
&&
Z_{\SU(2n)}^{[(2n+1)\square ;5\overline{\square} ;1 \overline {\text A})} (\vec \mu;\vec \nu;\tau_{\tilde A})= Z_{\USp(2n)}^{[(2n+6) \square]}\left(\frac{\tau_{\tilde A}}{2}+\vec \mu,
\nu_{1,2,3,4}-\frac{\tau_{\tilde A}}{2}
,\left(n-\frac{1}{2}\right) \tau_{\tilde A} +\nu_5\right)
\nonumber 
\\
&&
\times \,
\Gamma_h \left(n \tau_{\tilde A} \right)
\prod_{a=1}^{2n+1}\Gamma_h \left(\sum_{c=1}^{2n+1}  \mu_c- \mu_a,\nu_5+\mu_a\right)
\prod_{1\leq a <b \leq 4} \Gamma_h \left( (n-1) \tau_{\tilde A} + \nu_a + \nu_b \right).
\nonumber 
\\
\end{eqnarray}
This identity holds provided that 
the two relations
\begin{equation}
\label{3deffBC5}
(2n-2) \tau_{\tilde A} + \sum_{a=1}^{2n+1}\mu_a+\sum_{b=1}^5 \nu_b = 4 \omega
\,,\quad
(n-2) \tau_{\tilde A} +  \sum_{a=1}^4 \nu_a = 2 \omega\,,
\end{equation}
hold. The first relation is imposed by the KK monopole superpotential, while the second one is imposed by the tree level superpotential involving the 3d chiral fields.

We now fix the mass parameters as
\begin{equation}
\label{freezing3}
\nu_3 = \frac{\omega_1}{2} + \frac{\tau_{\tilde S}}{2}
\,,\quad
\nu_4= \frac{\omega_2}{2} + \frac{\tau_{\tilde S}}{2}
\,,\quad
\nu_5 =  \frac{\tau_{\tilde S}}{2}\,,
\end{equation}
where we are also re-defining $\tau_{\tilde A}$ as  $\tau_{\tilde S}$
as above.
In this way, after the application of the duplication formula (\ref{duplication}), we arrive to the relation
\begin{eqnarray}
\label{3dduplic5}
&&
Z_{\SU(2n)}^{[(2n+2)\square ;2\overline{\square} ;1 \overline {\text S}]} \left(\vec \mu, \omega-\frac{\tau_{\tilde S}}{2};\vec \nu;\tau_{\tilde S}\right)= Z_{\SO(2n+1)}^{[(2n+4)V]}\left(\frac{\tau_{\tilde S}}{2}+\vec \mu,
\vec \nu-\frac{\tau_{\tilde S}}{2}
,n \tau_{\tilde S} \right)
\nonumber \\
&&
\times \,
\prod_{a=1}^{2}
\Gamma_h\left(\omega-\frac{\tau_{\tilde S}}{2} +\nu_{a}\right) 
\prod_{a,b=1,2} \Gamma_h((2n-1)\tau_{\tilde S}+\nu_{a}+\nu_{b})
\prod_{a=1}^{2n+1} \Gamma_h\left(\sum_{c=1}^{2n+1}  \mu_c- \mu_a\right),
\nonumber \\
\end{eqnarray}
which holds provided that 
the mass parameters satisfy the relations
\begin{equation}
\label{3duplBC5}
\left(2n-\frac{1}{2}\right) \tau_{\tilde S} + \sum_{a=1}^{2n+1}\mu_a+\sum_{b=1}^2 \nu_b = 3 \omega\,,\quad
2(n-1) \tau_{\tilde S} + 2 \sum_{a=1}^2 \nu_a = 2 \omega\,.
\end{equation}
The first relation is compatible with the constraints enforced by the monopole superpotentials $Y_{\SU(2n-2)}^{(\text{bare})}$, while the second one is compatible with the superpotential deformation $\tilde S^{2n-2} \tilde Q_1^2 \tilde Q_2^2$. A further constraint on the mass parameters of the $\SU(2n)$ fundamentals concerns the $2n+2$-th mass parameter, which is fixed to $\omega-\frac{\tau_{\tilde S}}{2}$. Denoting the associated fields as $Q_{\tilde S}$, the final superpotential of the electric theory is 
\begin{equation}
W = \tilde S^{2n-2} \tilde Q_1^2 \tilde Q_2^2 +\tilde S Q_{\tilde S}^2+Y_{\SU(2n-2)}^{(\text{bare})}\,.
\end{equation}
On the dual side, we observe, using the arguments of the hyperbolic Gamma functions, the presence of  three singlets 
$\gamma = Q_{\tilde S} \tilde Q$, 
$X = \tilde S^{2n-1} \tilde Q^2$ and
$B = Q^{2n}$. Furthermore, by distinguishing the dual  $2n+4$ vectors 
as $2n+1$ fields $\varphi$, two fields $u$ and one field $t$, the constraint on the mass parameters are compatible with a superpotential
\begin{equation}
\label{WdualafterduplS2nm2}
W = Y_{\SO(2n+1)}^{+}+ B \varphi t  + 
X u^2 +  \gamma u  t+\varphi^{2n+1}\,.
\end{equation}
Observe that the linear monopole superpotentials is required to enforce the constraints on the mass parameters of the  $2n+4$ vectors $\varphi$, $u$ and $t$
\begin{equation}
\sum_{a=1}^2 m_{u_a} + m_t  + \sum_{i=1}^{2n+1} m_{\varphi_i} =3 \omega\,,
\end{equation}
which holds because of the constraints (\ref{3duplBC5}) 
\\
\\
We will prove below that this duality can be derived by deconfining the $\SU(2n)$ conjugate symmetric $\tilde S$ in terms of an $\SO(2n-1)$ gauge theory and then by confining the original $\SU(2n)$ gauge group. Such a procedure will be spelled out after discussing the $\SU(2n+1)$ case with the same superpotential. Indeed, as  we observe below,  the results for the even and for the odd cases can be unified.

%
%
%
%
%
%
\subsection{$\SU(2n+1)$ with $W= \tilde S^{2n-1}  \tilde Q_1^2 \tilde Q_2^2$}
\label{S2nm1odd}
%
%
%
%
%
%

Here we consider the duality between $\SU(2n+1)$ with $2n+2$ fundamentals,
five antifundamentals, a conjugate antisymmetric $\tilde A$, superpotential 
$W = \tilde A^{n-1} \tilde Q^55$, and $\USp(2n+2)$ with $2n+8$ fundamentals.
We can reduce the duality on $S^1$ by adding the contribution of the KK superpotential, avoiding the generation of an axial symmetry for both the phases.
At the level of the three sphere partition function we have the  identity 

\begin{eqnarray}
\label{3deffZ6}
&&
Z_{\SU(2n+1)}^{[(2n+2) \square ;5 \overline \square ;1 \overline{\text A}]} (\vec \mu;\vec \nu;\tau_{\tilde A})= Z_{\USp(2n+2)}^{[(2n+8)\square]}\left(\frac{\tau_{\tilde A}}{2}+\vec \mu,
\vec \nu-\frac{\tau_{\tilde A}}{2}
,\left(n+\frac{1}{2}\right) \tau_{\tilde A} \right)
\nonumber 
\\
&& \times \,
\prod_{a=1}^{2n+1}\Gamma_h \left(\sum_{c=1}^{2n+1}  \mu_c- \mu_a\right)
\prod_{1\leq a <b<c \leq 5} \Gamma_h \left( (n-1)\tau_{\tilde A} +\nu_a +\nu_b +\nu_c  \right),\nonumber 
\\
\end{eqnarray}
which holds provided that 
the two relations
\begin{equation}
\label{3deffBC6}
(2n-1) \tau_{\tilde A} + \sum_{a=1}^{2n+2}\mu_a+\sum_{b=1}^5 \nu_b = 4 \omega
\,,\quad
(n-2) \tau_{\tilde A} +\sum_{b=1}^5 \nu_b  = 2 \omega\,,
\end{equation}
hold. The first relation is imposed by the KK monopole superpotential, while the second one is imposed by the tree level superpotential involving the 3d chiral fields.
We now fix the mass parameters as
in (\ref{freezing2}) and, after the application of the duplication formula (\ref{duplication}), we arrive at
\begin{equation}
    \label{3dduplic6}
    \begin{aligned}
        &
        Z_{\SU(2n)}^{[(2n+3)\square ;2\overline \square;1 \overline{\text S}]} (\vec \mu, \omega-\frac{\tau_{\tilde S}}{2};\vec \nu;\tau_{\tilde S})
        =
        Z_{\SO(2n+1)}^{[(2n+5) \text{V}]}\left(\frac{\tau_{\tilde S}}{2}+\vec \mu,\vec \nu-\frac{\tau_{\tilde S}}{2},\left(n+\frac{1}{2} \right)\tau_{\tilde S}\right) \\
        &
        \times \,\prod_{a=1}^{2}\Gamma_h\left(\omega-\frac{\tau_{\tilde S}}{2} +\nu_{a} \right) \prod_{a,b=1,2} \Gamma_h\left(2n\tau_{\tilde S}+\nu_{a}+\nu_{b}\right)\prod_{a=1}^{2n+2} \Gamma_h\left(\sum_{c=1}^{2n+2}  \mu_c- \mu_a\right).
    \end{aligned}
\end{equation}
This identity holds provided that 
the mass parameters satisfy the relations
\begin{equation}
\label{3duplBC6}
\left(2n+\frac{1}{2}\right) \tau_{\tilde S} + \sum_{a=1}^{N+1}\mu_a+\sum_{b=1}^2 \nu_b = 3 \omega
\,,\quad
(2n-1) \tau_{\tilde S} +  2\sum_{a=1}^2 \nu_a = 2 \omega\,.
\end{equation}

The first relation is compatible with the constraints enforced by the monopole superpotentials $Y_{\SU(2n-1)}^{(\text{bare})}$, while the second one is compatible with the superpotential deformation $ \tilde S^{2n-1} \tilde Q_1^2 \tilde Q_2^2$.

A further constraint on the mass parameters of the $\SU(2n+1)$ fundamentals concerns the $2n+3$-th mass parameter, which is fixed to $\omega-\frac{\tau_{\tilde S}}{2}$. Denoting the associated fields as $Q_{\tilde S}$ the final superpotential of the electric theory is 
\begin{equation}
W = \tilde S^{2n-1} \tilde Q_1^2 \tilde Q_2^2+\tilde S Q_{\tilde S}^2+Y_{\SU(2n-1)}^{(\text{bare})}\,.
\end{equation}
On the dual side, we observe, using the arguments of the hyperbolic Gamma functions, the presence of  three singlets 
$\gamma = Q_{\tilde S} \tilde Q$, 
$X = \tilde S^{2n} \tilde Q^2$ and
$B = Q^{2n+1}$, formally identical to the ones obtained for the $\SU(2n)$ case above. 

Furthermore, by distinguishing the dual  $2n+5$ vectors 
as $2n+2$ fields $\varphi$, two fields $u$ and one field $t$, the constraint on the mass parameters are compatible with the superpotential
\begin{equation}
\label{WdualafterduplS2nm2bis}
W = Y_{\SO(2n+2)}^{+}+ B \varphi t  + 
X u^2 +  \gamma u  t+\varphi^{2n+2}\,.
\end{equation}
For this reason, we propose the following unified duality holding for generic $N$: 
\vspace{-1em}
\begin{center}
\scalebox{0.95}{
\begin{tabular}{|ccc|}
\hline
$\SU(N)$ with $1$  $\overline{\square \!  \square}$ $\tilde S$ & & $\SO(N+1)$ with $N+1$ $\square$ $\varphi$, $2$ $\square$ $u$, $1$ $\square$ $t$ \\
$N+1$ $\square$ $Q$, $1$  $\square$ $Q_{\tilde S} $, $2$ $\overline \square $  $\tilde Q$  & $\leftrightarrow$  & Singlets  $\gamma = Q_{\tilde S} \tilde Q$, 
$X = \tilde S^{N-1} \tilde Q^2$,
$B = Q^{N}$ \\
$ W = \tilde S^{N-2} \tilde Q_1^2 \tilde Q_2^2 +\tilde S Q_{\tilde S}^2+Y_{\SU(N-2)}^{(\text{bare})}
$
&& $W = Y_{\SO(N+1)}^{+}+ B \varphi t  + 
X u^2 +  \gamma u  t+\varphi^{N+1}$\\
\hline
\end{tabular}
}
\end{center}
Below we will provide a physical derivation of such a duality in terms of other fundamental dualities, by deconfining the conjugate symmetric tensor in terms of an $\SO(N+1)$ gauge group and then by confining the $\SU(N)$ gauge group. In this way we will be able to reproduce the expected dual superpotential as well.

%
%
%
%
%
%
\subsection{Proving the dualities through tensor deconfinement}
\label{decproofs}
%
%
%
%
%
%

We conclude the analysis by providing a proof of the dualities proposed via the application of the duplication formula above, in terms of more fundamental dualities that do not involve two index tensor matter fields.
The proof is based on the deconfinement of the  $\SU(N)$ (conjugate)  symmetric 
tensor $\tilde S$ in terms of a special orthogonal gauge group.
The basic confining duality, discussed originally in \cite{Benvenuti:2021nwt} relates an $\SO(N)$ gauge group with $N+1$ vectors $v$ and monopole superpotential $W = Y_{\SO(N)}^+$ to a WZ model, with a symmetric meson $S = v^2$ and $N$ baryons $q=v^N$ with superpotential $W = \det S + S q^2$.

In the following, we will use this confining duality in order to deconfine the $\SU(N)$
in the three cases proposed above. Such cases have the same field content but different superpotential.

\begin{itemize}
\item \underline{Case I: $W_{\text{ele}} = \det \tilde S  +\tilde S Q_{\tilde S}^2+Y_{\SU(N-2)}^{(\text{bare})} $}

\begin{figure}[h!]
    \centering
        \begin{minipage}{0.45\textwidth}
            \centering
            \makebox[\textwidth][c]{
            \begin{tikzpicture}[
                every node/.style={font=\footnotesize},
                box/.style={rectangle, draw, thick},
                box2/.style={rectangle, draw, dashed}
            ]
            \pgfmathsetmacro{\x}{2}
            \pgfmathsetmacro{\y}{1.5}
            \node[fill=SUcol,circle,draw,thick] (su) at (0,0) {$N$};
            \node[fill=SOcol,circle,draw,thick] (so) at (\x,0) {$N\!-\!1$};
            \node[box] (fond) at (-\x,0) {$N\!+\!1$};
            \node[box] (afond) at (0,\y) {$2$};
            \draw[<-,thick,>=stealth] (su) -- node[above] {$\tilde{P}$} (so);
            \draw[<-,thick,>=stealth] (su) -- node[left] {$\tilde Q$} (afond);
            \draw[->,thick,>=stealth] (su) -- node[above] {$Q$} (fond);
            \end{tikzpicture}
            }
        \end{minipage}
        \begin{minipage}{0.45\textwidth}
            \centering
            \makebox[\textwidth][c]{
            \begin{tikzpicture}[
                every node/.style={font=\footnotesize},
                box/.style={rectangle, draw, thick},
                box2/.style={rectangle, draw, dashed}
            ]
            \pgfmathsetmacro{\x}{2}
            \pgfmathsetmacro{\y}{1.5}
            \node[fill=SOcol,circle,draw,thick] (so) at (0,0) {$N\!-\!1$};
            \node[box] (fond) at (-\x,0) {$N\!+\!1$};
            \node[box] (sing) at (\x,0) {$1$};
            \draw[-,thick,>=stealth] (so) -- node[above] {$u$} (sing);
            \draw[-,thick,>=stealth] (so) -- node[above] {$\varphi$} (fond);
            \end{tikzpicture}
            }
        \end{minipage}
    \caption{The first quiver represents the charged matter content after we deconfined the conjugate symmetric tensor in presence of the electric deformation $W \supset \det \tilde{S}$. The second quiver is obtained after confining the $\SU(N)$ gauge node.}
    \label{figsymm1}
\end{figure}
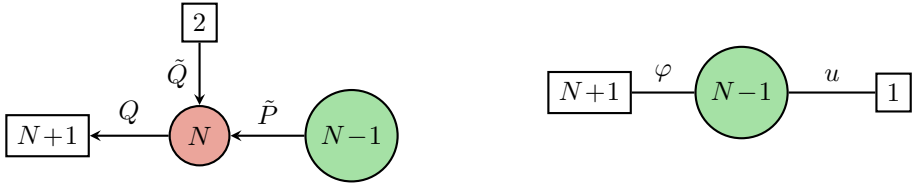

We start by deconfining the conjugate symmetric $\tilde S$ using an $\SO(N-1)$ gauge group. In this way we obtain an $\SU(N) \times \SO(N-1)$ quiver gauge theory, where the charged matter content is represented in the first quiver in Figure \ref{figsymm1}. The superpotential for this theory is
\begin{equation}
W = Y_{\SO(N-1)}^+ 
+ Y_{\SU(N)} \,.
\end{equation}
Then  we observe that the $\SU(N)$ gauge group is confining. The 
mesons of this confining duality are 
$\varphi =Q \tilde P$ and $M = Q \tilde Q $. There are a 
baryon $B =  Q^{N}$ and two anti baryons $u = \tilde P^{N-2} \tilde Q^{2}$
and $L= \tilde P^{N-1} \tilde Q$.
The meson $\varphi$ and the antibaryon $u$ are vectors of the leftover $\SO(N-1)$ 
gauge group, as stressed in the second quiver of Figure \ref{figsymm1}, while the fields $B$, $M$, and $L$ are singlets.
Observe that, by following the duality map,  the field $L$ is associated to the $\SU(N)$ gauge invariant combination $\tilde Q Q_{\tilde S}$, indeed the field $\tilde Q_S$ 
corresponds to the baryon $ \tilde P^{N-1} $.
This mapping is consistent with the constraint enforced by the monopole superpotential, $N/2 \, \tau_{\tilde S} = \omega$, which gives $m_{\tilde Q_S} = (N-1)\tau_{\tilde S}/2  =    \omega-\tau_{\tilde S}/2$.
In this way, we arrive to the final superpotential expected for this duality 
\begin{equation}
W = Y_{\SO(N-1)}^{+} + B u \varphi+\varphi^{N-1} M^2+B L M\,.
\end{equation}

\item  \underline{Case II: $W_{\text{ele}} =   \tilde S^{N-1} \tilde Q_2^2  +\tilde S Q_{\tilde S}^2+Y_{\SU(N-2)}^{(\text{bare})} $ }

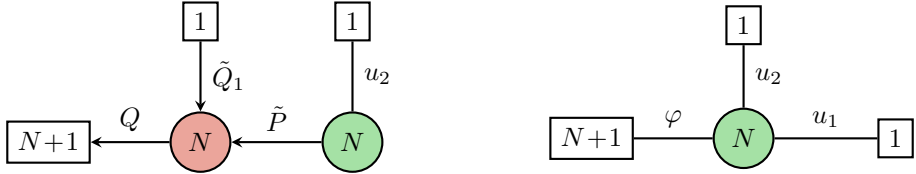
\begin{figure}[h!]
    \centering
        \begin{minipage}{0.45\textwidth}
            \centering
            \makebox[\textwidth][c]{
            \begin{tikzpicture}[
                every node/.style={font=\footnotesize},
                box/.style={rectangle, draw, thick},
                box2/.style={rectangle, draw, dashed}
            ]
            \pgfmathsetmacro{\x}{2}
            \pgfmathsetmacro{\y}{1.6}
            \node[fill=SUcol,circle,draw,thick] (su) at (0,0) {$N$};
            \node[fill=SOcol,circle,draw,thick] (so) at (\x,0) {$N$};
            \node[box] (fond) at (-\x,0) {$N\!+\!1$};
            \node[box] (afond) at (0,\y) {$1$};
            \node[box] (sing) at (\x,\y) {$1$};
            \draw[<-,thick,>=stealth] (su) -- node[above] {$\tilde{P}$} (so);
            \draw[<-,thick,>=stealth] (su) -- node[right] {$\tilde Q_1$} (afond);
            \draw[->,thick,>=stealth] (su) -- node[above] {$Q$} (fond);
            \draw[-,thick,>=stealth] (so) -- node[right] {$u_2$} (sing);
            \end{tikzpicture}
            }
        \end{minipage}
        \begin{minipage}{0.45\textwidth}
            \centering
            \makebox[\textwidth][c]{
            \begin{tikzpicture}[
                every node/.style={font=\footnotesize},
                box/.style={rectangle, draw, thick},
                box2/.style={rectangle, draw, dashed}
            ]
            \pgfmathsetmacro{\x}{2}
            \pgfmathsetmacro{\y}{1.5}
            \node[fill=SOcol,circle,draw,thick] (so) at (0,0) {$N$};
            \node[box] (fond) at (-\x,0) {$N\!+\!1$};
            \node[box] (sing) at (\x,0) {$1$};
            \node[box] (sing2) at (0,\y) {$1$};
            \draw[-,thick,>=stealth] (so) -- node[above] {$u_1$} (sing);
            \draw[-,thick,>=stealth] (so) -- node[above] {$\varphi$} (fond);
            \draw[-,thick,>=stealth] (so) -- node[right] {$u_2$} (sing2);
            \end{tikzpicture}
            }
        \end{minipage}
    \caption{The first quiver represents the charged matter content after we deconfined the conjugate symmetric tensor in presence of the electric deformation $W \supset \tilde{S}^{N-1}\tilde{Q}^2_2$. The second quiver is obtained after confining the $\SU(N)$ gauge node.}
    \label{figsymm2}
\end{figure}

In this case, we 
deconfine the conjugate symmetric $\tilde S$ using an $\SO(N)$ gauge group. We therefore obtain an $\SU(N) \times \SO(N)$ quiver gauge theory, where the charged matter content is represented in the first quiver in Figure \ref{figsymm2}. The superpotential for this theory is
\begin{equation}
W = Y_{\SO(N)}^+ 
+ Y_{\SU(N)} +
\sigma u_2^2 + \tilde \gamma \tilde P^{N}\,,
\end{equation}
where the singlet $\sigma$ is dual to the gauge invariant operator $\det S$.
Such a mapping can be proven by confining back the $\SO(N)$ gauge group and by integrating out the massive fields.

Then  we observe that the $\SU(N)$ gauge group is confining. The 
mesons of this confining duality are 
$\varphi =Q \tilde P$ and $M = Q \tilde Q_1 $. There are a 
baryon $B =  Q^{N}$ and two anti.baryons $u_1 = \tilde P^{N-2} \tilde Q^{2}$
and $B_1= \tilde P^{N} $.
The meson $\varphi$ and the antibaryon $u_1$ are vectors of the leftover $\SO(N)$ 
gauge group as stressed in the second quiver of Figure \ref{figsymm1}, while the fields $B$, $M$, and $\sigma$ are singlets.
The fields $B_1$ and $\gamma$ are massive and we integrate them out.

In this way, we arrive to the final superpotential expected for this duality 
\begin{equation}
W = Y_{\SO(N)}^{+} + \sigma u_2^2 + B u_1 \varphi+\varphi^N M\,.
\end{equation}

\item Case III: $W_{\text{ele}} = \tilde S^{N-2} \tilde Q_1^2 \tilde Q_2^2   +\tilde S Q_{\tilde S}^2+Y_{\SU(N-2)}^{(\text{bare})} $ 
\end{itemize}

\begin{figure}[h!]
    \centering
        \begin{minipage}{0.45\textwidth}
            \centering
            \makebox[\textwidth][c]{
            \begin{tikzpicture}[
                every node/.style={font=\footnotesize},
                box/.style={rectangle, draw, thick},
                box2/.style={rectangle, draw, dashed}
            ]
            \pgfmathsetmacro{\x}{2}
            \pgfmathsetmacro{\y}{1.6}
            \node[fill=SUcol,circle,draw,thick] (su) at (0,0) {$N$};
            \node[fill=SOcol,circle,draw,thick] (so) at (\x,0) {$N\!+\!1$};
            \node[box] (fond) at (-\x,0) {$N\!+\!1$};
            \node[box] (sing) at (\x,\y) {$2$};
            \draw[<-,thick,>=stealth] (su) -- node[above] {$\tilde{P}$} (so);
            \draw[->,thick,>=stealth] (su) -- node[above] {$Q$} (fond);
            \draw[-,thick,>=stealth] (so) -- node[right] {$u$} (sing);
            \end{tikzpicture}
            }
        \end{minipage}
        \begin{minipage}{0.45\textwidth}
            \centering
            \makebox[\textwidth][c]{
            \begin{tikzpicture}[
                every node/.style={font=\footnotesize},
                box/.style={rectangle, draw, thick},
                box2/.style={rectangle, draw, dashed}
            ]
            \pgfmathsetmacro{\x}{2}
            \pgfmathsetmacro{\y}{1.5}
            \node[fill=SOcol,circle,draw,thick] (so) at (0,0) {$N\!+\!1$};
            \node[box] (fond) at (-\x,0) {$N\!+\!1$};
            \node[box] (sing) at (\x,0) {$2$};
            \node[box] (sing2) at (0,\y) {$1$};
            \draw[-,thick,>=stealth] (so) -- node[above] {$u$} (sing);
            \draw[-,thick,>=stealth] (so) -- node[above] {$\varphi$} (fond);
            \draw[-,thick,>=stealth] (so) -- node[right] {$t$} (sing2);
            \end{tikzpicture}
            }
        \end{minipage}
    \caption{The first quiver represents the charged matter content after we deconfined the conjugate symmetric tensor in presence of the electric deformation $W \supset \tilde{S}^{N-2}\tilde{Q}^2_1\tilde{Q}^2_2$. The second quiver is obtained after confining the $\SU(N)$ gauge node.}
    \label{figsymm3}
\end{figure}
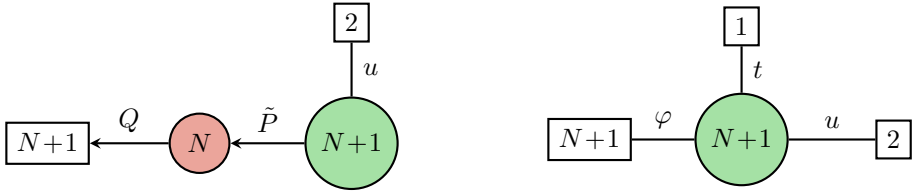

In this last case, we 
deconfine the conjugate symmetric $\tilde S$ using an $\SO(N+1)$ gauge group. We therefore obtain an $\SU(N) \times \SO(N+1)$ quiver gauge theory, where the charged matter content is represented in the first quiver in Figure \ref{figsymm3}. The superpotential for this theory is
\begin{equation}
W = Y_{\SO(N+1)}^+ 
+ Y_{\SU(N)} +
X u^2 + \tilde \gamma \tilde P^{N} u\,,
\end{equation}
where the singlets $X$ and $\tilde \gamma$ are  dual to the gauge invariant operators $X = S^{2n} \tilde Q^2$ and $\tilde Q Q_{\tilde S}$ respectively.

Then  we observe that the $\SU(N)$ gauge group is confining. The 
meson of this confining duality is 
$\varphi =Q \tilde P$. There are a 
baryon $B =  Q^{N}$ and an antibaryon  $t= \tilde P^{N} $.
The meson $\varphi$ and the antibaryon $t$ are vectors of the leftover $\SO(N+1)$ 
gauge group as stressed in the second quiver of Figure \ref{figsymm3}, while the fields $B$, $X$, and $\tilde \gamma$ are singlets.

In this way, we arrive to the final superpotential expected for this duality 
\begin{equation}
W = Y_{\SO(N+1)}^{+} + B \varphi t  + 
X u^2 +  \gamma u  t+\varphi^{N+1}\,.
\end{equation}
\\
\\
We conclude this section with a comment on the relation between the results found here and the reduction of the $\USp(2)$ duals discussed in Section \ref{sec:4d}.
The dual $\SO(N)$ theories found here can be further dualized to $\SO(3)$ gauge theories using the results 
of \cite{Aharony:2013kma}. The same result is obtained by considering the 4d $\USp(2)$ dual gauge theories on $S^1$,  freezing the mass parameters, following the duality dictionary and finally applying the duplication formula \eqref{duplication}.

\section{Conclusions}
\label{Sec:Conclusions}

In this paper we have studied dualities in various dimensions between special unitary and symplectic or orthogonal gauge groups. The dualities are obtained starting from 4d $\mathcal{N}=1$ $\SU(N)$ gauge theories with a conjugate antisymmetric, $N+1$ fundamentals and five antifundamentals. In the absence of superpotential, the gauge theory is dual to an $\USp(2) \times \SU(2)$ quiver and it has been shown to be in a mixed phase. We showed that baryonic deformations lead to an $\USp(2)$  magnetic free phase. The derivation has allowed to find intermediate phases corresponding to  $\USp(2M)$ SQCD with $2M+2$ fundamentals. We showed that the duality between the original theories and these last ones can be reduced to 2d, and we surveyed the various 2d dualities that originate from the 4d ones, mostly focusing on a single case, corresponding to $\SU(2n)$ with the deformation \eqref{WEevenDef2}.
We also studied the reduction to 3d, finding, upon the application of the duplication formula for the hyperbolic Gamma function, dualities between $\SU(N)$ with a symmetric tensor and monopole superpotential and $\SO(N)$ SQCD, that can be further shown to be dual to $\SO(3)$ with vectors.

Various generalizations and further studies are possible.
A first extension to our analysis consists of finding other examples of dualities between different types of gauge groups, similarly to other examples discussed in the literature (see e.g. \cite{Nii:2018bgf,Amariti:2025zgj}).
When reducing to 3d, we have found examples of dualities between $\SU(N)$ and $\SO(N)$ thanks to the application of the duplication formula. 
Similar examples of this kind have not been found here in 4d because of the limited amount of flavors. One possible extension would be to  
increase the amount of fundamentals and antifundamentals in order to search for 4d dualities between $\SU(N)$ and $\SO(N)$.
Note also that  mixed phases exist for more flavors as well \cite{Csaki:2004uj}, and it would be interesting to understand  the effect of the baryonic deformations considered here on such phases.
 One could also deform other examples of models with mixed phases, such
as those studied in \cite{Barnes:2005zn} with two gauge nodes.
 A final comment concerns the 2d dualities studied here. We found a 4d origin of a 2d duality already discussed in \cite{Jiang:2024ifv}, consisting of $\SU(N)$ 
with a conjugate antisymmetric, $N+1$ fundamentals and one antifundamental.
On the other hand, we did not find a 4d origin for a gauge/LG duality for $\SU(N)$ with $N+2$ fundamentals and a conjugate antisymmetric. 
Again, increasing the number of flavors in the 4d analysis is a necessary step in order to find a possible 4d explanation of this 2d duality.

\section*{Acknowledgments}
This work  has been supported in part by the Italian Ministero dell'Istruzione, Università e Ricerca (MIUR), in part by Istituto Nazionale di Fisica Nucleare (INFN) through the “Gauge Theories, Strings, Supergravity” (GSS) research project.

\bibliographystyle{JHEP}
\bibliography{ref.bib}
\end{document}